\documentclass[twocolumn]{aastex63}
\usepackage{graphicx}	
\usepackage{amsmath}	
\usepackage{amssymb}	
\usepackage{physics}
\usepackage{enumitem}
\usepackage{bbm}
\usepackage{color}
\usepackage{natbib}
\usepackage{gensymb}

\newcommand{\delSSFR}{$\Delta$\,SSFR\,}
\newcommand{\MI}{\mathrm{MI}}

\begin{document}

\title{What is Important? Morphological Asymmetries are Useful Predictors of Star Formation Rates of Star-forming Galaxies in SDSS Stripe 82}
\shortauthors{Yesuf, Ho, \& Faber}
\author{Hassen M. Yesuf}
\affiliation{Kavli Institute for Astronomy and Astrophysics, Peking University, Beijing 100871, China}
\affiliation{Kavli Institute for the Physics and Mathematics of the Universe, The University of Tokyo, Kashiwa, Japan 277-8583}
\author{Luis C. Ho}
\affiliation{Kavli Institute for Astronomy and Astrophysics, Peking University, Beijing 100871, China}
\affiliation{Department of Astronomy, School of Physics, Peking University, 
Beijing 100871, China}
\author{S.M. Faber}
\affiliation{UCO/Lick Observatory, Department of Astronomy and Astrophysics, University of California, Santa Cruz, CA 95064 USA}


\begin{abstract}

Morphology and structure of galaxies reflect their star formation and assembly histories. We use the framework of mutual information ($\MI$) to quantify interdependence among several structural variables and to rank them according to their relevance for predicting specific star formation rate (SSFR) by comparing the $\MI$ of the predictor variables with SSFR and penalizing variables that are redundant. We apply this framework to study $\sim 3,700$ face-on star-forming galaxies (SFGs) with varying degrees of bulge dominance and central concentration and with stellar mass $M_\star \approx 10^9\,M_\odot - 5\times 10^{11}\,M_\odot$ at redshift $z = 0.02-0.12$. We use the Sloan Digital Sky Survey (SDSS) Stripe 82 deep $i$-band imaging data, which improve measurements of asymmetry and bulge dominance indicators. We find that star-forming galaxies are a multi-parameter family. In addition to $M_\star$, asymmetry emerges as the most powerful predictor of SSFR residuals of SFGs, followed by bulge prominence/concentration. Star-forming galaxies with higher asymmetry and stronger bulges have higher SSFR at a given $M_\star$. The asymmetry reflects both irregular spiral arms and lopsidedness in seemingly isolated SFGs and structural perturbations by galaxy interactions or mergers.

\end{abstract}

\section{INTRODUCTION}

What are the fundamental sets of variables that independently describe galaxies? We still do not know. To give an analogy, for a single star, its mass, radius, luminosity, chemical composition, and rotation are fundamental variables. Stellar structure and evolution models describe the properties of a star at different stages, from its formation on and its evolution off the main sequence to its final demise as a stellar remnant. The models use the star's initial mass and composition as inputs and its luminosity and effective temperature (Hertzsprung-Russell (HR) diagram) as constraints. The theories of galaxy formation and evolution are not as mature as those of stellar evolution, and an ordering of input and output variables as clean as that for the star-forming galaxies has yet to be identified. 

On the observation side, analogous to the HR-diagram, the sets of galaxy properties that are strongly constraining and highly informative for developing models are also unclear. Significant progress, however, has been made in last few decades in understanding galaxies \citep[for recent reviews, see][]{Kennicutt+12,Kormendy+13,Conselice+14,Somerville+15,Naab+17,Wechsler+18,Sanchez+20,Tacconi+20}. We know that galaxies are born in dark matter halos; it is likely that some of the observable properties of galaxies will characterize the properties of halos, while others may reflect only the baryonic physics within halos. For example, the stellar mass ($M_\star$) is thought to be closely related to the halo mass \citep[$M_h$;][]{Wechsler+18}. One of the discoveries that improved our understanding of galaxies is the tight relationship (scatter of $\sim 0.3-0.4$ dex) between $M_\star$ and star formation rate \citep[SFR;][]{Brinchmann+04,Daddi+07,Elbaz+07,Noeske+07,Elbaz+11,Speagle+14}. By analogy to that of stars, this relation is called the star-forming main sequence (SFMS). It is often interpreted as indicating that galaxies exhibit a self-regulated quasi-equilibrium between external gas accretion, star formation, and gas outflow \citep{Bouche+10,Dave+11,Lilly+13, Forbes+14, Peng+14,Rodriguez-Puebla+16, Belfiore+19}. Just as scatter about the Zero Age Main Sequence (ZAMS) for stars revealed the existence of additional parameters for stars besides mass, so might an analysis of scatter about the SFMS reveal the existence of additional parameters besides $M_\star$ for galaxies. This hope is what motivates this investigation.

Galaxy size, which is often quantified by the half-light radius, $R_{50}$, has been proposed as the second important variable after $M_\star$ \citep{Omand+14,Chen+20, Lin+20a}. Several papers have studied the evolution of stellar population properties in the $R_{50}-M_\star$ plane \citep{Shen+03,vanderWel+09, Cappellari+13, Scott+17, Li+18, Lange+16, DEugenio+18, Barone+20}. For example, \citet{Omand+14} studied the trends of quiescent fraction on $R_{50}-M_\star$ plane, and showed that some SFGs coexist with quenched galaxies at the same $M_\star$. But the scatter about the mean $R_{50}-M_\star$ relation was found to have only a weak correlation with the residuals of the SFMS \citep{Lin+20a}. Furthermore, \citet{Barone+20} showed that the stellar ages and metallicities ($Z$) of SFGs depend additionally on $R_{50}$; the $M_\star-Z$ relation is well-known \citep[e.g.,][]{Tremonti+04,Gallazzi+05}. In particular, \citet{Barone+20} found that the stellar age correlates best with stellar surface mass density $\mu \propto M_\star/R_{50}^2$, and $Z$ correlates best with $M_\star/R_{50}$, which is proportional to the gravitational potential or the velocity dispersion ($\propto \sigma^2$). In addition, \citet{Chen+20} showed that $R_{50}-M_\star$ plane maps to $\Sigma_1-M_\star$ plane, where $\Sigma_1$ is the central mass density within 1\,kpc. $\Sigma_1$ is strongly correlated with $\sigma$ \citep{Fang+13, Yesuf+20a}. \citet{Chen+20} advocated that $R_{50}$ is an important second parameter in shaping the life histories of galaxies and the masses of their black holes. SFGs with larger radii at fixed $M_\star$ are expected to have lower black hole masses due to lower central densities \citep{Vandenbosch+16}. In NIHAO\footnote{Numerical Investigation of a Hundred Astrophysical Objects} galaxy simulations, just as $M_h$ maps onto $M_\star$, the halo viral radius ($R_\mathrm{vir}$) and halo concentration ($C_h$) map of galaxy radius $R_{50} \propto R_\mathrm{vir}\,C_h^{-0.7}$ \citep{Jiang+19}. The dependence on $C_h$ arises because higher $C_h$ halos accrete a larger fraction of their mass early thereby forming smaller and denser galaxies at the centers of halos. 

\subsection{Theoretical Views of the SFMS Residuals}

The residuals of the SFMS are thought to be linked to halo formation time and mass accretion rate, gas flows, burstiness of star formation, AGN feedback, or/and satellite quenching processes \citep{Lilly+13,Dekel+14a,Rodriguez-Puebla+16, Tacchella+16, Behroozi+19, Matthee+19}. The evolution timescales of galaxies along and across the SFMS are not well-constrained. Both short \citep[$\lesssim 1$\,Gyr;][]{Tacchella+16,Torrey+18,Caplar+19, WangLilly20a,WangLilly20b} and long \citep{Gladders+13, Abramson+16,Dressler+16, Matthee+19, Iyer+20} evolution timescales have been proposed.

Using zoom-in cosmological VELA simulations of massive galaxies at $z>1$, \citet{Tacchella+16} found that the high-redshift SFGs oscillate about the ridge SFMS on timescales $0.4\times t_H$, where $t_H$ is the Hubble time. The excursion upward is due to gas compaction, triggered possibly by minor mergers, counter-rotating streams, and/or violent disk instabilities. The motion downward is driven by central gas depletion by star formation and outflows. Simulated galaxies with \delSSFR $> 0$ tend to be compact blue nuggets with high gas fractions and short depletion times while those with \delSSFR $< 0$ have lower gas fractions and longer depletion times \citep{Tacchella+16}. In agreement with \citet{Tacchella+16}'s findings, observations show that \delSSFR depends on gas fraction and depletion time \citep[e.g.,][]{Genzel+15,Saintonge+17,Tacconi+18,Janowiecki+20,Ellison+20b,Morselli+20,Wang+20}.

Using EAGLE cosmological simulations, \citet{Matthee+19} found that the SFMS scatter at $z = 0.1$ originates from a combination of fluctuations on short ($< 0.2-2$\,Gyr) and long ($\sim 10$\,Gyr) timescales \citep[see also][]{Iyer+20,WangLilly20a,WangLilly20b}. Most of the SFMS scatter in the EAGLE simulations is driven by the long timescale fluctuations related to halo mass and halo formation time (i.e., assembly bias). For galaxies with $M_\star < 10^{10}\,M_\odot$, there is a clear trend that haloes that form later tend to host galaxies with higher SSFRs. While individual galaxies likely cross the whole SFMS multiple times during their evolution, distinct populations of SFGs fluctuate around median tracks associated with their halo properties such that those above/below the SFMS tend to be above/below the main sequence for much longer than 1\,Gyr (see Figure 5 in \citet{Matthee+19}). These authors also showed that the fluctuations on short timescales may be associated with self-regulation of star formation (feedback). At $M_\star \gtrsim 10^{10}\,M_\odot$, the black hole formation efficiency ($M_\mathrm{BH}/M_h$) was shown to correlate with \delSSFR. \citet{WangLilly20a,WangLilly20b} deduced that SFR fluctuations on timescale $< 1$\,Gyr may be due to variations in gas accretion rates.

\citet{Gladders+13} showed that SFHs of observed galaxies are well-characterized by a log-normal function in time, implying a slow evolution. \citet{Diemer+17} explored the connection between the SFHs, parameterized by the log-normal distribution, and several galaxy parameters such as halo mass and environment in Illustris simulations. Although the log-normal form ignores short-lived features such as starbursts and rapid quenching, the authors found that it fits very well the SFHs of majority of the simulated galaxies, and reproduces well how they evolve in $M_\star$-SSFR space. But it was a poor fit to galaxies that rapidly quenched after becoming satellites. Moreover, the authors found that SFHs in Illustris simulations are diverse, and are not determined by a single physical property of galaxies. The scatter between galaxy properties (e.g., halo mass and environment) and the log-normal (SFH) parameters is large. Nevertheless, the formation history of the halo has the strongest influence on peak time and width of the SFH, as also found in other work \citep[e.g.,][]{Behroozi+19}. \citet{Diemer+17}'s findings indicate that, at fixed $M_\star$, earlier-forming galaxies live in early forming halos, dense environments, and massive halos, host massive black holes, and have small galaxy sizes. Galaxies with high halo mass exhibit a short period of very intense star formation and then quench, whereas lower-mass haloes experience much more extended SFHs.

\subsection{Limited Observational Insights of the SFMS Residuals}

Observationally, the multivariate dependence of SSFR on variables other than $M_\star$ for SFGs has only been systematically quantified in few studies \citep[e.g.,][]{Reichard+09,Cibinel+19,Berti+20}. Furthermore, some of the few attempts have failed to find additional dependence, for example, on radius or environment \citep[e.g.,][]{Peng+10,Fang+18, Lin+20a}. However, the SSFR is correlated with several structural parameters when considering star-forming and quiescent galaxies together \citep[e.g.,][]{Strateva+01,Kauffmann+03b, Brinchmann+04, Franx+08, Wuyts+11,Cheung+12,Fang+13,Wake+12,Bell+12, Bluck+20,Yesuf+20a}. A clear picture that emerged from these studies is that $M_\star$ does not adequately predict the diversity of SFR in galaxies as a whole; how the mass is distributed across the galaxies (i.e., morphology) is more predictive. Which morphological variable, however, is superior in its ability to predict a galaxy’s SFH, and how it actually relates to physics of SF quenching are still unclear. According to previous studies, the best predictor might be the global S\'{e}rsic index \citep[$n_g$;][]{Bell+12} or velocity dispersion \citep[$\sigma$;][]{Wake+12,Cappellari+13, vanDokkum+15, Bluck+20} or central mass density within the half light radius \citep{Kauffmann+03b} or $\Sigma_1$\citep[][]{Cheung+12,Fang+13, vanDokkum+14,Woo+15, Barro+17,Whitaker+17}. All of the above properties have been associated to varying degrees with the relative importance of the ``bulge" or ``spheroid" in a galaxy's mass/light distribution \citep{Gadotti09,Gao+20,Luo+20,Sachdeva+20,Yesuf+20a}, which is reasonable since this in turn has been linked to black-hole mass \citep{Kormendy+13}, and thus to AGN quenching.  However, among these bulge-linked variables no single one has as yet stood out. Moreover, other studies have found that the asymmetry, clumpiness, and other substructures (bars and spiral arms) of the stellar disk are also linked to SFR \citep[][]{Conselice+00, Reichard+08, Wang+12,Kaviraj14, Bloom+17,YuHo19, Lin+20b,Yu+21}. For example, it is well-known that starburst galaxies are highly disturbed, gas-rich, and lie above SFMS \citep[e.g.,][]{Sanders+96,Veilleux+02,Yesuf+14,Larson+16,Shangguan+19}. Simulations also show that starbursts are triggered by galaxy mergers, which drive cold gas to the centers of galaxies \citep[e.g.,][]{Barnes+91,Mihos+96,Hopkins+06,Hani+20,Moreno+21}. Furthermore, \citet{Yu+21} found that spiral arm strength correlates well with \delSSFR and gas fraction. Arms are stronger above the SFMS and weaker below it. The authors argued that their results support the physical picture in which spiral arms are maintained by dynamical cooling provided by gas damping. In short, it is becoming apparent that the star formation properties of galaxies are influenced by several structural factors, but their overlaps and relative importance have not yet been measured quantitatively.

Furthermore, the aforementioned observational studies at least have one of the the following limitations: (1) the criterion for identifying the best parameter(s) is not optimal. In fact, most studies used qualitative or less rigorous criteria. On the other hand, some studies used more sophisticated machine learning techniques such as random forest but the ranking (the variable importance) from these techniques can be impacted by the presence of highly correlated covariates, which are not properly handled. (2) Deep images, which are essential to characterize substructures in galaxies, are not available for large samples of nearby galaxies. All the studies of nearby galaxies mentioned above are based on shallow images. Even in large-sample studies of distant galaxies based on deep Hubble images, there are not many studies that quantitatively or comprehensively ranked the relevance of substructures (variables of bulge prominence and disk asymmetry) to predict SFR. (3) Mixing galaxies of different $M_\star$ and SFR may result in correlations with structural variables that may not be causally linked to the SFR evolution because galaxies today are products of both recent and ancient evolution. Some of the aforementioned studies did not subdivide their samples into star-forming (SFGs) and quiescent galaxies (QGs) of different masses to assess the ranks of their structural variables.

\subsection{The Scope of This Paper}

This paper explores the residuals of the SFMS. Namely, it tries to answer which structural parameters best predict whether a star-forming galaxy is above or below the SFMS. Our new contributions are : (1) we adopt the statistical framework of mutual information ($\MI$) to rigorously quantify the inter-dependence among several structural variables and to rank their relevance to predicting \delSSFR, taking their inter-dependence into account. $\MI$ can accurately quantify non-monotonic trends as well as monotonic trends. A tight relation has high $\MI$, regardless of its shape. (2) We use deep imaging data in SDSS Stripe 82 to study a large sample of galaxies. The Stripe 82 data improve the reliability of measurements of variables such as asymmetry \citep{Bottrell+19}. (3) We focus on SFGs and rank the importance of 6 variables in addition to $M_\star$ in predicting their SFRs in narrow mass ranges. As a result of the improvements above, we find that, in SFGs, the residuals of SFMS depend on morphological variables such as asymmetry and central concentration. After $M_\star$, asymmetry is the best predictor of \delSSFR. Star-forming galaxies with higher asymmetry and higher central concentration have higher \delSSFR.

The rest of the paper is structured as follows: Section~\ref{sec:data} presents the data used in this study and the mutual information framework. Section~\ref{sec:results} presents our results. To interpret our results in the context of previous studies, section~\ref{sec:discussion} discusses the effects of mergers/interactions and diffuse gas accretion on SFR and galaxy structure. Section~\ref{sec:conclusion} presents the summary and conclusions of our work.

\section{DATA AND METHODOLOGY} \label{sec:data}

This section describes the Sloan Digital Sky Survey (SDSS) data, our sample selection, and the statistical framework of mutual information.

\subsection{The Sloan Digital Sky Survey}

\subsubsection{SDSS Stripe 82}\label{subsec: S82}

The Stripe 82 is a special region\footnote{It is a stripe along the Celestial Equator in the Southern Galactic Cap. It is located at $-1.25 \degree < \delta  < +1.25 \degree$  and $-50 \degree \le \mathrm{R.A} \le +60 \degree$, and covers an area of about 275 $\mathrm{deg^2}$} in SDSS, which was imaged multiple times. The co-added images in Stripe 82 reach $\sim 1.7-2$ magnitudes deeper than the single-epoch SDSS legacy images \citep{Annis+14,Fliri+16}. We use the bulge plus disk decomposition catalogues\footnote{http://orca.phys.uvic.ca/~cbottrel/share/Stripe82/Catalogs/} of galaxies in the Stripe 82 that were recently produced by \citet{Bottrell+19}. The authors provided quantitative morphological parameters for 16,908 galaxies that are a subset of the general SDSS DR7 spectroscopic sample, which were previously analyzed by \citet{Simard+11} in the same way. In both studies \citep[i.e.,][]{Simard+11,Bottrell+19}, structural variables were measured using the parametric surface-brightness decomposition \texttt{GIM2D} software \citep{Simard+02}. But \citet{Simard+11} used regular-depth SDSS images as their analysis was not focused on the Stripe 82. Next, we give a brief summary of the \texttt{GIM2D} model and its outputs that are relevant for this study. Detailed information can be found in the references above.

The two-dimensional galaxy model used by \texttt{GIM2D} has a maximum of 12 parameters which include the bulge-to-total ratio $B/T$, the bulge semi-major axis effective radius $r_e$, the bulge ellipticity $e$ ($e \equiv 1 - b/a$, where $a$ is the semi-major axis and $b$ is semi-minor axis), the disk exponential scale length $r_d$, the disk inclination angle $i$, and the S\'{e}rsic index $n$ \citep{Simard+02}. The \texttt{GIM2D} software uses a Bayesian Metropolis Monte Carlo sampling algorithm to constrain the model parameters. \citet{Bottrell+19} provide morphological parameters for every galaxy in their sample by fitting, using \texttt{GIM2D}, three models of varying complexity: a single-component S\'{e}rsic profile, a two-component profile with fixed bulge S\'{e}rsic index ($n_b=4$) plus an exponential disk, and a two-component free bulge S\'{e}rsic index plus an exponential disk.

The bulge or the single component of the surface brightness model used by \texttt{GIM2D} is a S\'{e}rsic profile of the following form:

\begin{equation}
I(r) = I_e \exp\left[-b_n\left((r/r_e)^{1/n}-1\right)\right]
\end{equation}

\noindent where $n$ is the S\'{e}rsic index, $I_e$ is the intensity at $r_e$, and $b_n$ is a constant set to $b_n =1.9992n - 0.3271$ so that $r_e$ is the projected radius enclosing half of the light \citep[e.g.,][]{Graham+05}. To differentiate the S\'{e}rsic index of the bulge from the galaxy-wide index, we denote the latter as $n_g$. 



\citet{Bottrell+19} performed the image decompositions simultaneously in pairs of bands, where one of the two bands is always the $r$-band. Some of the structural variables in pairs of bands are forced to be the same. The tied parameters are $r_e$, $e$, $n_b$, $r_d$, $i$, and bulge and disk position angles. $B/T$, the total flux, and centroid offsets are free to vary in each band of a pairwise fit. In addition, except the S\'{e}rsic index in the single component model, the structural variables for $ur$, $ir$, and $zr$ pairwise fits were fixed to values first derived in $gr$ decompositions. We adopt $ir$ band fits so that the measurements reflect more faithfully the stellar mass distribution galaxies and are not strongly biased by the dominant emission from young stars which account for a small fraction of the mass.

We use the effective surface mass density, $\mu$, and the central concentration, $C_1$, as additional bulge morphology indicators. The latter quantity is taken from \citet{Bottrell+19}'s catalog. We define $\mu = 0.5M_\star/(\pi R_{50}^2)$, where $M_\star$ is the total stellar mass, and $R_{50}$ is the $i$-band Petrosian half-light radius \citep[e.g.,][]{Kauffmann+03b}. $C_1$ is defined as follows \citep{Abraham+94,Trujillo+01,Graham+05, Simard+02, Bottrell+19}:

\begin{equation}\label{eq:C1}
C_{1} = \frac{\sum_{i,\,j \in E(\alpha r_{e})} I_{ij}}{\sum_{i,\,j \in E(r_{e})} I_{ij}}
\end{equation}

\noindent $E(r_e)$ is an elliptical isophote which encloses half of the total light, $E(\alpha r_e)$ is the isophote at a radius $\alpha \times r_e$, and $\alpha=0.3$. The subscript in $C_1$ emphasizes that the concentration is measured within one half-light radius.

\texttt{GIM2D} also computes several image indices from a residual image, which can be used characterize the substructures left after the best galaxy image model has been subtracted. We adopt the residual asymmetry indices $R_A$ as the standard measure of asymmetry. Here the residuals are remnants of a single-component S\'{e}rsic model.
We checked the other asymmetry indices give similar results. $R_A$ is the summed difference over all pixels between the residual image and its 180$\degree$ rotated residual image. This sum is normalized by the total flux and corrected for the background noise by similarly subtracting the background image and its 180$\degree$ rotated counterpart. It has the form :

\begin{equation}\label{eq:RA}
\begin{split}
R_A & = (R_A)_\mathrm{raw}-(R_A)_\mathrm{bkg}  \\
 & =  \frac{\sum_{i,j} 1/2\,\abs{R_{ij}-R_{ij}^{180}}}{\sum I_{ij}} - \frac{\sum_{i,j} 1/2\, \abs{B_{ij}-B_{ij}^{180}}}{\sum I_{ij}} \\
\end{split}
\end{equation}

\noindent where $R_{ij}$ is a residual flux value in the residual image of $(i,\,j)$th pixel, and the $R_{ij}^{180}$ is a pixel value in the residual image rotated by 180$\degree$. Similarly, the $B_{ij}$ is a background pixel value in the residual image, and $B_{ij}^{180}$ is a background pixel value in the rotated residual image. The background corrections term $(R_A)_\mathrm{bkg}$ is computed over pixels flagged as background pixels in the SExtractor segmentation image. $I_{ij}$ is the background-subtracted pixel flux. 

We use $R_A$ measured within 3 times the circular half-light radius as the fiducial measure of asymmetry \citep{Bottrell+19}, and denote it as $R_{A3}$. The difference between $R_{A2}$ and $R_{A1}$ indicates that the asymmetries mostly occur beyond $R_{50}$. We checked that using $R_{A2}$ instead does not change our conclusions. Furthermore, comparison of $R_{A}$ in different bands with SSFR indicates that the $R_{A}$ in $i$-band is not trivially reflecting SSFR traced by the galaxy color; $\Delta R_{A3}$ between $r$ and $i$ bands does not show trends with SSFR but $\Delta R_{A3}$ between $g$ and $r$ bands shows a significant correlation. \citet{Reichard+08} also showed that galaxies have similar distributions of $r-$ and $i$-band lopsidedness, and that the weak color dependence of lopsidedness implies that lopsidedness primarily traces the asymmetry in the underlying stellar mass distribution. We also checked that our main results do not change if we use $R_{A3}$ computed from residuals of a two-component S\'{e}rsic + disk model instead of our adopted single-component model. Similarly, we checked that our main results do not change if we adopt other asymmetry indicators: the residual asymmetry $A_z$ and the rotational asymmetry \citep[A;][]{Abraham+94,Abraham+96,Conselice+00}. 

\begin{figure}[!h]
\includegraphics[width=\linewidth=0.98]{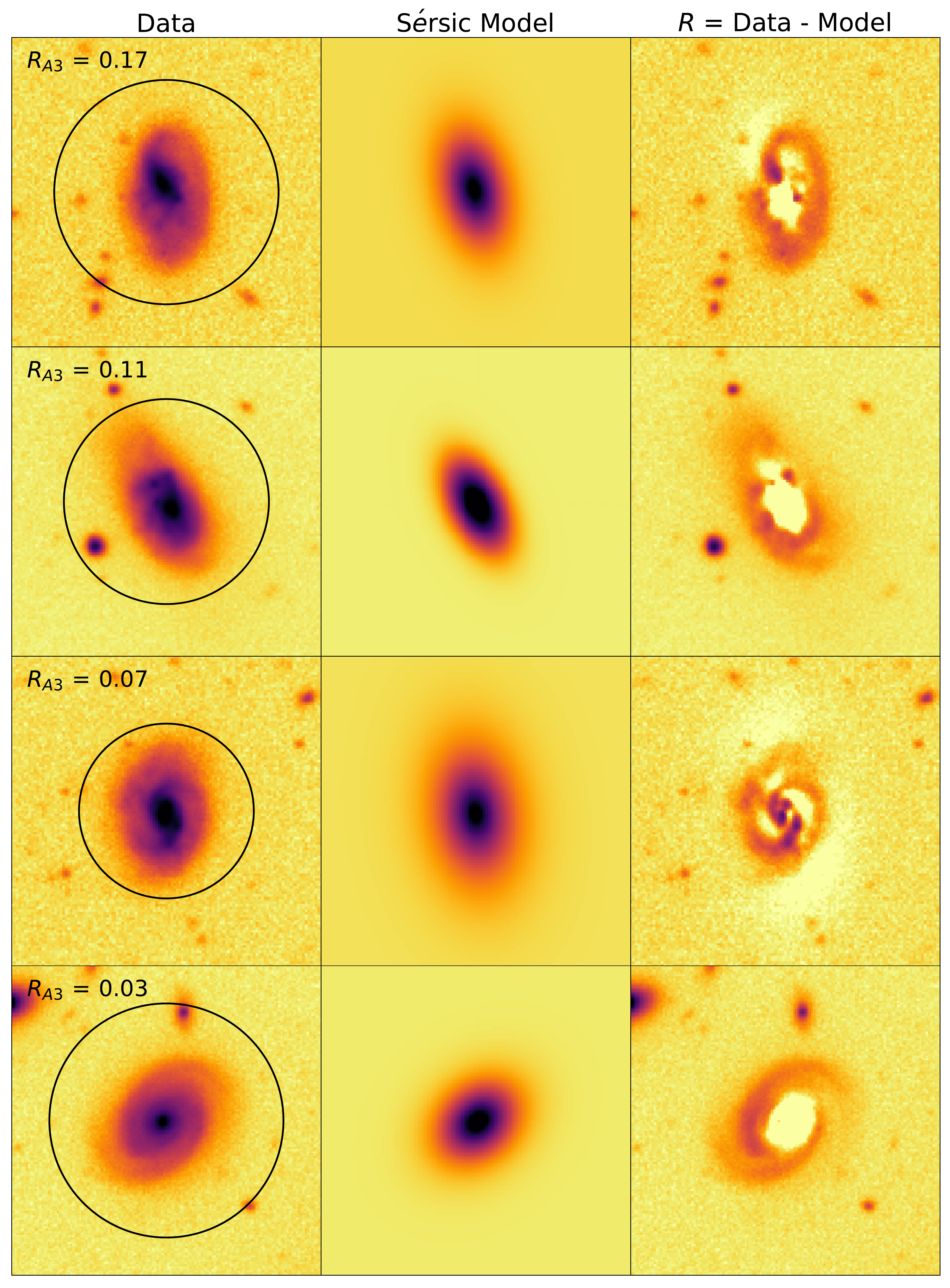}
\caption{Example galaxies with different residual asymmetries. The left images show the SDSS Stripe 82 $i$-band data. The top three galaxies illustrate asymmetry due to lopsidedness, mergers, or asymmetric spiral arms. The black circles are defined by three times the half-light radii. The middle images show the S\'{e}rsic models while the right images show the residuals after fitting the S\'{e}rsic models. The residual asymmetry within the three half-light radius, $R_{A3}$, is computed using the residual image and its rotated image according to equation~\ref{eq:RA}. We adopt $R_{A3}$ measured by \citet{Bottrell+19} as the fiducial measure of asymmetry. As discussed in section~\ref{subsec: S82}, we have checked that the details of how asymmetry is measured or defined do not affect our main conclusions. \label{fig:galsim}}
\end{figure}

Figure~\ref{fig:galsim} illustrates some example images, their fits, and residuals. The fits and residuals are generated using \texttt{GalSim} software \citep{Rowe+15} using \citet{Bottrell+19}'s best fit values, since the image outputs from \texttt{GIM2D} are not readily available.

To summarize, we take all the structural parameters ($R_{A}$, $\mu$, $C_1$, $n_g$, and $R_{50}$) from \citet{Bottrell+19}'s Stripe 82 catalog, and SDSS legacy catalogs ($\sigma$, $M_\star$, and SFR, as described next).

\subsubsection{Ancillary Data from SDSS Legacy Catalogs}

We use the publicly available Catalog Archive Server\footnote{http://skyserver.sdss.org/casjobs/ \\ We use the archive in the context of DR15 to retrieve various ancillary measurements in different catalogs. However, the sample of galaxies we use is restricted to those in DR7, because the \citet{Bottrell+19} sample is restricted to DR7. The following tables are queried: {\tt photoobjall, galSpecInfo}, and {\tt specDR7.}} to retrieve some of the measurements used in this work (e.g., stellar velocity dispersion $\sigma$). These data are supplemented with $M_\star$ and SFR taken from version 2 of the GALEX-SDSS-WISE Legacy Catalog\footnote{\textit{The Galaxy Evolution Explorer (GALEX)}-SDSS-\textit{Wide-field Infrared Survey Explorer (WISE)} legacy catalog can be found here: http://pages.iu.edu/$\sim$salims/gswlc/} \citep[GSWLC-2;][]{Salim+16,Salim+18}. $M_\star$ and SFR are derived by spectral energy distribution (SED) fitting of UV-optical photometry+ IR luminosity constraints using the Code Investigating GALaxy Emission \citep[CIGALE;][]{Noll+09}. \citet{Salim+18} estimated the total infrared luminosities from WISE 22\,$\mu$m or 12\,$\mu$m photometry using luminosity-dependent infrared templates of \citet{Chary+01} and calibrations derived from a subset of galaxies that have Herschel far-infrared photometry. For a narrow-line AGN, its IR luminosity is corrected using \ion{O}{3} 5007\,{\AA} line equivalent width before it is used in the SED fitting. Only $\sim 10\%$ of the current sample are narrow-line AGNs according their optical emission-line ratios \citep{Kewley+01}. Broad-line AGNs are not in the current sample since their SFRs are not estimated reliably. The SED fitting uses template superposition of two exponential star formation histories of an old stellar population (formed 10\,Gyr ago) and a younger population (100\,Myr to 5\,Gyr old). The young mass fraction varies between zero and 50\%. The stellar population models are calculated for four stellar metallicities (0.2 to 2.5\,$Z_\odot$) using \citet{Bruzual+03} models and assuming a \citet{Chabrier03} stellar initial mass function.


\subsection{Sample Selection}

\begin{figure}[ht]
\includegraphics[width=0.98\linewidth]{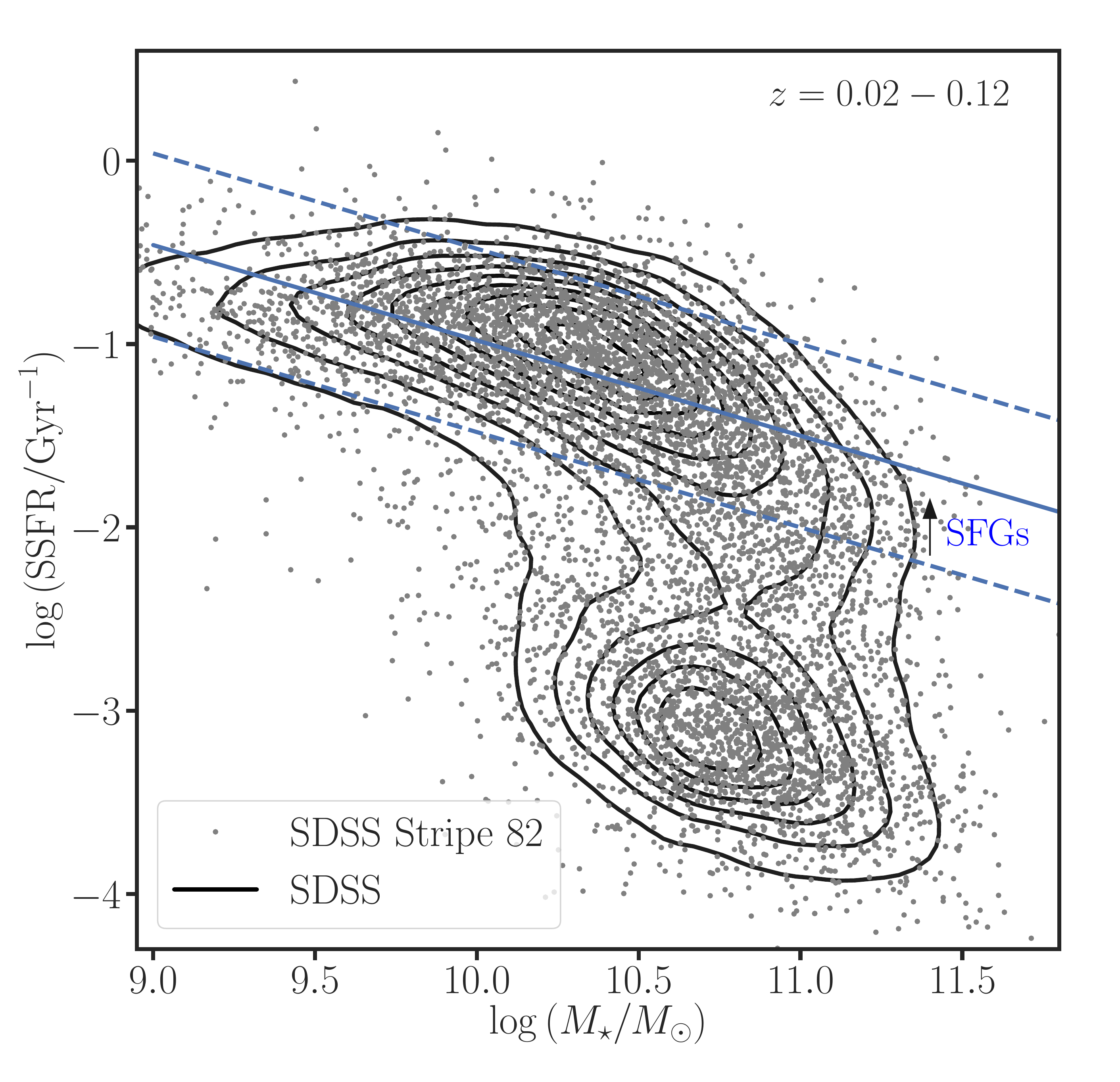}
\caption{Comparison of stellar mass and SSFR of Stripe 82 face-on galaxies (black points) at redshift $z=0.02-0.12$ with those of the general SDSS galaxies (contours) at similar redshifts. The blue solid line shows the best fit to the SFMS. As indicated by the blue dotted line, a simple offset of -0.5\,dex from the blue line is used to approximately define SFGs. \label{fig:MS}}
\end{figure}

From the Stripe 82 catalog of \citet{Bottrell+19}, we select 6,787 face-on galaxies at $0.02 < z < 0.12$ with good SFR measurements ($\mathrm{flag\_SED=0}$).
The axis-ratio ($b/a > 0.5$) cut aims to minimize dust and projection effects, and it removes $\sim 36\%$ of galaxies from the total sample of $\sim 10, 550$ at $0.02 < z < 0.12$. We focus our analysis on $\sim 3,700$ face-on SFGs in the sample in the entire mass range $M_\star \approx 10^9 - 5\times 10^{11}\,M_\odot$. We define SFGs as galaxies with $\Delta \log \,\mathrm{SSFR} > -0.5$\,dex. The difference from the ridgeline of the SFMS, $\Delta \log \,\mathrm{SSFR}$, is calculated by fitting a simple linear relation of the form  $\log \,\mathrm{(SSFR/yr^{-1})} = (\alpha-1)[\log\, (M_\star/M_\odot)-10.5] + \beta$ to face-on galaxies at $0.02 < z < 0.12$ in the legacy SDSS with $\mathrm{SSFR \, > \, 0.01\,Gyr^{-1}}$. We fix $\alpha=0.48$ based on the estimate of \citet{Speagle+14} and derive $\beta = -1.24$ from the median of the residuals. The fit is shown as the blue line in Figure~\ref{fig:MS}.

\subsection{Statistical Methods: Mutual Information}\label{subsec:MI}

The mutual information ($\MI$) of two random variables quantifies their mutual dependence, the amount of information one random variable contains about another \citep{CoverThomas12}. In other words, it is the reduction in the uncertainty of one random variable given the knowledge of the other. In contrast to the linear or monotonic correlation measures such Pearson's or Spearman's ($\rho$) correlation coefficient, $\MI$ is a more general measure which can precisely quantify nonlinear or/and nonmonotonic trends. In other words, it does not make assumptions on the distributional form of the variables or the nature of their dependence. We use $\MI$ to rigorously quantify the inter-dependence among several structural variables and to rank their relevance to predicting \delSSFR. A few previous studies have used $\MI$ to study other aspects of galaxies: spectral classification \citep{Slonim+01} or environmental effects \citep{Pandey+17, Alpaslan+20, Sarkar+20}. 

To define $\MI$ precisely, let $X$ and $Y$ be two continuous variables whose joint probability density function (PDF) is $p(x,y)$ and whose marginal PDFs are $p(x)$ and $p(y)$, respectively. The entropy $H$ of $X$ quantifies the amount of information required on the average to describe $X$ and it is defined as:

\begin{equation}\label{eq:H}
H\,(X) = -\int p(x) \log p(x)dx =  - \langle \log \, p(x) \rangle
\end{equation}

\noindent For example, let $X$ be a random Bernoulli variable, which takes the value 1 with probability $\theta$ and the value 0 with probability $1-\theta$. The Bernoulli distribution is a special case of the binomial distribution when the number of trail is $n = 1$. The entropy of $X$ is $H\,(X)=-\theta\log\,\theta-(1-\theta)\log(1-\theta)$. Therefore, $H(X) = 0$ when $\theta = 1$ or $\theta=0$, and $H(X) \ne 0$ when $\theta \in (0,1)$, reaching maximum at $\theta = 0.5$. Simply put, the outcome of a fair coin is the most uncertain; there is no advantage to be gained with a prior knowledge of $\theta$ in this case. But knowing $0.5 < \theta < 1$, one can still predict the outcome correctly more often than not. Likewise, the entropy of a normal random variable $Y$ is $H(Y)={\tfrac{1}{2}}(1+\log\,(2\pi\mathrm{Var(Y)}))$, where Var($Y$) is the variance of $Y$. In other words, if we know the mean and the variance of $Y$, how uncertain we are about the outcomes of $Y$ depends only on Var($Y$), as expected.

The $\MI$ of $X$ and $Y$ is the relative entropy or the Kullback-Leibler distance between $p(x,y)$ and $p(x)p(y)$.

\begin{equation}\label{eq:MI}
\begin{split}
I\,(X; Y) & = \int \int p(x,y) \log \frac{p(x,y)}{p(x)p(y)}dxdy \\
 & = H (X) - H(X|Y) \\
 & =  \langle \log \, \frac{p(x,y)}{p(x)p(y)} \rangle
\end{split}
\end{equation}

\noindent $H(X|Y)$ denotes the conditional entropy, which is the entropy of $X$ conditional on the knowledge of $Y$\footnote{The second equality of equation~\ref{eq:MI} can be easily shown using simple algebraic manipulations and the conditional probability formula $p(x|y) = p(x,y)/p(y)$ \citep{CoverThomas12}.}. In general, $H(X|Y) \le H(X)$. The $\MI$ $I(X; Y)$ corresponds to the intersection of the entropy of X with entropy of Y. $\MI$ is a nonnegative number and it is zero if and only if $X$ and $Y$ are independent (i.e., $p(x,y)=p(x)p(y)$). In the simple case when $p(x,y)$ is a bivariate normal distribution, $I\,(X,Y) = -1/2 \log (1-r^2)$, where $r$ is the Pearson's correlation coefficient.

Furthermore, $\MI$ can be used to subsequently select more relevant variables \citep[e.g.,][]{Estevez+09} by comparing the mutual information of the predictor variables with the response variable (in our case \delSSFR) and by penalizing variables that are redundant with already selected predictors. Let us denote the set of all predictor variables as $\textbf{F}$, the subset of already selected predictors variables as $\textbf{S}$, one of the already selected predictor as $f_s \in \textbf{S}$, and the response variable as $\mathcal{R}$. The score function that selects a new candidate variable $f_i$ is given by:

\begin{equation}\label{eq:score}
G  =  I (\mathcal{R}; f_i) - \beta \sum_{f_s \in \textbf{S}} \alpha(f_i, f_s)  \times I(f_i; f_s)
\end{equation}

\noindent $I(\mathcal{R}; f_i)$ quantifies the relative entropy between $\mathcal{R}$ and $f_i$ and thus is a measure of how closely related $f_i$ is to $\mathcal{R}$. Similarly, $I(f_i; f_s)$ quantifies the relative entropy between $f_i$ and one of the previously selected predictors, $f_s$. In other words, the first term on the right quantifies the relevance of a structural parameter to predicting \delSSFR while the second term quantifies the redundancy of this parameter with the previously selected structural parameters. The goal is to select variables that maximize this function (maximize relevance and minimize redundancy). The function $\alpha$ and the parameter $\beta$ attempt to balance both terms to the same scale. We use the normalized $\MI$ variable selection model proposed by \citet{Estevez+09}, which sets $\alpha (f_i, f_s) = 1/\mathrm{min} \{H(f_i),\ H(f_s)\}$ and $\beta =1/n(\textbf{S})$, where $n(\textbf{S})$ is the number of elements in set $\textbf{S}$. 

We use the \texttt{varrank} package \citep{Kratzer+18} in \texttt{R} to compute $\MI$ and $G$. \texttt{varrank} estimates $\MI$ using PDFs estimated from the counts of empirical frequencies (histograms) with a plug-in estimator based on entropy. It is a common and efficient approach to estimate $H$ by partitioning the range of a variable into bins of equal size, and approximating equation~\ref{eq:H} by finite sums. To that end, we adopt Scott's rule for an optimal construction of a histogram. According to this rule, which gives the optimal bin size for data drawn from the normal distribution, the number of bins of an equally spaced histogram of a continuous variable $X$ is given by $N_\mathrm{bin} = (\mathrm{max\{ X \}-min\{ X \}})/ (3.5\,\sigma_x N_x^{-1/3})$, where $N_x$ is the sample size and $\sigma_x$ is the sample standard deviation. 
Although the $\MI$ values and the exact rank orders change quantitatively (in detail) if we adopt instead the Doane or Sturges or Freedman-Diaconis rule, our conclusion that asymmetry and bulge prominence/concentration are important predictors of SSFR is broadly valid regardless of the rule used.

The \texttt{varrank} package implements both forward and backward sequential variable selection algorithms. In the forward selection, the first variable is selected 
by computing $I(\mathcal{R}:f_i)$ for all the variables and choosing the largest. Then, the following variables are selected sequentially by computing the score $G$ (equation~\ref{eq:score}) and selecting the variable with highest score at each step. The backward selection prunes the full set $\textbf{F}$ by minimizing equation~\ref{eq:score}. We use the forward selection scheme because it is less computationally-intensive.


\section{Results}\label{sec:results}
This section starts by quantifying the inter-dependence among galaxy parameters of SFGs and their ranking using the $\MI$ framework. The section ends with data visualizations in order to show that the results of the quantitative analysis are intuitively reasonable.

\begin{figure}
\includegraphics[width=0.98\linewidth]{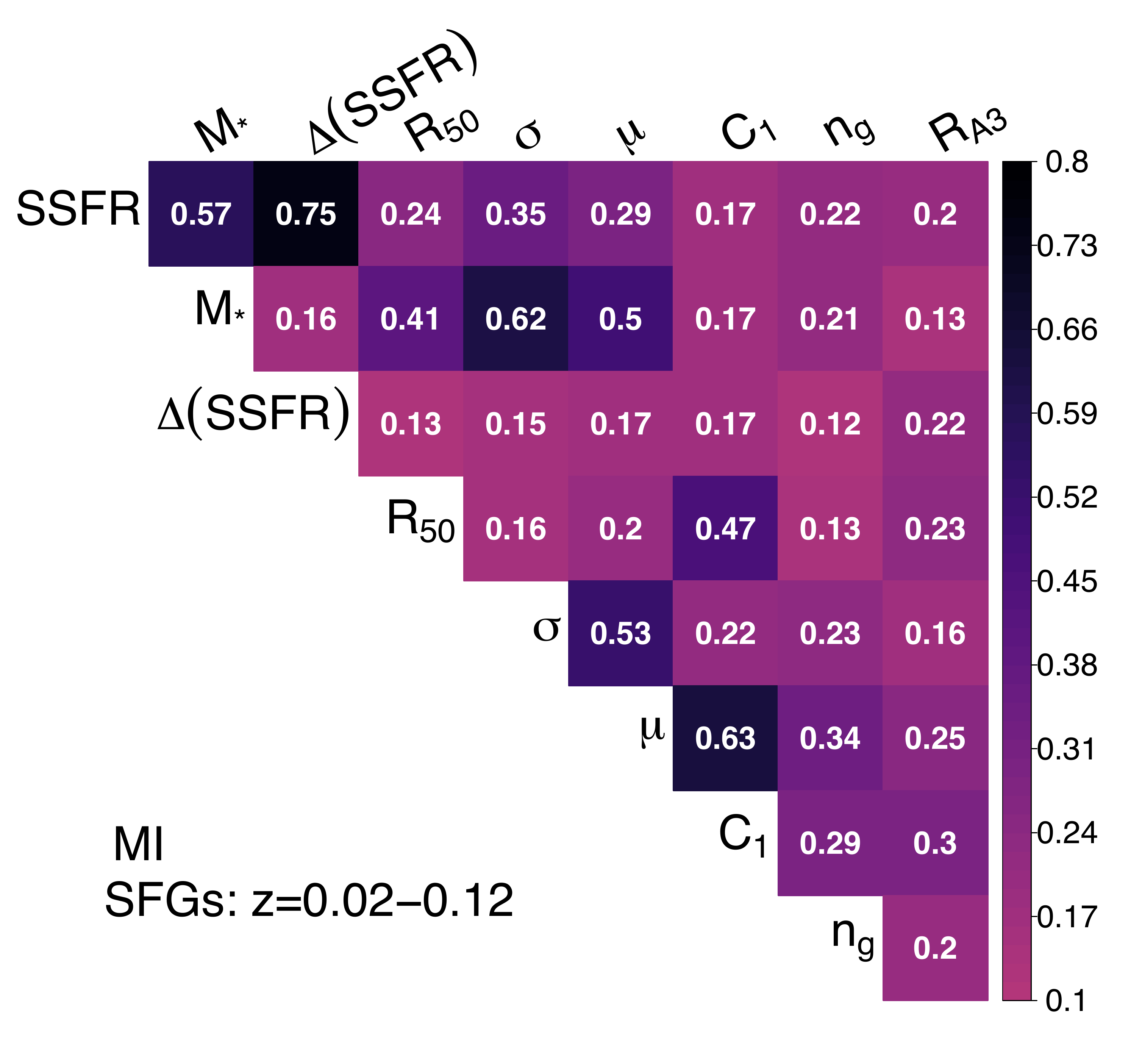}
\caption{The mutual information matrix for galaxy parameters of face-on star-forming galaxies in SDSS Stripe 82 at redshift
$0.02 < z < 0.12$. Some of the variables are highly correlated (dark colors) and carry similar information while others are more independent.
The variables are specific star formation rate (SSFR), stellar mass ($M_\star$), deviation from the SFMS (\delSSFR), half-light radius ($R_{50}$), stellar velocity dispersion ($\sigma$), mass surface density ($\mu$), central concentration ($C_1$), global S\'{e}rsic index ($n_g$), and residual asymmetry ($R_{A3}$)}. \label{fig:MIz}
\end{figure}

Figure~\ref{fig:MIz} shows the $\MI$ of galaxy parameters for the Stripe 82 face-on SFGs at $z=0.02-0.12$\,\footnote{We use \texttt{corrplot} in R to visualize the $\mathrm{MI}$ and scores matrices.}. The figure conveys several trends, most of which are not unexpected, but in a new way. First, $M_\star$ has the highest $\MI \,\approx 0.6$ with SSFR, as expected from our previous knowledge of the existence the tight SFMS. $R_{A3}$ has the lowest $\mathrm{MI} = 0.13$ with $M_\star$, and thus likely provides the most complementary information to $M_\star$ about processes related to SFR evolution. In fact, $R_{A3}$ has the highest $\mathrm{MI} = 0.22$ with \delSSFR. Although $\sigma$, $\mu$, and $R_{50}$ have high $\MI \, \approx 0.25-0.35$ with SSFR, they are redundant with $M_\star$ ($\MI \, \approx 0.4-0.6$). Note that $R_{A3}$ also has $\mathrm{MI} \approx 0.2$ with structural parameters such as $R_{50}$. We discussed in the introduction that previous studies suggested that $R_{50}$ might be the second important variable after $M_\star$. \citet{Lin+20a} found that $R_{50}$, though an important second structural variable, is not related to SSFR. In a moment, we will further show that $R_{A3}$ is a better predictor of \delSSFR than $R_{50}$, although the two parameters are related. In addition to recovering the well-known correlation between $M_\star$ and $R_{50}$, Figure~\ref{fig:MIz} indicates that $R_{50}$ has lower $\MI$ with most of the morphological indicators (e.g., $\mathrm{MI} = 0.13$ with $n_g$ or $\mathrm{MI} = 0.16$ with $\sigma$). In contrast,  $R_{50}$ and $C_1$ have high $\mathrm{MI} \approx 0.5$; they anti-correlate with each other (Spearman $\rho \approx -0.5$).

\begin{figure*}[ht]
\includegraphics[width=0.49\linewidth]{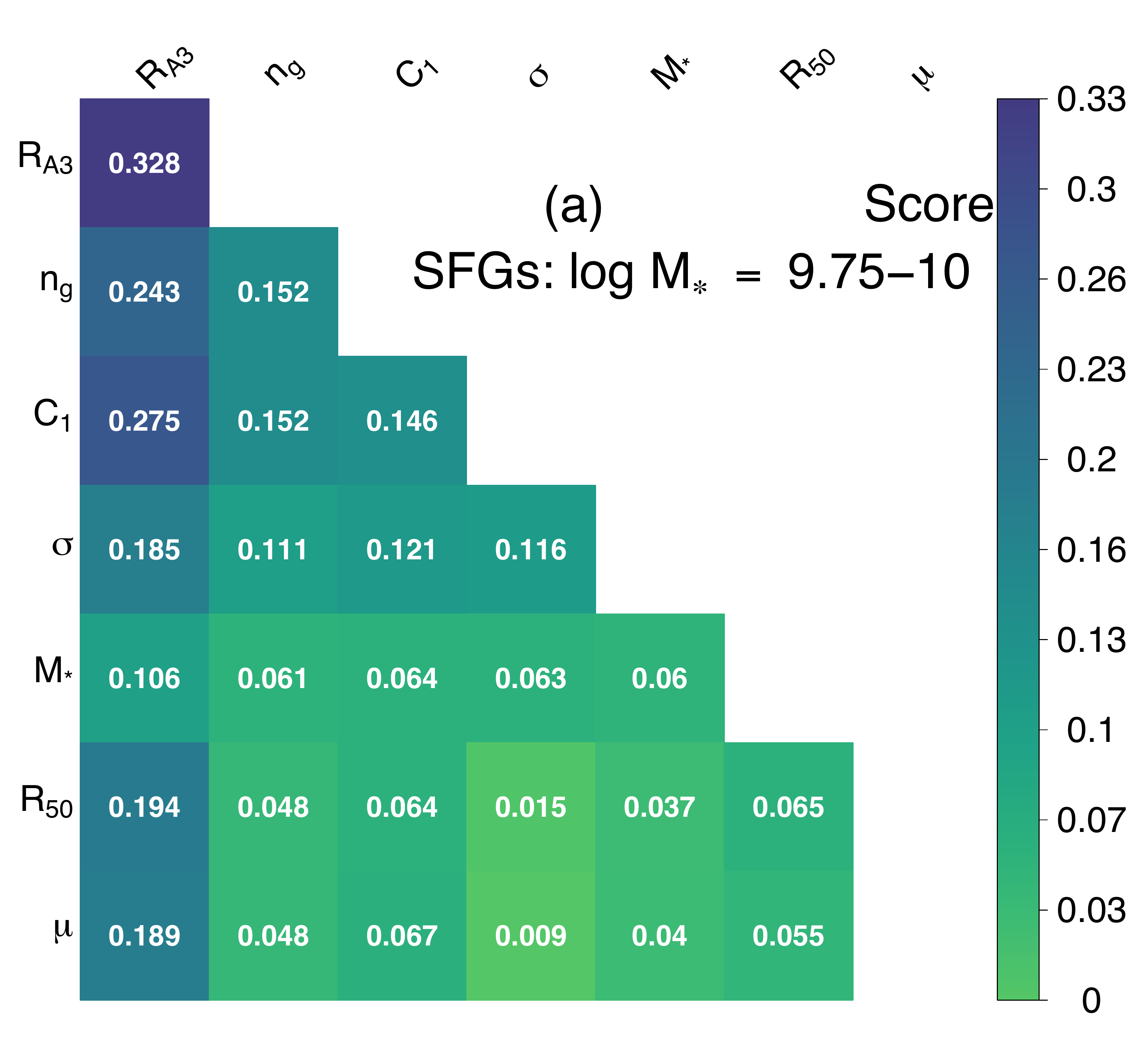}
\includegraphics[width=0.49\linewidth]{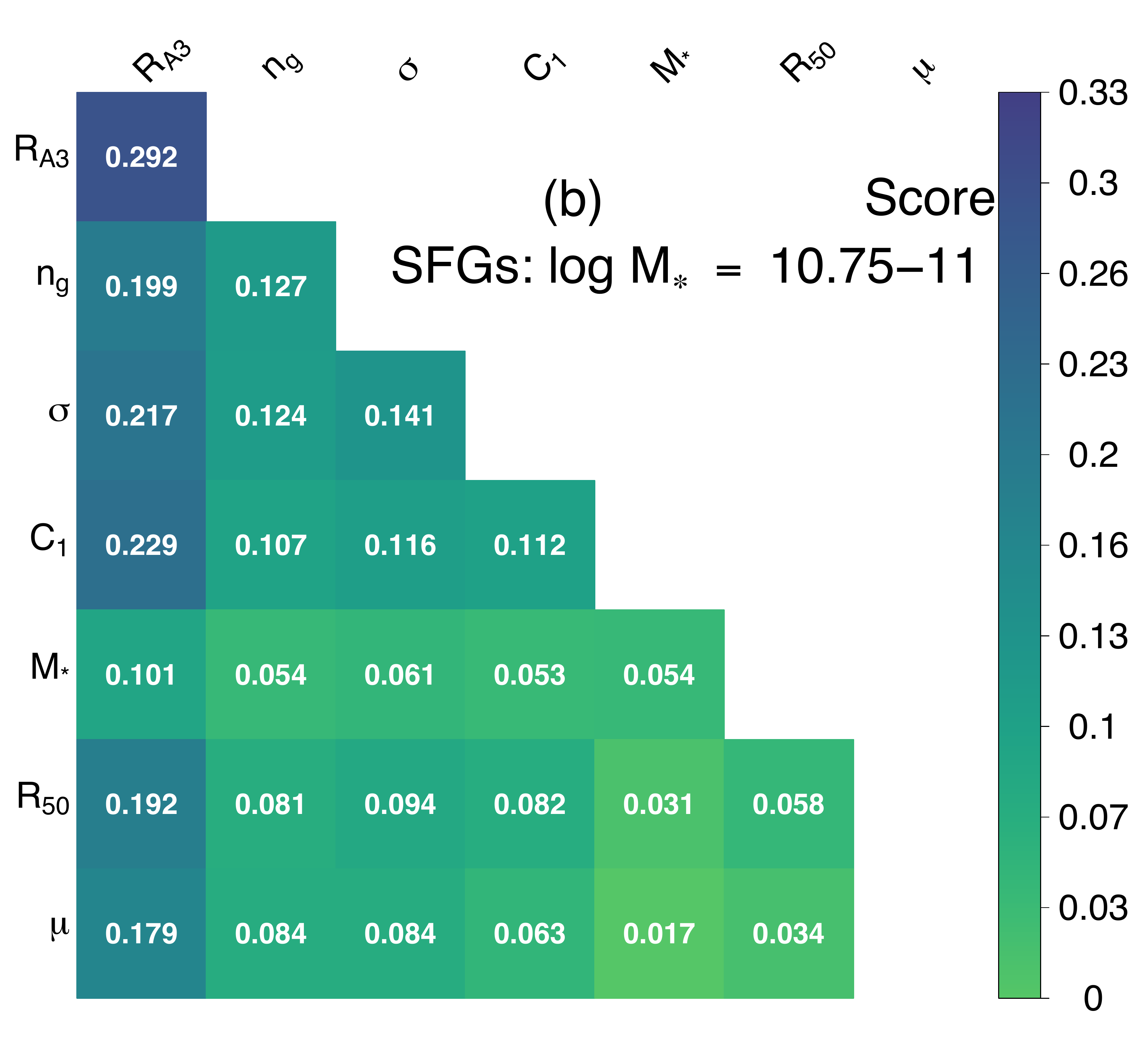}
\includegraphics[width=0.49\linewidth]{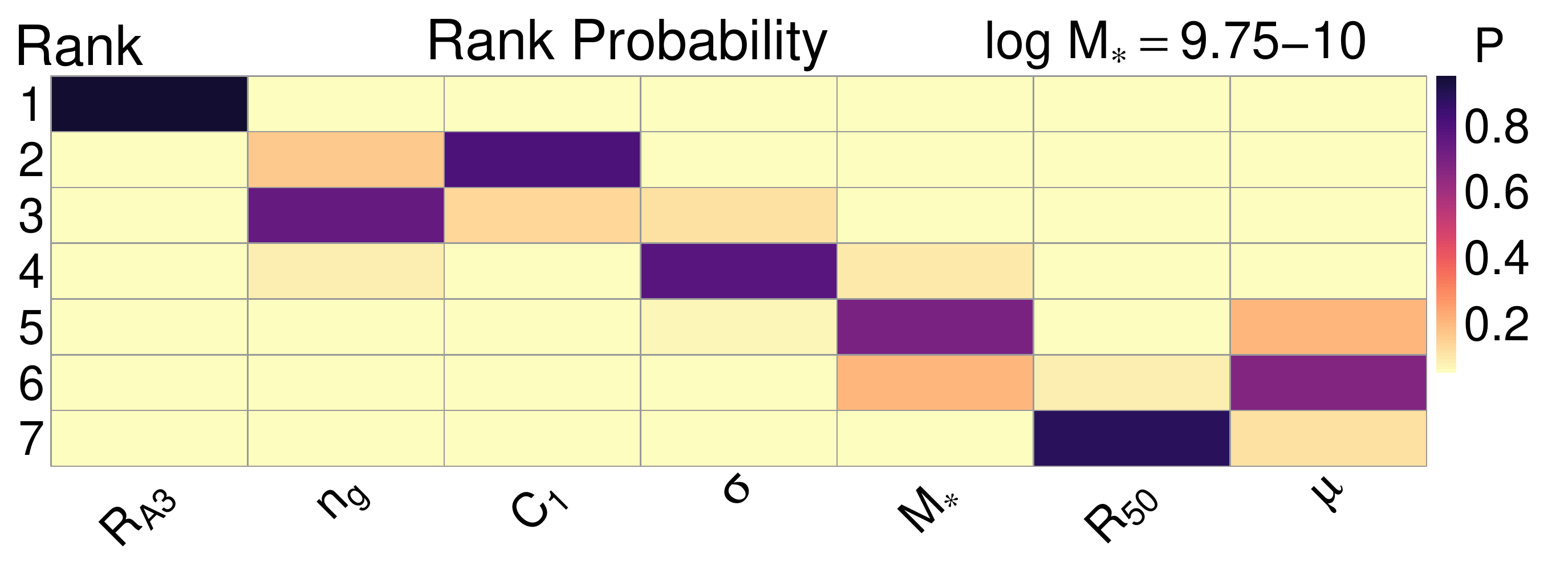}
\hfill
\includegraphics[width=0.49\linewidth]{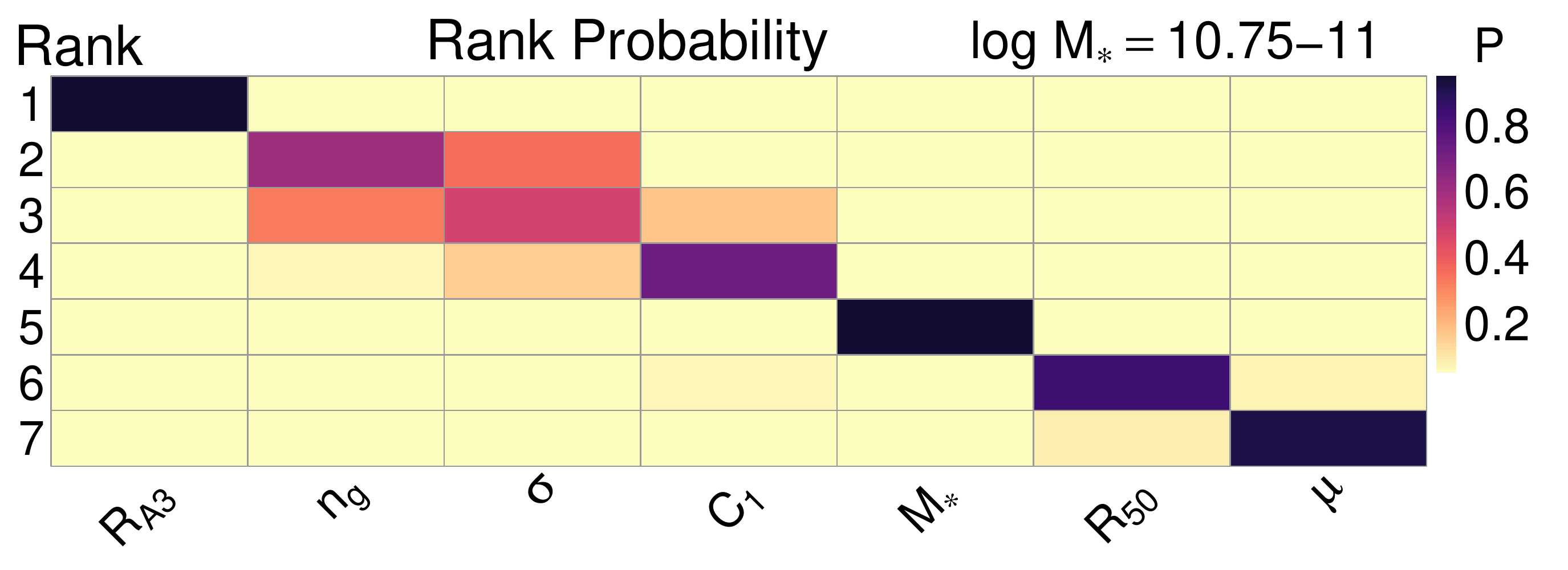}
\caption{The scores for predicting \delSSFR in Stripe 82 SFGs in two narrow mass ranges. The columns contain the $G$ scores (equation~\ref{eq:score}) at the successive variable selection steps. The first column shows the scores of the first step, the second column those of the second step, and so on. At each step, the variable with the highest score is selected (the diagonal values). The names of the variables on the left and top are ordered by a decreasing importance, after sequentially comparing the relevancy versus redundancy of information. A positive score indicates that a variable is more relevant than it is redundant; a negative score indicates otherwise. The variable selected at each step is the one with the highest score. The score at each selection step can be read from the diagonal elements. All diagonal values are positive indicating all variables are relevant for predicting SSFR. The rank probabilities shown in the bottom panels are estimated from 1000 score recalculations based on resampling 90\% of the data of a given mass range without a replacement.\label{fig:score_SFMz012}}
\end{figure*}

Having shown the high overlap among structural variables, next we take this interdependence (redundancy) into account and quantify the relevance of the variables to predicting \delSSFR in narrow mass ranges. Figure~\ref{fig:score_SFMz012} shows the $G$ scores computed according to equation~\ref{eq:score} in section~\ref{subsec:MI} for the SFG samples in two mass ranges. The process for computing this score matrix is different from Figure~\ref{fig:MIz}. The columns contain the scores at the successive selection steps. The first column shows the scores at the first step, the second column the scores at the second step, and so on. At each step, the variable with the highest score is selected (the diagonal values). So, the names of the variables down the left and across the top are ordered in decreasing relevance to \delSSFR. A positive score for a variable indicates its relevancy of information dominates over its redundancy; a negative score indicates otherwise. Note that the score is equivalent to $\MI$ for the first column, by definition, for the forward variable selection we adopted. To assess the stability of the rankings, we use resampling simulations to recalculate the scores and the ranks 1000 times for each stellar mass bins. The bottom panels of Figure~\ref{fig:score_SFMz012} show the rank probabilities calculated from resampling 90\% of the SFGs in a given $M_\star$, without a replacement.

For the SFGs with $\log\, M_\star/M_\odot = 9.75-10$ (Figure~\ref{fig:score_SFMz012}a), for example, $R_{A3}$ has the highest score ($\MI = 0.33$) with \delSSFR among all the variables and it is selected first. $C_1$ and $n_g$ are selected second and third and they have $G \approx 0.15$. Then, $\sigma$ is ranked fourth with $G \approx 0.12$. 
Although $\mu$ and $R_{50}$ have high $\MI$ with \delSSFR (see the first column in Figure~\ref{fig:score_SFMz012}a), $\mu$ and $R_{50}$ contain highly overlapped information with $C_1$ ($\mathrm{MI} \approx 0.8$) in this mass range. In contrast, $I(C_1; \sigma) \approx 0.2$ is about four times smaller than $I(C_1; \mu)$ and $I(C_1; R_{50}$) (see also Figure~\ref{fig:MIz} for the whole mass range). Thus, $\sigma$ contains the complementary information for predicting \delSSFR after $C_1$ and $n_g$. All structural variables have $G > 0$, indicating that the relevancy term dominates over the redundancy term. In other words, all variables are useful for predicting \delSSFR although some have minor contributions. Similarly, the scores for SFGs with $\log\, M_\star/M_\odot = 10.75-11$ are presented in Figure~\ref{fig:score_SFMz012}b. For this subsample, $R_{A3}$, $\sigma$, and $n_g$ are the top three variables.

\begin{figure*}[ht]
\includegraphics[width=0.98\linewidth]{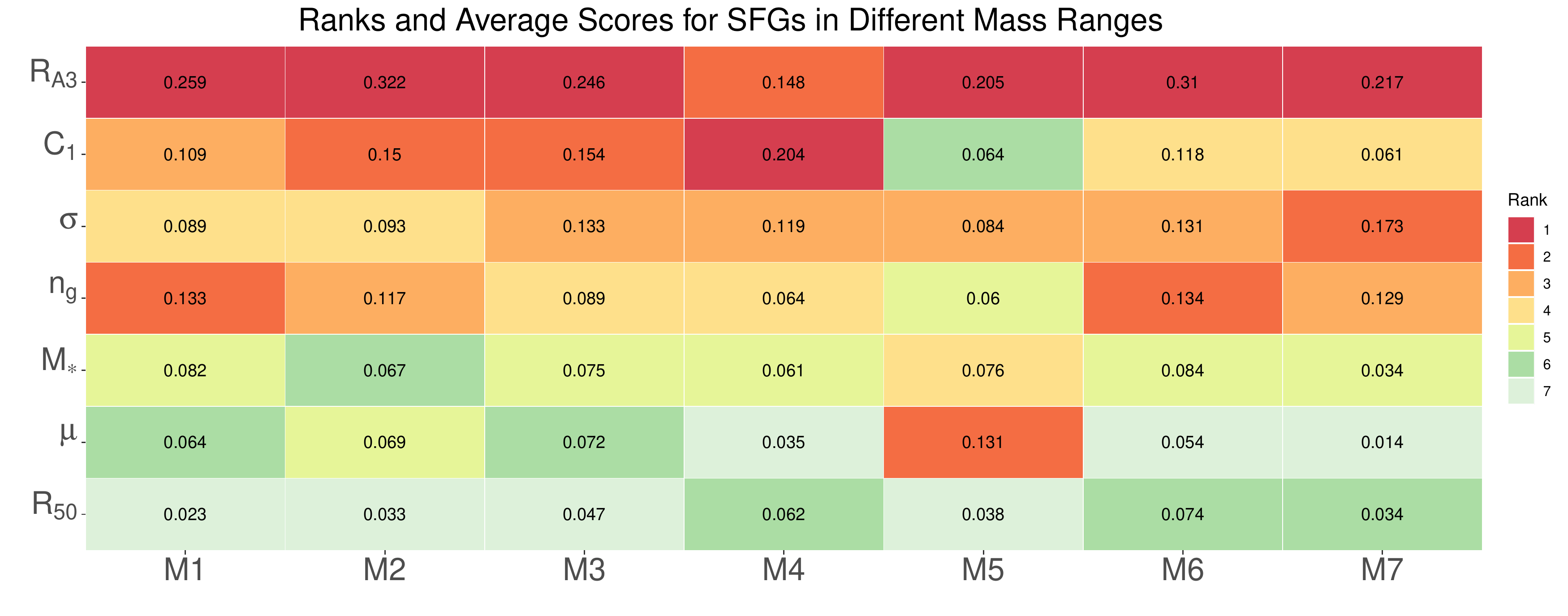}
\caption{The ranks of the variables based on average scores (the number in each box) for predicting \delSSFR of Stripe 82 SFGs in different narrow mass ranges from M1:\,$\log\,(M_\star/M_\odot) = 9.5-9.75$ to M7:\,$\log\,(M_\star/M_\odot) = 11-11.25$ with $\Delta M_\star$ increasing by 0.25\,dex. The average scores are calculated using the scores of 1000 resampled galaxies in each mass, similar to Figure~\ref{fig:score_SFMz012}. \label{fig:rankM}}
\end{figure*}

Figure~\ref{fig:rankM} summarizes the rankings of the $G$ scores of the various variables as a function of $M_\star$. $R_{A3}$ is the highest ranked variable in 6 of the 7 subsamples of narrow stellar mass (0.25\,dex) ranges between $\log\, M_\star/M_\odot = 9.5-11.25$. Besides, $C_1$ and $n_g$ are generally among the top three parameters at low mass ($\log\, M_\star/M_\odot < 10.5$), while $\sigma$ and $n_g$ are more important than $C_1$ at high $M_\star$. The results are substantially the same if we bin the sample by $M_\star$ of 0.5\, dex. We have also checked the results are similar if we screen out satellites from the samples and redo the analysis for centrals only (see the appendix).

\begin{figure*}
\includegraphics[width=0.98\linewidth]{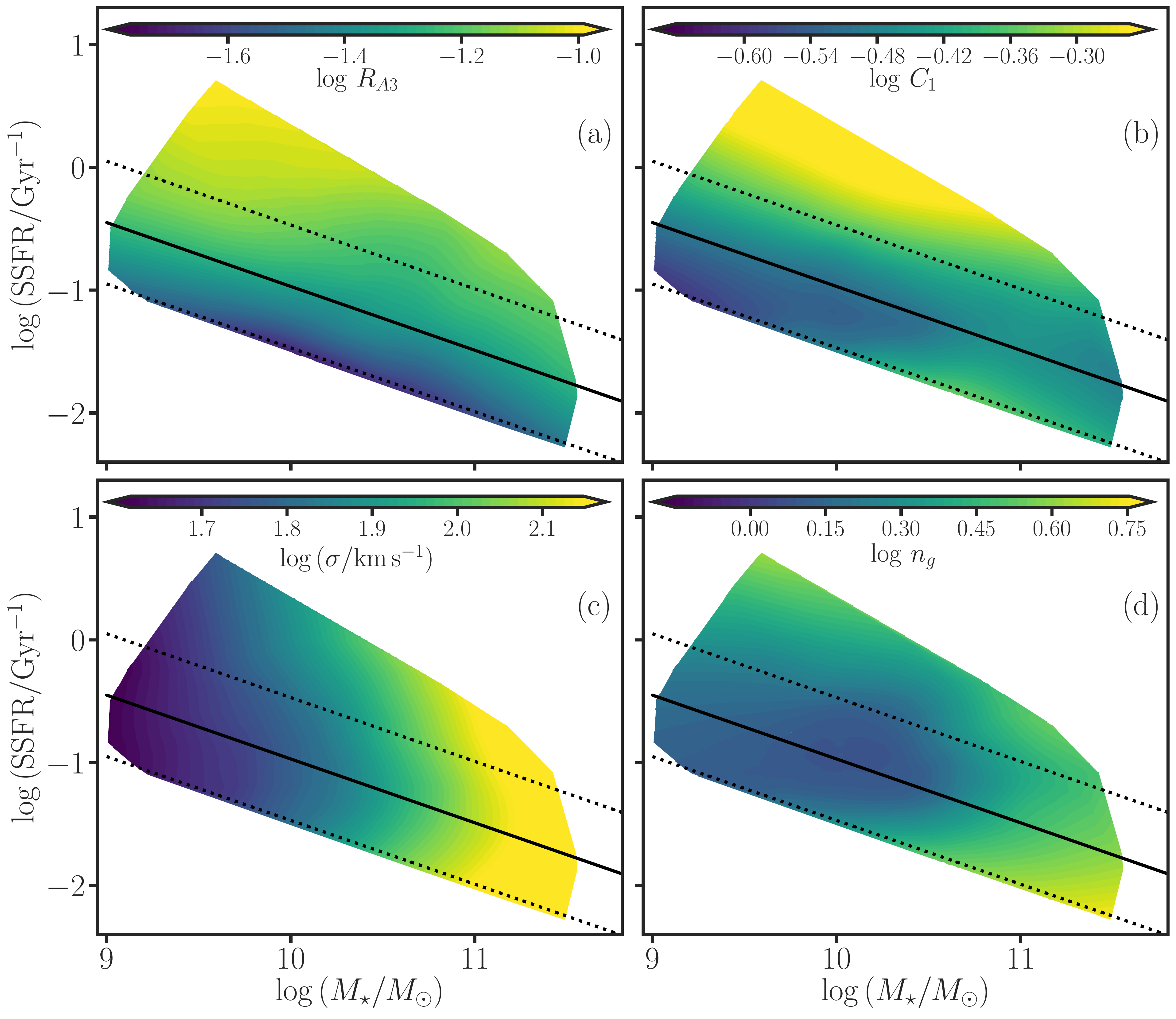}
\caption{Trends of residual asymmetry $R_{A3}$, central concentration within the half-light radius $C_1$,  global S\'{e}rsic index $n_g$, and the stellar velocity dispersion $\sigma$ in $M_\star$-SSFR space for SDSS Stripe 82 face-on bulges at $z =0.02-0.12$. The data are LOESS-smoothed to reveal mean trends. The black lines denote the ridge-line of the SFMS and its $\pm 0.5$\,dex offsets. Morphological asymmetry $R_{A3}$ is most strongly linked to \delSSFR, followed by three bulge-related parameters (concentration, velocity dispersion, and global S\'{e}rsic index).\label{fig:RA_exp}}
\end{figure*}

Figure~\ref{fig:RA_exp} visualizes trends of $R_{A3}$, $C_1$, $\sigma$, and $n_g$ on $M_\star$ versus SSFR space. In each panel, the figure presents the data as LOESS-smoothed curve color-coded by one of the structural variables above, to emphasize average trends. We use the \texttt{loess} python package \citep{Cappellari+13} to smooth the data by fitting a local linear regression. We use a smoothing span of 30\% and set the ranges of the colorbars such that they enclose $10-90$ percentile ranges of the data. The main points we want the reader to notice in this figure are: (1) none of the structural variables correlate linearly or very strongly with the residual of SSFR about the mean SFMS. In other words, they are not strong predictors of \delSSFR individually. However, as shown in panel (a) $R_{A3}$ is the best on average (spearman $\rho \approx 0.4$, see the plot of $R_{A3}$ vs. \delSSFR in the appendix). At fixed $M_\star$, SFGs with high $R_{A3}$ have high \delSSFR on average; 2) the SFMS is populated by galaxies with diverse structures, even at a given $M_\star$. There are significant numbers of SFGs with high $C_1$, $\sigma$, and $n_g$. This would have been clearer if we had instead plotted the sample as individual points without the LOESS smoothing. Anyway, later plots of \delSSFR with $R_{A3}$, $C_1$, and $\sigma$ also show this point (see the appendix); (3) the patterns of structural variables vary with $M_\star$, which is why their ranking in Figure~\ref{fig:rankM} depends on mass. SFGs with high $C_1$ have high \delSSFR at fixed $M_\star$ (and $R_{A3}$) especially at low $M_\star$ (see panels (b) and the appendix). At the high $M_\star$ end, the average relationship between \delSSFR and $\sigma$ and $n_g$ is curved; SFGs with high $\sigma$ and $n_g$ have sightly lower \delSSFR on average (panels (c) and (d)). The relationship of \delSSFR with $\sigma$, however, is not monotonic and has a large spread -- $\sigma$ increases both above and below the (black) ridgeline (see panel (c) and the appendix). For SFGs, $\sigma$ shows a strong correlation with $M_\star$, while $R_{A3}$ shows a weak correlation with $M_\star$.

We have also checked the correlations or trends in Figure~\ref{fig:MIz} and Figure~\ref{fig:score_SFMz012} are not completely driven by a small number of peculiar objects with particularly high \delSSFR by restricting the sample to within 0.5\,dex of the SFMS ridgeline. Our results are still valid for this sample and are not confined just to the highest star-forming galaxies. Likewise, excluding SFGs that are dominated by narrow-line AGNs ($\sim 10\%$) does not change the main conclusions. In summary, our analysis indicates that the offset from the ridge of SFMS depends on asymmetry ($R_{A3}$) and bulge prominence variables such as $C_1$, $\sigma$, and/or $n_g$. 

\begin{figure*}[ht]
\includegraphics[width=0.98\linewidth]{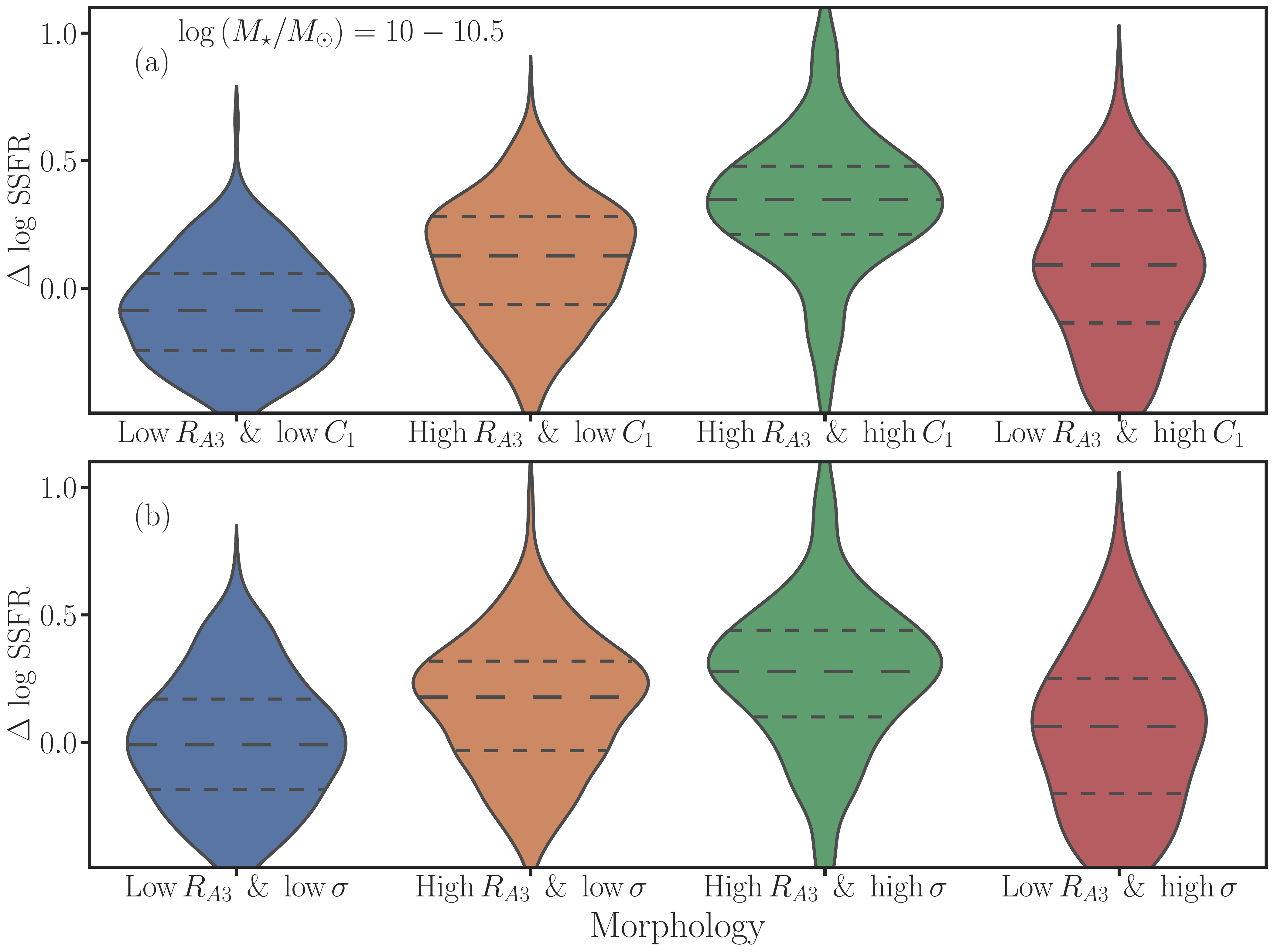}
\caption{The distributions of deviations of SSFR, $\Delta\,\mathrm{SSFR}$, from the SFMS for four sets of SFGs binned by $R_{A3}$ and $C_1$, or $\sigma$. The mass range of the subsample is $\log\, M_\star/M_\odot = 10-10.5$. The violin plots (VPs) show the kernel density estimates of the distributions and the dashed lines denote the median and quantiles of the distributions for each bin. The widths of the VPs span the values of the PDFs of \delSSFR and their mirror images. More asymmetric galaxies (orange and green VPs) are more likely to have enhanced SSFR than less asymmetric and lower $C_1$ (or lower $\sigma$)  SFGs (blue VP). The SFR enhancements fade as the asymmetries disappear in concentrated or high $\sigma$ SFGs (red VP). These trends may indicate evolution of the SFR that is driven by a quenching mechanism that involves burst of star formation and morphological disturbances.\label{fig:RAVd_M10prof}}
\end{figure*}

The violin plots in Figures~\ref{fig:RAVd_M10prof} further illustrate the point above using SFGs with $\log\,M_\star/M_\odot = 10.0-10.5$ and $z = 0.02-0.12$. Figure~\ref{fig:RAVd_M10prof}a plots the probability density function (PDF) of \delSSFR after subdividing the sample into four bins by $R_{A3}$ and $C_1$ (using their medians for the whole SFG sample). The widths of the violin plots span the values of the PDFs of \delSSFR and their mirror images, which are the reflected duplications of the \delSSFR PDFs across the vertical lines that split the violin plots into halves\footnote{We use the \texttt{seaborn} python package to make the violin plots.}. The plots show that the distributional properties (e.g., median, quartiles, and skewness) of \delSSFR depend both on $R_{A3}$ and $C_1$. The distribution of \delSSFR of SFGs with high $R_{A3}$ shifts to high values (i.e., SSFR is relatively enhanced by $\sim 0.15-0.2$\,dex) compared to SFGs that have low $R_{A3}$ and $C_1$. The bin where both $R_{A3}$ and $C_1$ are high has the highest median \delSSFR. The distribution of \delSSFR shifts relatively to lower values when the SFGs have lower $R_{A3}$ but higher $C_1$. This distribution is also wider and flatter than that of high $R_{A3}$ and high $C_1$ SFGs. Figure~\ref{fig:RAVd_M10prof}b shows similar trends for the distribution of \delSSFR split by $R_{A3}$ and $\sigma$. The trends presented in Figure~\ref{fig:RAVd_M10prof} are similar for different mass ranges, and for the pairing of $R_{A3}$ with $n_g$, $\mu$, or $R_{50}$.

\begin{figure*}[ht]
\includegraphics[width=0.98\linewidth]{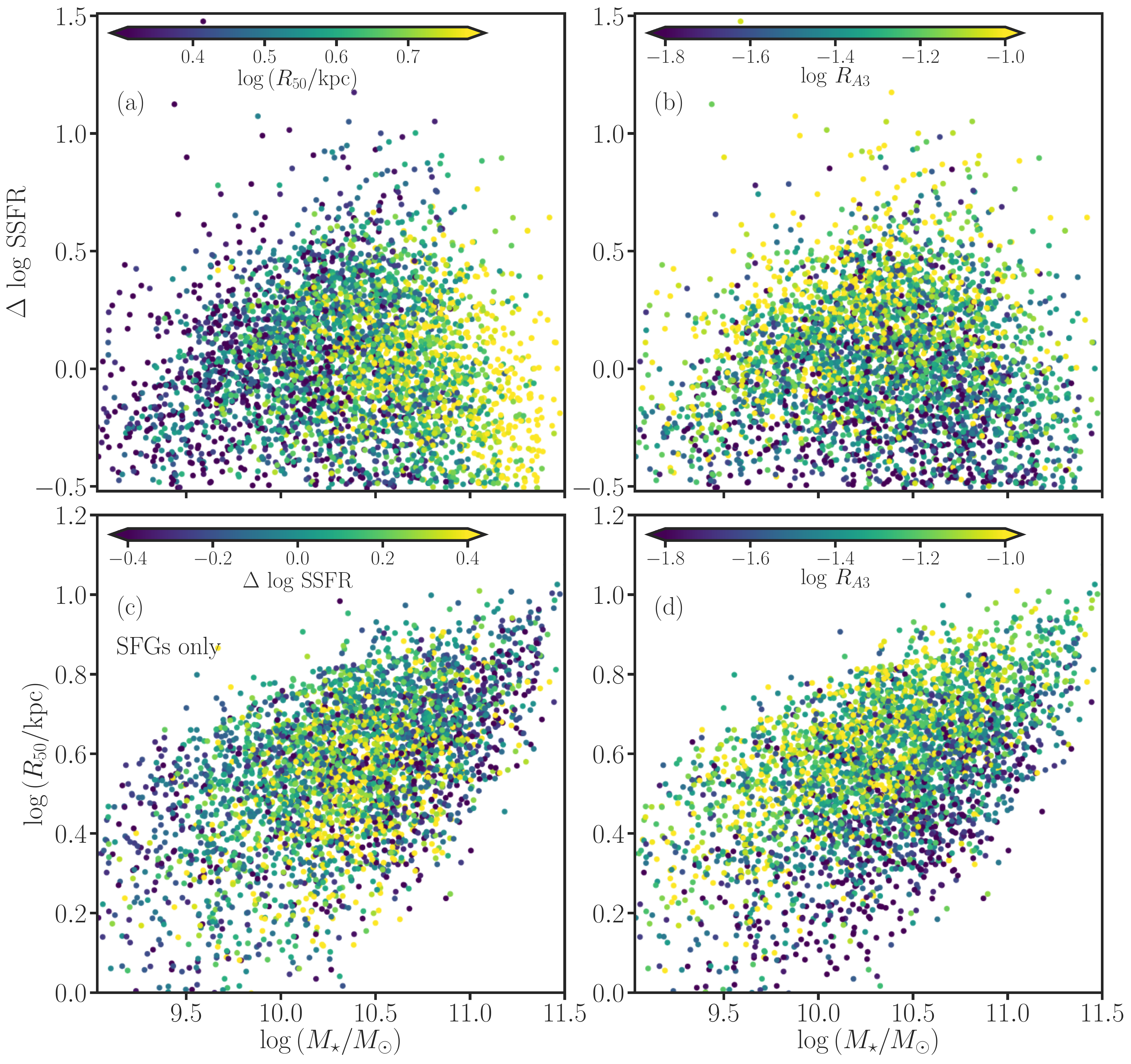}
\caption{Top: the $M_\star-$\delSSFR relation color-coded by $R_{50}$ in panel (a) and by $R_{A3}$ panel (b). Bottom: the $R_{50}$-$M_\star$ relation is color-coded by \delSSFR in panel (c) and by $R_{A3}$ in panel (d). There is only a weak trend between $R_{50}$ and \delSSFR. On the other hand, the pattern of $R_{A3}$ is correlated $R_{50}$ and \delSSFR; there is a redundancy between $R_{50}$ and $R_{A3}$. And the latter is a better predictor of \delSSFR. \label{fig:M_R50}}
\end{figure*}

In the introduction, we discussed the context in which $R_{50}$ is the second parameter \citep[e.g.,][]{Omand+14,Chen+20, Lin+20a}. In comparison, our $\MI$ analysis in this section indicates that $R_{A3}$ is the second useful parameter for predicting \delSSFR and that $R_{50}$ is one of the least important parameters in this regard. To visually compare $R_{A3}$ and $R_{50}$ in the context of $M_\star-$\delSSFR, Figure~\ref{fig:M_R50} shows the $M_\star-$\delSSFR plot color-coded by $R_{50}$ in panel (a) and by $R_{A3}$ in panel (b), and the $R_{50}-M_\star$ relation color-coded by \delSSFR in panel (c) and by $R_{A3}$ in panel (d). There is only a weak trend between $R_{50}$ and \delSSFR at a fixed $M_\star$. Namely, SFGs with \delSSFR $\gtrsim 0.5$\,dex and $\log\,M_\star/M_\odot \approx 10-11$ are smaller than SFGs that lie within 0.3\,dex of the SFMS ridgeline at a similar $M_\star$. The trend between $R_{50}$ and \delSSFR is not as strong as that of $R_{A3}$ and \delSSFR. Figure~\ref{fig:M_R50}(a)\,\&\,(b) confirms that the second parameter in the context of $M_\star-$\delSSFR is $R_{A3}$ and not $R_{50}$. Likewise, there is no clear trend of \delSSFR on $R_{50}-M_\star$ relation (panel (c)). In contrast, the pattern of $R_{A3}$ is correlated with $R_{50}-M_\star$ (panel (d)), as expected from the $\MI$ values in Figure~\ref{fig:MIz}. In other words, small galaxies at a fixed $M_\star$ are on average less asymmetric than big galaxies, and $R_{50}$ is redundant with $R_{A3}$ (and is also highly redundant with $C_1$). Although $R_{A3}$ is the best predictor of \delSSFR, Figure~\ref{fig:M_R50} also shows our earlier point that the correlation between \delSSFR and $R_{A3}$ is only moderate.

\begin{figure*}
\centering
\includegraphics[width=0.95\linewidth]{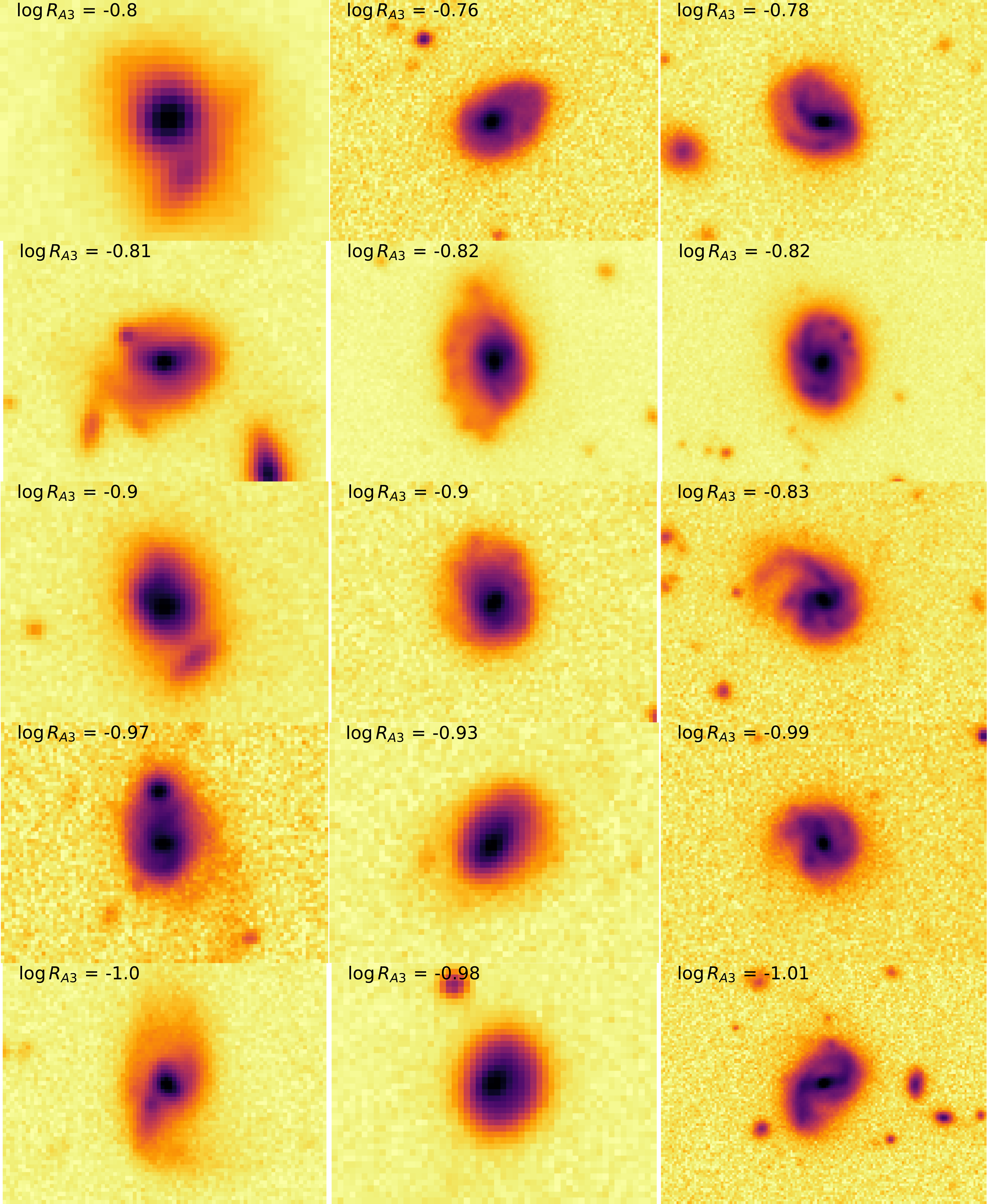}
\caption{Example $i$-band Stripe 82 images of star-forming galaxies with high asymmetries. All galaxies have $\log\,M_\star/M_\odot = 10-10.5$, and $z = 0.05-0.1$. The images are displayed using asinh scaling (with $a=0.05$ parameter), which is logarithmic at high surface brightness and linear at low surface brightness. They images show that the asymmetries arise from mergers/interactions, lopsidedness, or asymmetric spiral arms. \label{fig:cutout_highRA2}}
\end{figure*}

To summarize, we analyzed the multivariate SDSS Stripe 82 data of $\sim 3,700$ face-on SFGs using the framework of mutual information. We find that galaxies are a multi-parameter family; the variation in SFRs of Stripe 82 galaxies is not fully explained by one or two variables. The most predictive variable for \delSSFR after $M_\star$ is asymmetry, $R_{A3}$. This is true in most ($\sim 60-85$\%) mass slices (0.25 dex) in the range $3 \times 10^{9}\,M_\odot$ to $2 \times 10^{11}\,M_\odot$ (see also the Appendix that repeats the analysis for centrals only). In few cases when $C_1$ is the most predictive variable, $R_{A3}$ is ranked second. From inspection of individual images, we see that $R_{A3}$ reflects multiple factors, including irregular clumps and spiral arms comprised of young stars, lopsidedness in seemingly isolated galaxies, and structural perturbations by galaxy interactions or mergers (see Figure~\ref{fig:cutout_highRA2}). After asymmetry ($R_{A3}$), $C_1$, $n_g$, and $\sigma$ are, overall, the highest ranked variables. These variables predominate at different masses. The two next-leading parameters for galaxies below $< 3 \times 10^{10}\,M_\odot$ are likely $C_1$ and $n_g$, whereas the two next-leading variables terms for massive galaxies are $\sigma$ and $n_g$. Our analysis does not indicate that $R_{50}$ is the second important parameter after $M_\star$ for predicting \delSSFR in SFGs.

\section{Discussion}\label{sec:discussion}
 
This section discusses the possible role of recent gas accretion in explaining the observed link between galaxy structure and $\Delta\,\mathrm{SSFR}$. Gas might be accreted directly into galaxies or brought in by mergers. Both dynamical perturbations or direct gas accretions can induce asymmetries, drive gas to the center, elevate SFR, and build a central concentration of stars. Mergers and gas accretions are regulated by galaxy environments, which might also explain the link between star formation and galaxy structure. Observational studies like this one do not yet have measurements of all potentially important parameters (e.g., gas contents and flows, and environmental parameters) for a particular galaxy sample. We thus synthesize available results from previous studies to assemble a preliminary picture that can be tested in future work.

\subsection{The Role of wet Mergers and Interactions on Asymmetry and SFR Enhancements}

It is generally observed that galaxy mergers and interactions impact the asymmetry and structure of galaxies \citep[e.g.,][]{DePropris+07, Casteels+14, Patton+16, Cibinel+19} and also lead to central SFR enhancements \citep{Barrera-Ballesteros+15, Knapen+15, Chown+19, Thorp+19, Pan+19, Steffen+21}. For example, at $z \approx 0.1$, high \delSSFR galaxies such as starbursts or luminous infrared galaxies (LIRGs) are highly disturbed merger remnants \citep{Sanders+96,Veilleux+02,Yesuf+14,Ellison+13,Larson+16, Pawlik+16,Shangguan+19}. Likewise, most high redshift starbursts and LIRGs are also mergers or irregular galaxies \citep{Kartaltepe+12, Cibinel+19}. At $0.2 \le z \le 2$, \citet{Cibinel+19} found that the \delSSFR distribution of mergers spans the whole range of SFMS, but it is skewed toward high \delSSFR, with the median being \delSSFR $\sim 0.2-0.25$. Almost all galaxies with \delSSFR $> 0.6$ (i.e., starbursts) in \citet{Cibinel+19}'s sample are ongoing mergers. The majority of these galaxies are morphologically disturbed, late-stage mergers close to the coalescence phase. Using the DEEP2 survey, \citet{Lin+08} classified wet, dry, and mixed mergers according to the colors of the individual components in close pairs. These authors showed that wet mergers dominate merger events since $z \sim 1.2$, reaching $\sim 40\%-60\%$. Dry and mixed mergers, however, become more important over time, especially at $z \lesssim 0.2$. 

The fraction of asymmetric galaxies which are disturbed by mergers is not constrained well observationally, especially for asymmetries caused by minor mergers. Using the GAMA\footnote{Galaxy And Mass Assembly (GAMA) survey \citep{Driver+11}.} survey, \citet{Casteels+14} found a significant increase in mean asymmetries of mergers for projected separations less than the sum of the individual galaxys' Petrosian 90\% radii. For major merger pairs (with mass ratios $>$1/4), both galaxies show high asymmetries. In contrast, in minor merger pairs (with mass ratios $<$1/4), the lower mass companion is highly asymmetric, but the massive galaxy is less affected. The authors also found that the fraction of highly asymmetric galaxies that are ongoing major mergers is only $\sim 2-4$\%. In comparison, the minor merger rate (mass ratios of $1/4-1/10$) is $\sim 3-7$ times higher than the major merger rate \citep[e.g.,][]{Lotz+11,Rodriguez-Gomez+15}. Therefore, roughly $\sim 10-30$\% of asymmetric galaxies might be the results of minor mergers. \citet{Kaviraj14} inferred that around 40\% of the star formation activity in the $z < 0.07$ spiral galaxies is triggered by minor mergers. These authors used SDSS Stripe 82 imaging to visually classify bright ($r <16.8$) galaxies into spirals and E/S0s, and then calculated the SSFR enhancements in disturbed spirals by comparing them with a sample of undisturbed spirals. Although the merger rates are quite uncertain, it is indisputable that some of the observed asymmetries/disturbances in SFGs are linked to galaxy mergers or interactions \citep{DePropris+07,Casteels+14, Kaviraj14}.

\citet{Reichard+09} investigated the link between lopsidedness\footnote{The authors defined lopsidedness as the radially averaged $m=1$ Fourier amplitude between the radii enclosing 50\% and 90\% of the galaxy light.}, central SSFR, metallicity, and the presence of AGN using shallow SDSS images. These authors also found strong links between lopsidedness and recent/ongoing central star formation: the more lopsideded a galaxy, the younger its stellar population. Starburst galaxies are on average the most lopsided. The observed correlation between lopsidedness and SFR \citep{Rudnick+00,Reichard+09} suggests a mutual triggering event.

In general, lopsidedness is common in SFGs and may be triggered by several phenomena \citep{Richter+94,Rix+95,Zaritsky+97,Bournaud+05, Mapelli+08, Reichard+08,JogCombes09,Zaritsky+13} including mergers and interactions. Asymmetric gas accretion, gravitational instability, and/or asymmetric (off-center) dark matter halo are some of the other physical mechanisms that explain the origin of the lopsidedness in galaxies. The simultaneous occurrence of the $m=1$ lopsided asymmetry and the $m=2$ spiral arm modes is also common \citep{Bournaud+05,vanEymeren+11,Zaritsky+13}. The observation that lopsidedness is more frequent in late-type galaxies than in early-type galaxies was also used as evidence against the sole merger origin \citep{Bournaud+05}; galaxy mergers/interactions transform galaxies into early-types after the interactions. 

Previous estimates, albeit very uncertain, indicate that the morphological asymmetries are fairly long-lived \citep[$\sim 0.2-2$\,Gyr][]{Lotz+10a,Lotz+10b,Whitney+21}. Using simulations of mergers of varying mass ratios and gas fractions, \citet{Lotz+10a} found that the timescale for detecting a merger with high asymmetry depends strongly on the gas fraction of the primary galaxy. For a major mergers, the authors found that the timescale is $\sim 0.3-0.7$\,Gyr for $f_\mathrm{gas} \sim 20-40\%$. For minors mergers, it may be $< 0.06$\,Gyr for $f_\mathrm{gas} \lesssim 20\%$ or similar to that of the major mergers for higher $f_\mathrm{gas}$. Using the Illustris TNG300 simulations, \citet{Whitney+21} recently found that the mean merger timescale for mergers identified using the concentration, asymmetry, and smoothness (CAS) system is $0.56^{+0.23}_{-0.18}$\,Gyr. Furthermore, the lack of strong correlation between the strength of lopsidedness with the tidal interaction parameter or the presence of nearby companions indicates that lopsidedness must be a long-lived event if it is caused by galaxy interactions/mergers \citep[$\sim 1-2$ Gyr;][]{Wilcots+04, Bournaud+05, Saha+07, JogCombes09, vanEymeren+11,Yozin+14,Ghosh+21}. Incidentally, the time gas takes to fall in from the halo virial radius to the position of a simulated galaxy is also in the range $\sim 0.1-2$\,Gyr \citep{Nelson+15}. 

We have attempted to quantify the relationship between mergers and asymmetry for our galaxies. A preliminary visual inspection and a neighbor count for our sample indicates that about 65\% of SFGs with $R_{A3} > 0.1$ in our sample do not have close neighbors within $10-100$\,kpc, and that $\gtrsim 80$\% the SFGs with $R_{A3} > 0.1$ are not conspicuous merger remnants. Most of these galaxies are asymmetric due to asymmetric spiral arms and/or lopsidedness (see also Figure~\ref{fig:cutout_highRA2}). Moreover, we find that the anti-correlation between $R_{A3}$ and the first nearest neighbor distance is weak ($\rho \approx -0.05$). Furthermore, $\sim 70$\% of galaxies with $R_{A3} > 0.05$ (the median for SFGs) do not also have close neighbors within $10-100$\,kpc. We defer more quantitative analysis of different modes of asymmetry and their relationship with the environment for future work. In general, our visual estimates agree with published studies suggesting that observed asymmetries are due to multiple causes with some preponderance of the evidence suggesting that most asymmetries are not caused by mergers. With better estimates and understanding of the asymmetry timescale, the observed correlation between \delSSFR and asymmetry in SFGs may help constrain the evolution timescales of galaxies across SFMS. As discussed in the introduction, different galaxy evolution models predict drastically different SFR variability timescales \citep[e.g.,][]{Iyer+20}. Overall, the existing evidence points to fairly long-term variations in SFR, $\sim 0.2-2$\,Gyr, for merger perturbations and similarly for perturbations due to diffuse gas accretion.

\subsection{The Relationship of Gas, Asymmetry, and SFR}

Numerous observations indicate that SFR is linked to or regulated by the amount of gas in galaxies \citep[e.g.,][]{Kennicutt+98, Genzel+15, Saintonge+17, Catinella+18,Yesuf+20b}. Although directly observing gas accretion is difficult, the observed short gas depletion timescale ($\sim 1-2$\,Gyr) indicates the need for a continuous gas accretion to sustain the observed cosmic star formation \citep[e.g.,][]{Sancisi+08,Sanchez+14}. Current observational estimates indicate that gas accretion rates from minor mergers are not enough ($\sim 0.1-0.3\,M_\odot$yr$^{-1}$) to sustain the observed SFR in local galaxies \citep{Sancisi+08,DiTeodoro+14}. Furthermore, galaxy simulations indicate that gas infall rates onto dark matter halos are dominated by the diffuse component over the merger contribution, at least for galaxies in low mass halos $M_h < 10^{13}\,M_\odot$\footnote{Based on existing group catalogs \citep{Tinker+20,Yang+21}, most galaxies in our sample live in low mass halos below this limit.}\citep{Fakhouri+10,vandeVoort+11,LHuillier+12, Wright+21}. Some simulations indicate that the variation in gas accretion rates may offer a physical explanation for the stratification of the SFMS; at fixed $M_\star$, simulated central galaxies in haloes experiencing high gas accretion rates preferentially lie above the SFMS \citep{Sanchez+18,Wright+21}. While the details are not yet well understood, asymmetric but diffuse gas accretion rates of a few $M_\odot$\,yr$^{-1}$ along cosmological filaments may plausibly explain the ubiquity of asymmetries in strongly star-forming galaxies \citep{Keres+05,Bournaud+05,Sancisi+08,JogCombes09,Cen+14}.

Consistent with the possibility of recent gas accretion, \citet{Wang+11} found a correlation between the asymmetry of the stellar disk and the \ion{H}{1} mass fraction in SDSS galaxies with $M_\star >10^{10}\,M_\odot$ and $z < 0.05$. The asymmetric galaxies have both enhanced SSFR (as traced by $NUV-r$ color) and gas content. Note that the typical galaxies in \citet{Wang+11}'s sample are not mergers, and their asymmetry measurements are based on shallow SDSS images. Moreover, \citet{Espada+11} found tentative evidence that isolated galaxies with high FIR emission ($> 10^{10} L_\odot$) show higher \ion{H}{1} asymmetry than isolated galaxies with low FIR emission. The prevalent \ion{H}{1} asymmetry in isolated galaxies cannot be due to recent galaxy interactions \citep[see also][]{Baldwin+80, Richter+94,Haynes+98, Matthews+98,Vulcani+18}. In comparison, (post-)mergers also on average exhibit elevated atomic and molecular gas amounts compared to isolated galaxies \citep{Lisenfeld+11,Stark+13,Silverman+15,Ellison+18,Violino+18}. Using the GOALS\footnote{The Great Observatories All-sky LIRG Survey} sample, \citet{Shangguan+19} found that luminous infrared galaxies (LIRGs) have higher gas fractions than those of main-sequence SFGs. The most gas-rich LIRGs are late-stage mergers. The merger stage for the GOALS galaxies was determined from HST images. In addition, a small but significant difference between the \ion{H}{1} asymmetry distributions of close pairs and isolated galaxies was also observed \citep{Bok+19}. In short, existing data suggest that both diffuse gas accretion and wet mergers are contributing to the asymmetry and \delSSFR enhancements.

Furthermore, it is well known that mergers and disturbed galaxies also exhibit low gas-phase metallicities compared to isolated galaxies of similar $M_\star$ \citep{Ellison+08, Reichard+09,HwangHsiang+19,Thorp+19,Bustamante+20}. This is interpreted as metallicity dilution by recent gas accretion \citep{Forbes+14,Torrey+19,Collacchioni+20}.
\citet{Reichard+09} found that lopsided galaxies have lower gas phase metallicity than the average mass-metallicity relation at a given $M_\star$. \citet{HwangHsiang+19} argued that the low-metallicity ionized gas in their sample originates from a combination of galaxy interactions, mergers, and gas accretions from the circum-galactic media. They used the SDSS MaNGA\footnote{Mapping Nearby Galaxies at Apache Point Observatory} data of late-type SFGs to identify regions in which the gas-phase metallicity is low compared to the expectation from the mass-metallicity relation. The authors found that the incidence rate of the low-metallicity gas is significantly higher in low-mass ($M < 3 \times 10^{10}\,M_\odot$) and asymmetric SFGs, which are mostly not interacting or mergers. In short, the observed metallicity and asymmetry connection indirectly supports the role of diffuse gas accretion in some galaxies in establishing the observed relationship between \delSSFR and asymmetry.

In summary, morphological asymmetries are commonly observed in galaxies, and they are associated with SFR enhancements. Many processes such as galaxy merging, tidal interactions, and asymmetric gas accretion can cause asymmetric stellar disks. The dominant process has yet to be identified. Gas coming into galaxies from both direct accretion and from mergers is certainly important. Based on synthesis of previous results, we consider the alternative that diffuse gas accretion can lead to both asymmetry in the stellar distribution and SFR enhancement. Future observational constraints on the gas content of our sample are needed to distinguish between (1) enhanced gas supply from diffuse accretion, which increases both \delSSFR and asymmetry versus (2) enhanced \delSSFR at fixed gas fraction in asymmetric galaxies relative to normal galaxies.

\subsection{Signature of Compaction and Movement Across the SFMS?}

In accordance with the compaction scenario \citep{Dekel+14b,Zolotov+15,Tacchella+16}, a galaxy on the SFMS may experience some trigger (merger, interaction, or some jolt) and/or a fresh infusion of gas, the gas falls to the center, SFR in the center goes up, thereby enhancing the central stellar concentration. Then, the galaxy exhausts it fuel or the fuel is ejected by feedback, and the galaxy traverses the SFMS quickly and drops below it. 

Are galaxies actually crossing the SFMS in large numbers? This would be difficult with major mergers, which not only are rare but also permanently alter the galaxy structure; mergers move mass to the center and destroy disks. Because SFGs with both bulges and disks are prevalent, the disks would need to rebuild. In contrast, diffuse gas variations inject gas without major perturbations. It is at least plausible that variations in diffuse infall might move galaxies up and down in \delSSFR without perturbing their morphologies. The prevalence of asymmetry (lopsideddness and spiral arms) in relatively isolated galaxies might indicate that galaxies probably cross the SFMS in large numbers in $\sim 1-2$\,Gyr timescales. But if long-term perturbations related to dark matter halos are dominating, then galaxies do not bob up and down about the SFMS on short timescales -- they stay above or below the SFMS for long times \citep{Abramson+16,Diemer+17, Matthee+19,Iyer+20,Berti+20}. 

In general, \delSSFR is anti-correlated with the clustering amplitude of SFGs measured on the scales of $\gtrsim1$ Mpc \citep{Li+08,Berti+20}. Galaxies above the ridge of the SFMS are less clustered than those below it, at a fixed $M_\star$, implying a very long-lived evolutionary effect. On scales $\lesssim 100$\,kpc, however, \citet{Li+08} found that the clustering amplitude for SFGs is not monotonic (i.e., it is V-shaped) with SSFR; it increases with increasing SSFR for SSFR between $\sim 0.1-1$\,Gyr$^{-1}$ and decreases with SSFR between $\sim 0.01-0.1$\,Gyr$^{-1}$ \citep[see also][]{Gunawardhana+18}. \citet{Li+08} interpreted this trend as a signature of tidal interactions, which led to enhancement in star formation and morphological transformation. Furthermore, \citet{Wuyts+11} found that the S\'{e}rsic index $n$ does not vary monotonically with \delSSFR. It also shows a reversal in the high end of \delSSFR. We find similar trends in our analysis for various morphological variables, the most remarkable being the central concentration, $C_1$, for galaxies with $M_\star  \lesssim 3 \times 10^{10}\,M_\odot$. Figure~\ref{fig:RA_exp}b shows that $C_1$ is lowest on the SFMS, but it increases above and below the SFMS. Note that some previous studies did not find compact SFGs at the high end of \delSSFR \citep[e.g.,][]{Morselli+19, Cook+20}. 

Although gas compaction events are expected to be rare in the nearby universe \citep{Dekel+14b}, where gas fractions are low and mergers are infrequent, some local galaxies (e.g., starburst and post-starburst galaxies) may still experience such events \citep{McIntosh+14,Yesuf+14,Yesuf+20b, Wang+20}. Recently, \citet{Wang+20} studied how \ion{H}{1} inside galaxies fuels star formation in late-type disk galaxies and found trends that are generally consistent with the compaction model. In particular, they found that \delSSFR is well correlated with the inner \ion{H}{1} surface density and inner \ion{H}{1} mass-to-stellar mass ratio. The highest gas density and the fastest depletion of \ion{H}{1} within the stellar disks are found in the most compact SFGs, at a given $M_\star$. \citet{WangLilly20a,WangLilly20b} have modeled the SFR fluctuations of MaNGA galaxies and concluded that the stellar populations and line strengths of these local galaxies are consistent with fairly rapid fluctuations of SFR in a Gyr or less timescale.

In summary, bobbing of galaxies is an important open question that will probably require a combination of kinematic modeling, stellar population modeling, and environmental statistics to finally resolve. If there is rapid bobbing, then galaxies at the bottom of SFMS are changing into galaxies at the top, and vice versa. In our opinion, the current data do not rule out bobbing of at least some galaxies. Hard-to-change parameters like $\sigma$ and $R_{50}$ are fairly constant across the SFMS. $R_{A3}$ and $C_1$ do change, but they are light-based ($i$-band). They might be effected by recent star formation that may fade on short timescales or that might have contributed only a small fraction of the total mass. Mass-based concentration measurements of galaxies, therefore, will be valuable to constrain the timescales of fluctuations about the SFMS ridgeline.

\subsection{Future Work: Comparison with Simulations and Improved Observations}

It is very challenging to infer galaxy evolutionary histories from the snapshot at $z=0$ alone, which includes the end result of recent and ancient evolution. We hope that the quantitative approaches and results presented in our work inspire detailed comparison with emerging state-of-the-art cosmological simulations. Simulations now have capabilities to generate more realistic mock images that mimic observations \citep[e.g.,][]{Snyder+15,Correa+17,Genel+18, Rodriguez-Gomez+19,Bignone+20}. They reproduce galaxy properties that are in broad in agreement with observations. Comparison with simulations will offer valuable insights about the underlying physical processes that gave rise to the complex trends observed in local galaxies \citep[for excellent case studies, see][]{Cortese+19, Rodriguez-Gomez+19}. 

The current work does not examine the impact of environment on star formation and galaxy structure. Numerous observational and theoretical studies, however,  show that environment is important in shaping galaxy encounters, star formation, morphology and gas properties of galaxies \citep[e.g.,][]{Dressler+80,Postman+84,Balogh+98, Hashimoto+00,Angiras+06,Boselli+06, Blanton+09, Lin+08, Ellison+10,Peng+10,Thomas+10,Wetzel+12,Kampczyk+13, Woo+15, vandeVoort+17, Hwang+19}. For example, \citet{vandeVoort+17} used EAGLE simulations to study the environmental dependence of gas accretion onto galaxies. They found a strong suppression of gas accretion rates in massive haloes and in dense environments, especially for satellite galaxies at smaller halo-centric distances. The SFRs of both centrals and satellites show similar behavior to their gas accretion rates. The authors therefore concluded that the environmental suppression of gas accretion could directly lead to the quenching of star formation. On the other hand, low halo mass galaxies ($M_h < 3 \times 10^{11}\,M_\odot$) are primarily fed by cold, filamentary gas streams, especially if they live in low-density environments \citep{Keres+05,Keres+09,Fakhouri+10, vandeVoort+17}. The majority of galaxies in our sample have low halo masses (are centrals) and live in the field or in poor groups \citep{Tinker+20,Yang+21}. Therefore, both cold gas accretion and merging are viable mechanisms for explaining the observed correlation between \delSSFR and asymmetry.

The structural parameters ($R_{A}$, $\mu$, $C_1$, $n_g$, and $R_{50}$) may not be sufficiently representative. In the future, more parameters such as the Gini coefficient and the second-order moment of the 20\% brightest pixels \citep[$G-M_{20}$;][]{Lotz+04} should be included. Apparently, in addition to morphological parameters based on high quality images, measurements of environment, and cold gas are vital to disentangle the physical processes that resulted in the diversity of SFHs and structures of galaxies today. Hopefully, future studies will include all these parameters in multivariate analysis like this one.

\section{Summary and Conclusions}\label{sec:conclusion} 

We use the statistical framework of mutual information to quantify the inter-dependence among several structural variables and to rank their relevance to predicting \delSSFR, controlling for the redundancy within the selected variables. We apply this framework to study $\sim 3,700$ face-on SFGs in SDSS Stripe 82 survey. The deep $i$-band imaging data from this survey result in more reliable measurements of structural variables such as asymmetry \citep{Bottrell+19}. 

Our main conclusions are:

\begin{itemize}

\item Multiple structural variables are helpful for predicting SFR of SFGs, although some have minor contributions. After $M_\star$, morphological asymmetry ($R_{A3}$) is the most important predictor of variations in SSFR on the SFMS. SFGs with higher asymmetry have higher \delSSFR. The asymmetry reflects both asymmetric spiral arms and lopsidedness in seemingly isolated galaxies and structural perturbations by galaxy mergers or interactions.
\item The exact ranks of the variables may depend on how the galaxy sample is subdivided (e.g., by $M_\star$). After asymmetry ($R_{A3}$), $C_1$, $n_g$, and $\sigma$ are, overall, the highest ranked variables. The two next-leading parameters for galaxies below $< 3 \times 10^{10}\,M_\odot$ are likely $C_1$ and $n_g$, whereas the two next-leading variables terms for massive galaxies are $\sigma$ and $n_g$.
\item The SFMS is populated by galaxies with diverse structure, even at constant $M_\star$. There are significant numbers of bulge-dominated, concentrated, and/or compact SFGs with similar morphologies as QGs of similar $M_\star$. SFGs with enhanced SSFR, some of which are starbursts, are likely more concentrated and/or asymmetric.
\item Two interpretations are developed to explain the positive correlation between high asymmetry $R_{A3}$ and higher star formation rate on the SFMS. One is that extra star formation is driven by mergers and tidal encounters. The other is that it is driven by enhanced diffuse gas accretion. Present data do not permit a clear choice between these two accretion modes and, indeed, both may be operating in different cases. However, the two modes may have different timescales: mergers and tidal perturbations have short lifetimes, in which case the residence times above and below the SFMS would be brief. Diffuse accretion fluctuations might be longer-lived and causes galaxies to remain above and below the SFMS for longer periods. It is hoped that improved galaxy simulations together with more extensive environmental data will explain the origin of galaxies' SFMS residuals.
\end{itemize}

We speculate that the central concentration of gas and star formation through these two channels also drives bulge build up. Some SFGs probably compactify by a mechanism that is linked with enhanced star formation and morphological asymmetries. At the moment, we do not know how much of the bulge mass is built recently, perhaps $\sim 5-30$\% of the total mass \citep{Rudnick+00,Thomas+06, Kaviraj+07, Mendel+13}. The build up of the bulge, then, may generate its own agent for quenching. But establishing this picture would require much more evidence than we presented. We hope future work will examine it in detail.

\acknowledgments

We are very thankful to the anonymous referee for the helpful comments and suggestions that significantly improved the presentation of the paper. We acknowledge discussions with David Koo and Yifei Luo and we thank them very much for their inputs for the interpretation of our results. We also thank Conner Bottrell, Hua Gao, and John Silverman for useful comments and suggestions. 

LCH was supported by the National Key R \& D Program of China (2016YFA0400702) and the National Science Foundation of China (11721303, 11991052). H. Yesuf was supported by The Research Fund for International Young Scientists of NSFC (11950410492). S.~M Faber acknowledges support from NSF (AST-0808133 and AST-1615730).

Funding for SDSS has been provided by the Alfred P. Sloan Foundation, the Participating Institutions, the National Science Foundation, and the U.S. Department of Energy Office of Science. The SDSS-III web site is http://www.sdss3.org/.  SDSS-III is managed by the Astrophysical Research Consortium for the Participating Institutions of the SDSS-III Collaboration including the University of Arizona, the Brazilian Participation Group, Brookhaven National Laboratory, Carnegie Mellon University, University of Florida, the French Participation Group, the German Participation Group, Harvard University, the Instituto de Astrofisica de Canarias, the Michigan State/Notre Dame/JINA Participation Group, Johns Hopkins University, Lawrence Berkeley National Laboratory, Max Planck Institute for Astrophysics, Max Planck Institute for Extraterrestrial Physics, New Mexico State University, New York University, Ohio State University, Pennsylvania State University, University of Portsmouth, Princeton University, the Spanish Participation Group, University of Tokyo, University of Utah, Vanderbilt University, University of Virginia, University of Washington, and Yale University.

\software{astropy \citep{astropy13, astropy18}, corrplot \citep{corrplot21}, GalSim \citep{Rowe+15}, GIM2D \citep{Simard+02}, loess \citep{Cappellari+13}, matplotlib \citep{Hunter07}, seaborn \citep{Waskom21}, varrank \citep{Kratzer+18}}

\begin{thebibliography}{}
\expandafter\ifx\csname natexlab\endcsname\relax\def\natexlab#1{#1}\fi
\bibitem[{{Abraham} {et~al.}(1996){Abraham}, {Tanvir}, {Santiago}, {Ellis},
  {Glazebrook}, \& {van den Bergh}}]{Abraham+96}
{Abraham}, R.~G., {Tanvir}, N.~R., {Santiago}, B.~X., {et~al.} 1996, \mnras,
  279, L47

\bibitem[{{Abraham} {et~al.}(1994){Abraham}, {Valdes}, {Yee}, \& {van den
  Bergh}}]{Abraham+94}
{Abraham}, R.~G., {Valdes}, F., {Yee}, H.~K.~C., \& {van den Bergh}, S. 1994,
  \apj, 432, 75

\bibitem[{{Abramson} {et~al.}(2016){Abramson}, {Gladders}, {Dressler},
  {Oemler}, {Poggianti}, \& {Vulcani}}]{Abramson+16}
{Abramson}, L.~E., {Gladders}, M.~D., {Dressler}, A., {et~al.} 2016, \apj, 832,
  7

\bibitem[{{Alpaslan} \& {Tinker}(2020)}]{Alpaslan+20}
{Alpaslan}, M., \& {Tinker}, J.~L. 2020, \mnras, 496, 5463

\bibitem[{{Angiras} {et~al.}(2006){Angiras}, {Jog}, {Omar}, \&
  {Dwarakanath}}]{Angiras+06}
{Angiras}, R.~A., {Jog}, C.~J., {Omar}, A., \& {Dwarakanath}, K.~S. 2006,
  \mnras, 369, 1849

\bibitem[{{Annis} {et~al.}(2014){Annis}, {Soares-Santos}, {Strauss}, \& {Becker}}]{Annis+14} 
  {Annis}, J., {Soares-Santos}, M.,  {Strauss}, M. A, {Becker}, A. C., {et~al.} 2014, \apj, 794, 120


\bibitem[{{Astropy Collaboration} {et~al.}(2013){Astropy Collaboration},
  {Robitaille}, {Tollerud}, {Greenfield}, {Droettboom}, {Bray}, {Aldcroft},
  {Davis}, {Ginsburg}, {Price-Whelan}, {Kerzendorf}, {Conley}, {Crighton},
  {Barbary}, {Muna}, {Ferguson}, {Grollier}, {Parikh}, {Nair}, {Unther},
  {Deil}, {Woillez}, {Conseil}, {Kramer}, {Turner}, {Singer}, {Fox}, {Weaver},
  {Zabalza}, {Edwards}, {Azalee Bostroem}, {Burke}, {Casey}, {Crawford},
  {Dencheva}, {Ely}, {Jenness}, {Labrie}, {Lim}, {Pierfederici}, {Pontzen},
  {Ptak}, {Refsdal}, {Servillat}, \& {Streicher}}]{astropy13}
{Astropy Collaboration}, {Robitaille}, T.~P., {Tollerud}, E.~J., {et~al.} 2013,
  \aap, 558, A33

\bibitem[{{Astropy Collaboration} {et~al.}(2018){Astropy Collaboration},
  {Price-Whelan}, {Sip{\H{o}}cz}, {G{\"u}nther}, {Lim}, {Crawford}, {Conseil},
  {Shupe}, {Craig}, {Dencheva}, {Ginsburg}, {VanderPlas}, {Bradley},
  {P{\'e}rez-Su{\'a}rez}, {de Val-Borro}, {Aldcroft}, {Cruz}, {Robitaille},
  {Tollerud}, {Ardelean}, {Babej}, {Bach}, {Bachetti}, {Bakanov}, {Bamford},
  {Barentsen}, {Barmby}, {Baumbach}, {Berry}, {Biscani}, {Boquien}, {Bostroem},
  {Bouma}, {Brammer}, {Bray}, {Breytenbach}, {Buddelmeijer}, {Burke},
  {Calderone}, {Cano Rodr{\'\i}guez}, {Cara}, {Cardoso}, {Cheedella}, {Copin},
  {Corrales}, {Crichton}, {D'Avella}, {Deil}, {Depagne}, {Dietrich}, {Donath},
  {Droettboom}, {Earl}, {Erben}, {Fabbro}, {Ferreira}, {Finethy}, {Fox},
  {Garrison}, {Gibbons}, {Goldstein}, {Gommers}, {Greco}, {Greenfield},
  {Groener}, {Grollier}, {Hagen}, {Hirst}, {Homeier}, {Horton}, {Hosseinzadeh},
  {Hu}, {Hunkeler}, {Ivezi{\'c}}, {Jain}, {Jenness}, {Kanarek}, {Kendrew},
  {Kern}, {Kerzendorf}, {Khvalko}, {King}, {Kirkby}, {Kulkarni}, {Kumar},
  {Lee}, {Lenz}, {Littlefair}, {Ma}, {Macleod}, {Mastropietro}, {McCully},
  {Montagnac}, {Morris}, {Mueller}, {Mumford}, {Muna}, {Murphy}, {Nelson},
  {Nguyen}, {Ninan}, {N{\"o}the}, {Ogaz}, {Oh}, {Parejko}, {Parley}, {Pascual},
  {Patil}, {Patil}, {Plunkett}, {Prochaska}, {Rastogi}, {Reddy Janga},
  {Sabater}, {Sakurikar}, {Seifert}, {Sherbert}, {Sherwood-Taylor}, {Shih},
  {Sick}, {Silbiger}, {Singanamalla}, {Singer}, {Sladen}, {Sooley},
  {Sornarajah}, {Streicher}, {Teuben}, {Thomas}, {Tremblay}, {Turner},
  {Terr{\'o}n}, {van Kerkwijk}, {de la Vega}, {Watkins}, {Weaver}, {Whitmore},
  {Woillez}, {Zabalza}, \& {Astropy Contributors}}]{astropy18}
{Astropy Collaboration}, {Price-Whelan}, A.~M., {Sip{\H{o}}cz}, B.~M., {et~al.}
  2018, \aj, 156, 123


\bibitem[{{Baldwin} {et~al.}(1980){Baldwin}, {Lynden-Bell}, \&
  {Sancisi}}]{Baldwin+80}
{Baldwin}, J.~E., {Lynden-Bell}, D., \& {Sancisi}, R. 1980, \mnras, 193, 313

\bibitem[{{Balogh} {et~al.}(1998){Balogh}, {Schade}, {Morris}, {Yee},
  {Carlberg}, \& {Ellingson}}]{Balogh+98}
{Balogh}, M.~L., {Schade}, D., {Morris}, S.~L., {et~al.} 1998, \apjl, 504, L75

\bibitem[{{Barnes} \& {Hernquist}(1991)}]{Barnes+91}
{Barnes}, J.~E., \& {Hernquist}, L.~E. 1991, \apjl, 370, L65

\bibitem[{{Barone} {et~al.}(2020){Barone}, {D'Eugenio}, {Colless}, \&
  {Scott}}]{Barone+20}
{Barone}, T.~M., {D'Eugenio}, F., {Colless}, M., \& {Scott}, N. 2020, \apj,
  898, 62

\bibitem[{{Barrera-Ballesteros} {et~al.}(2015){Barrera-Ballesteros},
  {S{\'a}nchez}, {Garc{\'\i}a-Lorenzo}, {Falc{\'o}n-Barroso}, {Mast},
  {Garc{\'\i}a-Benito}, {Husemann}, {van de Ven}, {Iglesias-P{\'a}ramo},
  {Rosales-Ortega}, {P{\'e}rez-Torres}, {M{\'a}rquez}, {Kehrig}, {Marino},
  {Vilchez}, {Galbany}, {L{\'o}pez-S{\'a}nchez}, {Walcher}, \& {Califa
  Collaboration}}]{Barrera-Ballesteros+15}
{Barrera-Ballesteros}, J.~K., {S{\'a}nchez}, S.~F., {Garc{\'\i}a-Lorenzo}, B.,
  {et~al.} 2015, \aap, 579, A45

\bibitem[{{Barro} {et~al.}(2017){Barro}, {Faber}, {Koo}, {Dekel}, {Fang},
  {Trump}, {P{\'e}rez-Gonz{\'a}lez}, {Pacifici}, {Primack}, {Somerville},
  {Yan}, {Guo}, {Liu}, {Ceverino}, {Kocevski}, \& {McGrath}}]{Barro+17}
{Barro}, G., {Faber}, S.~M., {Koo}, D.~C., {et~al.} 2017, \apj, 840, 47

\bibitem[{{Behroozi} {et~al.}(2019){Behroozi}, {Wechsler}, {Hearin}, \&
  {Conroy}}]{Behroozi+19}
{Behroozi}, P., {Wechsler}, R.~H., {Hearin}, A.~P., \& {Conroy}, C. 2019,
  \mnras, 488, 3143

\bibitem[{{Belfiore} {et~al.}(2019){Belfiore}, {Vincenzo}, {Maiolino}, \&
  {Matteucci}}]{Belfiore+19}
{Belfiore}, F., {Vincenzo}, F., {Maiolino}, R., \& {Matteucci}, F. 2019,
  \mnras, 487, 456

\bibitem[{{Bell} {et~al.}(2012){Bell}, {van der Wel}, {Papovich}, {Kocevski},
  {Lotz}, {McIntosh}, {Kartaltepe}, {Faber}, {Ferguson}, {Koekemoer}, {Grogin},
  {Wuyts}, {Cheung}, {Conselice}, {Dekel}, {Dunlop}, {Giavalisco},
  {Herrington}, {Koo}, {McGrath}, {de Mello}, {Rix}, {Robaina}, \&
  {Williams}}]{Bell+12}
{Bell}, E.~F., {van der Wel}, A., {Papovich}, C., {et~al.} 2012, \apj, 753, 167

\bibitem[{{Berti} {et~al.}(2020){Berti}, {Coil}, {Hearin}, \&
  {Behroozi}}]{Berti+20}
{Berti}, A.~M., {Coil}, A.~L., {Hearin}, A.~P., \& {Behroozi}, P.~S. 2020,\aj, 161, 49

\bibitem[{{Bignone} {et~al.}(2020){Bignone}, {Pedrosa}, {Trayford}, {Tissera},
  \& {Pellizza}}]{Bignone+20}
{Bignone}, L.~A., {Pedrosa}, S.~E., {Trayford}, J.~W., {Tissera}, P.~B., \&
  {Pellizza}, L.~J. 2020, \mnras, 491, 3624

\bibitem[{{Blanton} \& {Moustakas}(2009)}]{Blanton+09}
{Blanton}, M.~R., \& {Moustakas}, J. 2009, \araa, 47, 159

\bibitem[{{Bloom} {et~al.}(2017){Bloom}, {Fogarty}, {Croom}, {Schaefer},
  {Bryant}, {Cortese}, {Richards}, {Bland-Hawthorn}, {Ho}, {Scott},
  {Goldstein}, {Medling}, {Brough}, {Sweet}, {Cecil}, {L{\'o}pez-S{\'a}nchez},
  {Glazebrook}, {Parker}, {Allen}, {Goodwin}, {Green}, {Konstantopoulos},
  {Lawrence}, {Lorente}, {Owers}, \& {Sharp}}]{Bloom+17}
{Bloom}, J.~V., {Fogarty}, L.~M.~R., {Croom}, S.~M., {et~al.} 2017, \mnras,
  465, 123

\bibitem[{{Bluck} {et~al.}(2020){Bluck}, {Maiolino}, {Piotrowska}, {Trussler},
  {Ellison}, {S{\'a}nchez}, {Thorp}, {Teimoorinia}, {Moreno}, \&
  {Conselice}}]{Bluck+20}
{Bluck}, A. F.~L., {Maiolino}, R., {Piotrowska}, J.~M., {et~al.} 2020, \mnras,
  499, 230

\bibitem[{{Bok} {et~al.}(2019){Bok}, {Blyth}, {Gilbank}, \& {Elson}}]{Bok+19}
{Bok}, J., {Blyth}, S.~L., {Gilbank}, D.~G., \& {Elson}, E.~C. 2019, \mnras,
  484, 582

\bibitem[{{Boselli} \& {Gavazzi}(2006)}]{Boselli+06}
{Boselli}, A., \& {Gavazzi}, G. 2006, \pasp, 118, 517

\bibitem[{{Bottrell} {et~al.}(2019){Bottrell}, {Simard}, {Mendel}, \&
  {Ellison}}]{Bottrell+19}
{Bottrell}, C., {Simard}, L., {Mendel}, J.~T., \& {Ellison}, S.~L. 2019,
  \mnras, 486, 390

\bibitem[{{Bouch{\'e}} {et~al.}(2010){Bouch{\'e}}, {Dekel}, {Genzel}, {Genel},
  {Cresci}, {F{\"o}rster Schreiber}, {Shapiro}, {Davies}, \&
  {Tacconi}}]{Bouche+10}
{Bouch{\'e}}, N., {Dekel}, A., {Genzel}, R., {et~al.} 2010, \apj, 718, 1001

\bibitem[{{Bournaud} {et~al.}(2005){Bournaud}, {Combes}, {Jog}, \&
  {Puerari}}]{Bournaud+05}
{Bournaud}, F., {Combes}, F., {Jog}, C.~J., \& {Puerari}, I. 2005, \aap, 438,
  507

\bibitem[{{Brinchmann} {et~al.}(2004){Brinchmann}, {Charlot}, {White},
  {Tremonti}, {Kauffmann}, {Heckman}, \& {Brinkmann}}]{Brinchmann+04}
{Brinchmann}, J., {Charlot}, S., {White}, S.~D.~M., {et~al.} 2004, \mnras, 351,
  1151
 
\bibitem[{{Bruzual} \& {Charlot}(2003)}]{Bruzual+03}
{Bruzual}, G. \&  {Charlot}, S. 2003, \mnras, 344, 1000

\bibitem[{{Bustamante} {et~al.}(2020){Bustamante}, {Ellison}, {Patton}, \&
  {Sparre}}]{Bustamante+20}
{Bustamante}, S., {Ellison}, S.~L., {Patton}, D.~R., \& {Sparre}, M. 2020,
  \mnras, 494, 3469

\bibitem[{{Caplar} \& {Tacchella}(2019)}]{Caplar+19}
{Caplar}, N., \& {Tacchella}, S. 2019, \mnras, 487, 3845

\bibitem[{{Cappellari} {et~al.}(2013){Cappellari}, {McDermid}, {Alatalo},
  {Blitz}, {Bois}, {Bournaud}, {Bureau}, {Crocker}, {Davies}, {Davis}, {de
  Zeeuw}, {Duc}, {Emsellem}, {Khochfar}, {Krajnovi{\'c}}, {Kuntschner},
  {Morganti}, {Naab}, {Oosterloo}, {Sarzi}, {Scott}, {Serra}, {Weijmans}, \&
  {Young}}]{Cappellari+13}
{Cappellari}, M., {McDermid}, R.~M., {Alatalo}, K., {et~al.} 2013, \mnras, 432,
  1862

\bibitem[{{Casteels} {et~al.}(2014){Casteels}, {Conselice}, {Bamford},
  {Salvador-Sol{\'e}}, {Norberg}, {Agius}, {Baldry}, {Brough}, {Brown},
  {Drinkwater}, {Driver}, {Graham}, {Bland-Hawthorn}, {Hopkins}, {Kelvin},
  {L{\'o}pez-S{\'a}nchez}, {Loveday}, {Robotham}, \&
  {V{\'a}zquez-Mata}}]{Casteels+14}
{Casteels}, K. R.~V., {Conselice}, C.~J., {Bamford}, S.~P., {et~al.} 2014,
  \mnras, 445, 1157

\bibitem[{{Catinella} {et~al.}(2018){Catinella}, {Saintonge}, {Janowiecki},
  {Cortese}, {Dav{\'e}}, {Lemonias}, {Cooper}, {Schiminovich}, {Hummels},
  {Fabello}, {Ger{\'e}b}, {Kilborn}, \& {Wang}}]{Catinella+18}
{Catinella}, B., {Saintonge}, A., {Janowiecki}, S., {et~al.} 2018, \mnras, 476,
  875

\bibitem[{{Cen}(2014)}]{Cen+14}
{Cen}, R. 2014, \apjl, 789, L21

\bibitem[{{Chabrier}(2003)}]{Chabrier03}
{Chabrier}, G. 2003, \pasp, 115, 763
 
 \bibitem[{{Chary} \& {Elbaz}(2001)}]{Chary+01}
{Chary}, R. \& {Elbaz}, D., 2001, \apj, 556, 562
 
 
\bibitem[{{Chen} {et~al.}(2020){Chen}, {Faber}, {Koo}, {Somerville}, {Primack},
  {Dekel}, {Rodr{\'\i}guez-Puebla}, {Guo}, {Barro}, {Kocevski}, {van der Wel},
  {Woo}, {Bell}, {Fang}, {Ferguson}, {Giavalisco}, {Huertas-Company}, {Jiang},
  {Kassin}, {Lin}, {Liu}, {Luo}, {Luo}, {Pacifici}, {Pandya}, {Salim}, {Shu},
  {Tacchella}, {Terrazas}, \& {Yesuf}}]{Chen+20}
{Chen}, Z., {Faber}, S.~M., {Koo}, D.~C., {et~al.} 2020, \apj, 897, 102

\bibitem[{{Cheung} {et~al.}(2012){Cheung}, {Faber}, {Koo}, {Dutton}, {Simard},
  {McGrath}, {Huang}, {Bell}, {Dekel}, {Fang}, {Salim}, {Barro}, {Bundy},
  {Coil}, {Cooper}, {Conselice}, {Davis}, {Dom{\'\i}nguez}, {Kassin},
  {Kocevski}, {Koekemoer}, {Lin}, {Lotz}, {Newman}, {Phillips}, {Rosario},
  {Weiner}, \& {Willmer}}]{Cheung+12}
{Cheung}, E., {Faber}, S.~M., {Koo}, D.~C., {et~al.} 2012, \apj, 760, 131

\bibitem[{{Chown} {et~al.}(2019){Chown}, {Li}, {Athanassoula}, {Li}, {Wilson},
  {Lin}, {Mo}, {Parker}, \& {Xiao}}]{Chown+19}
{Chown}, R., {Li}, C., {Athanassoula}, E., {et~al.} 2019, \mnras, 484, 5192

\bibitem[{{Cibinel} {et~al.}(2019){Cibinel}, {Daddi}, {Sargent}, {Le Floc'h},
  {Liu}, {Bournaud}, {Oesch}, {Amram}, {Calabr{\`o}}, {Duc}, {Pannella},
  {Puglisi}, {Perret}, {Elbaz}, \& {Kokorev}}]{Cibinel+19}
{Cibinel}, A., {Daddi}, E., {Sargent}, M.~T., {et~al.} 2019, \mnras, 485, 5631

\bibitem[{{Collacchioni} {et~al.}(2020){Collacchioni}, {Lagos}, {Mitchell},
  {Schaye}, {Wisnioski}, {Cora}, \& {Correa}}]{Collacchioni+20}
{Collacchioni}, F., {Lagos}, C. D.~P., {Mitchell}, P.~D., {et~al.} 2020,
  \mnras, 495, 2827

\bibitem[{{Conselice}(2014)}]{Conselice+14}
{Conselice}, C.~J. 2014, \araa, 52, 291

\bibitem[{{Conselice} {et~al.}(2000){Conselice}, {Bershady}, \&
  {Jangren}}]{Conselice+00}
{Conselice}, C.~J., {Bershady}, M.~A., \& {Jangren}, A. 2000, \apj, 529, 886

\bibitem[{{Cook} {et~al.}(2020){Cook}, {Cortese}, {Catinella}, \&
  {Robotham}}]{Cook+20}
{Cook}, R. H.~W., {Cortese}, L., {Catinella}, B., \& {Robotham}, A. 2020,
  \mnras, 493, 5596

\bibitem[{{Correa} {et~al.}(2017){Correa}, {Schaye}, {Clauwens}, {Bower},
  {Crain}, {Schaller}, {Theuns}, \& {Thob}}]{Correa+17}
{Correa}, C.~A., {Schaye}, J., {Clauwens}, B., {et~al.} 2017, \mnras, 472, L45

\bibitem[{{Cortese} {et~al.}(2019){Cortese}, {van de Sande}, {Lagos},
  {Catinella}, {Davies}, {Croom}, {Brough}, {Bryant}, {Lawrence}, {Owers},
  {Richards}, {Sweet}, \& {Bland -Hawthorn}}]{Cortese+19}
{Cortese}, L., {van de Sande}, J., {Lagos}, C.~P., {et~al.} 2019, \mnras, 485,
  2656

\bibitem[{Cover \& Thomas(2012)}]{CoverThomas12}
Cover, T.~M., \& Thomas, J.~A. 2012, Elements of Information Theory (John Wiley
  and Sons)

\bibitem[{{Daddi} {et~al.}(2007){Daddi}, {Dickinson}, {Morrison}, {Chary},
  {Cimatti}, {Elbaz}, {Frayer}, {Renzini}, {Pope}, {Alexander}, {Bauer},
  {Giavalisco}, {Huynh}, {Kurk}, \& {Mignoli}}]{Daddi+07}
{Daddi}, E., {Dickinson}, M., {Morrison}, G., {et~al.} 2007, \apj, 670, 156

\bibitem[{{Dav{\'e}} {et~al.}(2011){Dav{\'e}}, {Oppenheimer}, \&
  {Finlator}}]{Dave+11}
{Dav{\'e}}, R., {Oppenheimer}, B.~D., \& {Finlator}, K. 2011, \mnras, 415, 11

\bibitem[{{De Propris} {et~al.}(2007){De Propris}, {Conselice}, {Liske},
  {Driver}, {Patton}, {Graham}, \& {Allen}}]{DePropris+07}
{De Propris}, R., {Conselice}, C.~J., {Liske}, J., {et~al.} 2007, \apj, 666,
  212

\bibitem[{{Dekel} \& {Burkert}(2014)}]{Dekel+14b}
{Dekel}, A., \& {Burkert}, A. 2014, \mnras, 438, 1870

\bibitem[{{Dekel} \& {Mandelker}(2014)}]{Dekel+14a}
{Dekel}, A., \& {Mandelker}, N. 2014, \mnras, 444, 2071

\bibitem[{{D'Eugenio} {et~al.}(2018){D'Eugenio}, {Colless}, {Groves}, {Bian},
  \& {Barone}}]{DEugenio+18}
{D'Eugenio}, F., {Colless}, M., {Groves}, B., {Bian}, F., \& {Barone}, T.~M.
  2018, \mnras, 479, 1807

\bibitem[{{Di Teodoro} \& {Fraternali}(2014)}]{DiTeodoro+14}
{Di Teodoro}, E.~M., \& {Fraternali}, F. 2014, \aap, 567, A68

\bibitem[{{Diemer} {et~al.}(2017){Diemer}, {Sparre}, {Abramson}, \&
  {Torrey}}]{Diemer+17}
{Diemer}, B., {Sparre}, M., {Abramson}, L.~E., \& {Torrey}, P. 2017, \apj, 839,
  26

\bibitem[{{Dressler}(1980)}]{Dressler+80}
{Dressler}, A. 1980, \apj, 236, 351

\bibitem[{{Dressler} {et~al.}(2016){Dressler}, {Kelson}, {Abramson},
  {Gladders}, {Oemler}, {Poggianti}, {Mulchaey}, {Vulcani}, {Shectman},
  {Williams}, \& {McCarthy}}]{Dressler+16}
{Dressler}, A., {Kelson}, D.~D., {Abramson}, L.~E., {et~al.} 2016, \apj, 833,
  251

\bibitem[{{Driver} {et~al.}(2011){Driver}, {Hill}, {Kelvin}, {Robotham},
  {Liske}, {Norberg}, {Baldry}, {Bamford}, {Hopkins}, {Loveday}, {Peacock},
  {Andrae}, {Bland-Hawthorn}, {Brough}, {Brown}, {Cameron}, {Ching}, {Colless},
  {Conselice}, {Croom}, {Cross}, {de Propris}, {Dye}, {Drinkwater}, {Ellis},
  {Graham}, {Grootes}, {Gunawardhana}, {Jones}, {van Kampen}, {Maraston},
  {Nichol}, {Parkinson}, {Phillipps}, {Pimbblet}, {Popescu}, {Prescott},
  {Roseboom}, {Sadler}, {Sansom}, {Sharp}, {Smith}, {Taylor}, {Thomas},
  {Tuffs}, {Wijesinghe}, {Dunne}, {Frenk}, {Jarvis}, {Madore}, {Meyer},
  {Seibert}, {Staveley-Smith}, {Sutherland}, \& {Warren}}]{Driver+11}
{Driver}, S.~P., {Hill}, D.~T., {Kelvin}, L.~S., {et~al.} 2011, \mnras, 413,
  971

\bibitem[{{Elbaz} {et~al.}(2007){Elbaz}, {Daddi}, {Le Borgne}, {Dickinson},
  {Alexander}, {Chary}, {Starck}, {Brandt}, {Kitzbichler}, {MacDonald},
  {Nonino}, {Popesso}, {Stern}, \& {Vanzella}}]{Elbaz+07}
{Elbaz}, D., {Daddi}, E., {Le Borgne}, D., {et~al.} 2007, \aap, 468, 33

\bibitem[{{Elbaz} {et~al.}(2011){Elbaz}, {Dickinson}, {Hwang},
  {D{\'\i}az-Santos}, {Magdis}, {Magnelli}, {Le Borgne}, {Galliano},
  {Pannella}, {Chanial}, {Armus}, {Charmandaris}, {Daddi}, {Aussel}, {Popesso},
  {Kartaltepe}, {Altieri}, {Valtchanov}, {Coia}, {Dannerbauer}, {Dasyra},
  {Leiton}, {Mazzarella}, {Alexander}, {Buat}, {Burgarella}, {Chary}, {Gilli},
  {Ivison}, {Juneau}, {Le Floc'h}, {Lutz}, {Morrison}, {Mullaney}, {Murphy},
  {Pope}, {Scott}, {Brodwin}, {Calzetti}, {Cesarsky}, {Charlot}, {Dole},
  {Eisenhardt}, {Ferguson}, {F{\"o}rster Schreiber}, {Frayer}, {Giavalisco},
  {Huynh}, {Koekemoer}, {Papovich}, {Reddy}, {Surace}, {Teplitz}, {Yun}, \&
  {Wilson}}]{Elbaz+11}
{Elbaz}, D., {Dickinson}, M., {Hwang}, H.~S., {et~al.} 2011, \aap, 533, A119

\bibitem[{{Ellison} {et~al.}(2018){Ellison}, {Catinella}, \&
  {Cortese}}]{Ellison+18}
{Ellison}, S.~L., {Catinella}, B., \& {Cortese}, L. 2018, \mnras, 478, 3447

\bibitem[{{Ellison} {et~al.}(2013){Ellison}, {Mendel}, {Scudder}, {Patton}, \&
  {Palmer}}]{Ellison+13}
{Ellison}, S.~L., {Mendel}, J.~T., {Scudder}, J.~M., {Patton}, D.~R., \&
  {Palmer}, M. J.~D. 2013, \mnras, 430, 3128

\bibitem[{{Ellison} {et~al.}(2008){Ellison}, {Patton}, {Simard}, \&
  {McConnachie}}]{Ellison+08}
{Ellison}, S.~L., {Patton}, D.~R., {Simard}, L., \& {McConnachie}, A.~W. 2008,
  \aj, 135, 1877

\bibitem[{{Ellison} {et~al.}(2010){Ellison}, {Patton}, {Simard}, {McConnachie},
  {Baldry}, \& {Mendel}}]{Ellison+10}
{Ellison}, S.~L., {Patton}, D.~R., {Simard}, L., {et~al.} 2010, \mnras, 407,
  1514

\bibitem[{{Ellison} {et~al.}(2020){Ellison}, {Thorp}, {Lin}, {Pan}, {Bluck},
  {Scudder}, {Teimoorinia}, {S{\'a}nchez}, \& {Sargent}}]{Ellison+20b}
{Ellison}, S.~L., {Thorp}, M.~D., {Lin}, L., {et~al.} 2020, \mnras, 493, L39

\bibitem[{{Espada} {et~al.}(2011){Espada}, {Verdes-Montenegro}, {Huchtmeier},
  {Sulentic}, {Verley}, {Leon}, \& {Sabater}}]{Espada+11}
{Espada}, D., {Verdes-Montenegro}, L., {Huchtmeier}, W.~K., {et~al.} 2011,
  \aap, 532, A117

\bibitem[{Est{\'e}vez {et~al.}(2009)Est{\'e}vez, Tesmer, Perez, \&
  Zurada}]{Estevez+09}
Est{\'e}vez, P.~A., Tesmer, M., Perez, C.~A., \& Zurada, J.~M. 2009, IEEE
  Transactions on neural networks, 20, 189

\bibitem[{{Fakhouri} \& {Ma}(2010)}]{Fakhouri+10}
{Fakhouri}, O., \& {Ma}, C.-P. 2010, \mnras, 401, 2245

\bibitem[{{Fang} {et~al.}(2013){Fang}, {Faber}, {Koo}, \& {Dekel}}]{Fang+13}
{Fang}, J.~J., {Faber}, S.~M., {Koo}, D.~C., \& {Dekel}, A. 2013, \apj, 776, 63

\bibitem[{{Fang} {et~al.}(2018){Fang}, {Faber}, {Koo}, {Rodr{\'\i}guez-Puebla},
  {Guo}, {Barro}, {Behroozi}, {Brammer}, {Chen}, {Dekel}, {Ferguson},
  {Gawiser}, {Giavalisco}, {Kartaltepe}, {Kocevski}, {Koekemoer}, {McGrath},
  {McIntosh}, {Newman}, {Pacifici}, {Pandya}, {P{\'e}rez-Gonz{\'a}lez},
  {Primack}, {Salmon}, {Trump}, {Weiner}, {Willner}, {Acquaviva}, {Dahlen},
  {Finkelstein}, {Finlator}, {Fontana}, {Galametz}, {Grogin}, {Gruetzbauch},
  {Johnson}, {Mobasher}, {Papovich}, {Pforr}, {Salvato}, {Santini}, {van der
  Wel}, {Wiklind}, \& {Wuyts}}]{Fang+18}
{Fang}, J.~J., {Faber}, S.~M., {Koo}, D.~C., {et~al.} 2018, \apj, 858, 100

\bibitem[{{Fliri} \& {Trujillo}(2016)}]{Fliri+16}
{Fliri}, J., \& {Trujillo}, I. 2016, \mnras, 456, 1359

\bibitem[{{Forbes} {et~al.}(2014){Forbes}, {Krumholz}, {Burkert}, \&
  {Dekel}}]{Forbes+14}
{Forbes}, J.~C., {Krumholz}, M.~R., {Burkert}, A., \& {Dekel}, A. 2014, \mnras,
  443, 168

\bibitem[{{Franx} {et~al.}(2008){Franx}, {van Dokkum}, {F{\"o}rster Schreiber},
  {Wuyts}, {Labb{\'e}}, \& {Toft}}]{Franx+08}
{Franx}, M., {van Dokkum}, P.~G., {F{\"o}rster Schreiber}, N.~M., {et~al.}
  2008, \apj, 688, 770

\bibitem[{{Gadotti}(2009)}]{Gadotti09}
{Gadotti}, D.~A. 2009, \mnras, 393, 1531

\bibitem[{{Gallazzi} {et~al.}(2005){Gallazzi}, {Charlot}, {Brinchmann},
  {White}, \& {Tremonti}}]{Gallazzi+05}
{Gallazzi}, A., {Charlot}, S., {Brinchmann}, J., {White}, S. D.~M., \&
  {Tremonti}, C.~A. 2005, \mnras, 362, 41

\bibitem[{{Gao} {et~al.}(2020){Gao}, {Ho}, {Barth}, \& {Li}}]{Gao+20}
{Gao}, H., {Ho}, L.~C., {Barth}, A.~J., \& {Li}, Z.-Y. 2020, \apjs, 247, 20

\bibitem[{{Genel} {et~al.}(2018){Genel}, {Nelson}, {Pillepich}, {Springel},
  {Pakmor}, {Weinberger}, {Hernquist}, {Naiman}, {Vogelsberger}, {Marinacci},
  \& {Torrey}}]{Genel+18}
{Genel}, S., {Nelson}, D., {Pillepich}, A., {et~al.} 2018, \mnras, 474, 3976

\bibitem[{{Genzel} {et~al.}(2015){Genzel}, {Tacconi}, {Lutz}, {Saintonge},
  {Berta}, {Magnelli}, {Combes}, {Garc{\'\i}a-Burillo}, {Neri}, {Bolatto},
  {Contini}, {Lilly}, {Boissier}, {Boone}, {Bouch{\'e}}, {Bournaud}, {Burkert},
  {Carollo}, {Colina}, {Cooper}, {Cox}, {Feruglio}, {F{\"o}rster Schreiber},
  {Freundlich}, {Gracia-Carpio}, {Juneau}, {Kovac}, {Lippa}, {Naab}, {Salome},
  {Renzini}, {Sternberg}, {Walter}, {Weiner}, {Weiss}, \& {Wuyts}}]{Genzel+15}
{Genzel}, R., {Tacconi}, L.~J., {Lutz}, D., {et~al.} 2015, \apj, 800, 20

\bibitem[{{Ghosh} {et~al.}(2021){Ghosh}, {Saha}, {Jog}, {Combes}, \& {Di
  Matteo}}]{Ghosh+21}
{Ghosh}, S., {Saha}, K., {Jog}, C.~J., {Combes}, F., \& {Di Matteo}, P. 2021,
  arXiv e-prints, arXiv:2105.05270

\bibitem[{{Gladders} {et~al.}(2013){Gladders}, {Oemler}, {Dressler},
  {Poggianti}, {Vulcani}, \& {Abramson}}]{Gladders+13}
{Gladders}, M.~D., {Oemler}, A., {Dressler}, A., {et~al.} 2013, \apj, 770, 64

\bibitem[{{Graham} \& {Driver}(2005)}]{Graham+05}
{Graham}, A.~W., \& {Driver}, S.~P. 2005, \pasa, 22, 118

\bibitem[{{Gunawardhana} {et~al.}(2018){Gunawardhana}, {Norberg}, {Zehavi},
  {Farrow}, {Loveday}, {Hopkins}, {Davies}, {Wang}, {Alpaslan},
  {Bland-Hawthorn}, {Brough}, {Holwerda}, {Owers}, \&
  {Wright}}]{Gunawardhana+18}
{Gunawardhana}, M.~L.~P., {Norberg}, P., {Zehavi}, I., {et~al.} 2018, \mnras,
  479, 1433

\bibitem[{{Hani} {et~al.}(2020){Hani}, {Gosain}, {Ellison}, {Patton}, \&
  {Torrey}}]{Hani+20}
{Hani}, M.~H., {Gosain}, H., {Ellison}, S.~L., {Patton}, D.~R., \& {Torrey}, P.
  2020, \mnras, 493, 3716

\bibitem[{{Hashimoto} \& {Oemler}(2000)}]{Hashimoto+00}
{Hashimoto}, Y., \& {Oemler}, Augustus, J. 2000, \apj, 530, 652

\bibitem[{{Haynes} {et~al.}(1998){Haynes}, {Hogg}, {Maddalena}, {Roberts}, \&
  {van Zee}}]{Haynes+98}
{Haynes}, M.~P., {Hogg}, D.~E., {Maddalena}, R.~J., {Roberts}, M.~S., \& {van
  Zee}, L. 1998, \aj, 115, 62

\bibitem[{{Hopkins} {et~al.}(2006){Hopkins}, {Hernquist}, {Cox}, {Di Matteo},
  {Robertson}, \& {Springel}}]{Hopkins+06}
{Hopkins}, P.~F., {Hernquist}, L., {Cox}, T.~J., {et~al.} 2006, \apjs, 163, 1

\bibitem[{Hunter(2007)}]{Hunter07}
Hunter, J.~D. 2007, Computing in Science \& Engineering, 9, 90

\bibitem[{{Hwang} {et~al.}(2019{\natexlab{a}}){Hwang}, {Barrera-Ballesteros},
  {Heckman}, {Rowlands}, {Lin}, {Rodriguez-Gomez}, {Pan}, {Hsieh},
  {S{\'a}nchez}, {Bizyaev}, {S{\'a}nchez Almeida}, {Thilker}, {Lotz}, {Jones},
  {Nair}, {Andrews}, \& {Drory}}]{HwangHsiang+19}
{Hwang}, H.-C., {Barrera-Ballesteros}, J.~K., {Heckman}, T.~M., {et~al.}
  2019{\natexlab{a}}, \apj, 872, 144

\bibitem[{{Hwang} {et~al.}(2019{\natexlab{b}}){Hwang}, {Shin}, \&
  {Song}}]{Hwang+19}
{Hwang}, H.~S., {Shin}, J., \& {Song}, H. 2019{\natexlab{b}}, \mnras, 489, 339

\bibitem[{{Iyer} {et~al.}(2020){Iyer}, {Tacchella}, {Genel}, {Hayward},
  {Hernquist}, {Brooks}, {Caplar}, {Dav{\'e}}, {Diemer}, {Forbes}, {Gawiser},
  {Somerville}, \& {Starkenburg}}]{Iyer+20}
{Iyer}, K.~G., {Tacchella}, S., {Genel}, S., {et~al.} 2020, \mnras, 498, 430

\bibitem[{{Janowiecki} {et~al.}(2020){Janowiecki}, {Catinella}, {Cortese},
  {Saintonge}, \& {Wang}}]{Janowiecki+20}
{Janowiecki}, S., {Catinella}, B., {Cortese}, L., {Saintonge}, A., \& {Wang},
  J. 2020, \mnras, 493, 1982

\bibitem[{{Jiang} {et~al.}(2019){Jiang}, {Dekel}, {Kneller}, {Lapiner},
  {Ceverino}, {Primack}, {Faber}, {Macci{\`o}}, {Dutton}, {Genel}, \&
  {Somerville}}]{Jiang+19}
{Jiang}, F., {Dekel}, A., {Kneller}, O., {et~al.} 2019, \mnras, 488, 4801

\bibitem[{{Jog} \& {Combes}(2009)}]{JogCombes09}
{Jog}, C.~J., \& {Combes}, F. 2009, \physrep, 471, 75

\bibitem[{{Kampczyk} {et~al.}(2013){Kampczyk}, {Lilly}, {de Ravel}, {Le
  F{\`e}vre}, {Bolzonella}, {Carollo}, {Diener}, {Knobel}, {Kova{\v{c}}},
  {Maier}, {Renzini}, {Sargent}, {Vergani}, {Abbas}, {Bardelli}, {Bongiorno},
  {Bordoloi}, {Caputi}, {Contini}, {Coppa}, {Cucciati}, {de la Torre},
  {Franzetti}, {Garilli}, {Iovino}, {Kneib}, {Koekemoer}, {Lamareille}, {Le
  Borgne}, {Le Brun}, {Leauthaud}, {Mainieri}, {Mignoli}, {Pello}, {Peng},
  {Perez Montero}, {Ricciardelli}, {Scodeggio}, {Silverman}, {Tanaka}, {Tasca},
  {Tresse}, {Zamorani}, {Zucca}, {Bottini}, {Cappi}, {Cassata}, {Cimatti},
  {Fumana}, {Guzzo}, {Kartaltepe}, {Marinoni}, {McCracken}, {Memeo}, {Meneux},
  {Oesch}, {Porciani}, {Pozzetti}, \& {Scaramella}}]{Kampczyk+13}
{Kampczyk}, P., {Lilly}, S.~J., {de Ravel}, L., {et~al.} 2013, \apj, 762, 43

\bibitem[{{Kartaltepe} {et~al.}(2012){Kartaltepe}, {Dickinson}, {Alexander},
  {Bell}, {Dahlen}, {Elbaz}, {Faber}, {Lotz}, {McIntosh}, {Wiklind}, {Altieri},
  {Aussel}, {Bethermin}, {Bournaud}, {Charmandaris}, {Conselice}, {Cooray},
  {Dannerbauer}, {Dav{\'e}}, {Dunlop}, {Dekel}, {Ferguson}, {Grogin}, {Hwang},
  {Ivison}, {Kocevski}, {Koekemoer}, {Koo}, {Lai}, {Leiton}, {Lucas}, {Lutz},
  {Magdis}, {Magnelli}, {Morrison}, {Mozena}, {Mullaney}, {Newman}, {Pope},
  {Popesso}, {van der Wel}, {Weiner}, \& {Wuyts}}]{Kartaltepe+12}
{Kartaltepe}, J.~S., {Dickinson}, M., {Alexander}, D.~M., {et~al.} 2012, \apj,
  757, 23

\bibitem[{{Kauffmann} {et~al.}(2003){Kauffmann}, {Heckman}, {White}, {Charlot},
  {Tremonti}, {Peng}, {Seibert}, {Brinkmann}, {Nichol}, {SubbaRao}, \&
  {York}}]{Kauffmann+03b}
{Kauffmann}, G., {Heckman}, T.~M., {White}, S. D.~M., {et~al.} 2003, \mnras,
  341, 54

\bibitem[{{Kaviraj}(2014)}]{Kaviraj14}
{Kaviraj}, S. 2014, \mnras, 440, 2944

\bibitem[{{Kaviraj} {et~al.}(2007){Kaviraj}, {Schawinski}, {Devriendt},
  {Ferreras}, {Khochfar}, {Yoon}, {Yi}, {Deharveng}, {Boselli}, {Barlow},
  {Conrow}, {Forster}, {Friedman}, {Martin}, {Morrissey}, {Neff},
  {Schiminovich}, {Seibert}, {Small}, {Wyder}, {Bianchi}, {Donas}, {Heckman},
  {Lee}, {Madore}, {Milliard}, {Rich}, \& {Szalay}}]{Kaviraj+07}
{Kaviraj}, S., {Schawinski}, K., {Devriendt}, J.~E.~G., {et~al.} 2007, \apjs,
  173, 619

\bibitem[{{Kennicutt}(1998)}]{Kennicutt+98}
{Kennicutt}, Robert~C., J. 1998, \apj, 498, 541

\bibitem[{{Kennicutt} \& {Evans}(2012)}]{Kennicutt+12}
{Kennicutt}, R.~C., \& {Evans}, N.~J. 2012, \araa, 50, 531

\bibitem[{{Kere{\v{s}}} {et~al.}(2009){Kere{\v{s}}}, {Katz}, {Fardal},
  {Dav{\'e}}, \& {Weinberg}}]{Keres+09}
{Kere{\v{s}}}, D., {Katz}, N., {Fardal}, M., {Dav{\'e}}, R., \& {Weinberg},
  D.~H. 2009, \mnras, 395, 160

\bibitem[{{Kere{\v{s}}} {et~al.}(2005){Kere{\v{s}}}, {Katz}, {Weinberg}, \&
  {Dav{\'e}}}]{Keres+05}
{Kere{\v{s}}}, D., {Katz}, N., {Weinberg}, D.~H., \& {Dav{\'e}}, R. 2005,
  \mnras, 363, 2

\bibitem[{{Kewley} {et~al.}(2001) {Kewley}, {Dopita}, {Sutherland}, {Heisler},\&  {Trevena}}]{Kewley+01}
{Kewley}, L.~J.,  {Dopita}, M.~A.,  {Sutherland}, R.~S., {et~al.} 2001 \apj, 556, 121

\bibitem[{{Knapen} {et~al.}(2015){Knapen}, {Cisternas}, \&
  {Querejeta}}]{Knapen+15}
{Knapen}, J.~H., {Cisternas}, M., \& {Querejeta}, M. 2015, \mnras, 454, 1742

\bibitem[{{Kormendy} \& {Ho}(2013)}]{Kormendy+13}
{Kormendy}, J., \& {Ho}, L.~C. 2013, \araa, 51, 511

\bibitem[{Kratzer \& Furrer(2018)}]{Kratzer+18}
Kratzer, G., \& Furrer, R. 2018, arXiv preprint arXiv:1804.07134

\bibitem[{{Lange} {et~al.}(2016){Lange}, {Moffett}, {Driver}, {Robotham},
  {Lagos}, {Kelvin}, {Conselice}, {Margalef-Bentabol}, {Alpaslan}, {Baldry},
  {Bland-Hawthorn}, {Bremer}, {Brough}, {Cluver}, {Colless}, {Davies},
  {H{\"a}u{\ss}ler}, {Holwerda}, {Hopkins}, {Kafle}, {Kennedy}, {Liske},
  {Phillipps}, {Popescu}, {Taylor}, {Tuffs}, {van Kampen}, \&
  {Wright}}]{Lange+16}
{Lange}, R., {Moffett}, A.~J., {Driver}, S.~P., {et~al.} 2016, \mnras, 462,
  1470

\bibitem[{{Larson} {et~al.}(2016){Larson}, {Sanders}, {Barnes}, {Ishida},
  {Evans}, {U}, {Mazzarella}, {Kim}, {Privon}, {Mirabel}, \&
  {Flewelling}}]{Larson+16}
{Larson}, K.~L., {Sanders}, D.~B., {Barnes}, J.~E., {et~al.} 2016, \apj, 825,
  128

\bibitem[{{L'Huillier} {et~al.}(2012){L'Huillier}, {Combes}, \&
  {Semelin}}]{LHuillier+12}
{L'Huillier}, B., {Combes}, F., \& {Semelin}, B. 2012, \aap, 544, A68

\bibitem[{{Li} {et~al.}(2008){Li}, {Kauffmann}, {Heckman}, {Jing}, \&
  {White}}]{Li+08}
{Li}, C., {Kauffmann}, G., {Heckman}, T.~M., {Jing}, Y.~P., \& {White}, S.
  D.~M. 2008, \mnras, 385, 1903

\bibitem[{{Li} {et~al.}(2018){Li}, {Mao}, {Cappellari}, {Ge}, {Long}, {Li},
  {Mo}, {Li}, {Zheng}, {Bundy}, {Thomas}, {Brownstein}, {Roman Lopes}, {Law},
  \& {Drory}}]{Li+18}
{Li}, H., {Mao}, S., {Cappellari}, M., {et~al.} 2018, \mnras, 476, 1765

\bibitem[{{Lilly} {et~al.}(2013){Lilly}, {Carollo}, {Pipino}, {Renzini}, \&
  {Peng}}]{Lilly+13}
{Lilly}, S.~J., {Carollo}, C.~M., {Pipino}, A., {Renzini}, A., \& {Peng}, Y.
  2013, \apj, 772, 119

\bibitem[{{Lin} {et~al.}(2008){Lin}, {Patton}, {Koo}, {Casteels}, {Conselice},
  {Faber}, {Lotz}, {Willmer}, {Hsieh}, {Chiueh}, {Newman}, {Novak}, {Weiner},
  \& {Cooper}}]{Lin+08}
{Lin}, L., {Patton}, D.~R., {Koo}, D.~C., {et~al.} 2008, \apj, 681, 232

\bibitem[{{Lin} {et~al.}(2020{\natexlab{a}}){Lin}, {Li}, {Du}, {Wang}, {Xiao},
  {Bureau}, {Fraser-McKelvie}, {Masters}, {Lin}, {Wake}, \& {Hao}}]{Lin+20b}
{Lin}, L., {Li}, C., {Du}, C., {et~al.} 2020{\natexlab{a}}, \mnras, 499, 1406

\bibitem[{{Lin} {et~al.}(2020{\natexlab{b}}){Lin}, {Faber}, {Koo}, {Salim},
  {Dutton}, {Fang}, {Jiang}, {Lee}, {Rodr{\'\i}guez-Puebla}, {Wel}, {Guo},
  {Barro}, {Primack}, {Dekel}, {Chen}, {Luo}, {Pandya}, {Somerville},
  {Ferguson}, {Kassin}, {Koekemoer}, {Grogin}, {Galametz}, {Santini},
  {Nayyeri}, {Stefanon}, {Dahlen}, {Mobasher}, \& {Hao}}]{Lin+20a}
{Lin}, L., {Faber}, S.~M., {Koo}, D.~C., {et~al.} 2020{\natexlab{b}}, \apj,
  899, 93

\bibitem[{{Lisenfeld} {et~al.}(2011){Lisenfeld}, {Espada}, {Verdes-Montenegro},
  {Kuno}, {Leon}, {Sabater}, {Sato}, {Sulentic}, {Verley}, \&
  {Yun}}]{Lisenfeld+11}
{Lisenfeld}, U., {Espada}, D., {Verdes-Montenegro}, L., {et~al.} 2011, \aap,
  534, A102

\bibitem[{{Lotz} {et~al.}(2011){Lotz}, {Jonsson}, {Cox}, {Croton}, {Primack},
  {Somerville}, \& {Stewart}}]{Lotz+11}
{Lotz}, J.~M., {Jonsson}, P., {Cox}, T.~J., {et~al.} 2011, \apj, 742, 103

\bibitem[{{Lotz} {et~al.}(2010{\natexlab{a}}){Lotz}, {Jonsson}, {Cox}, \&
  {Primack}}]{Lotz+10a}
{Lotz}, J.~M., {Jonsson}, P., {Cox}, T.~J., \& {Primack}, J.~R.
  2010{\natexlab{a}}, \mnras, 404, 590

\bibitem[{{Lotz} {et~al.}(2010{\natexlab{b}}){Lotz}, {Jonsson}, {Cox}, \&
  {Primack}}]{Lotz+10b}
{Lotz}, J.~M., {Jonsson}, P., {Cox}, T.~J., \& {Primack}, J.~R. 2010{\natexlab{b}}, \mnras, 404, 575


\bibitem[{{Lotz} {et~al.}(2004){Lotz}, {Primack},\& {Madau}}]{Lotz+04}
{Lotz}, J.~M.,  {Primack}, J.~R., {Madau},P., {et~al.} 2004, \apj,128,163


\bibitem[{{Luo} {et~al.}(2020){Luo}, {Faber}, {Rodr{\'\i}guez-Puebla}, {Woo},
  {Guo}, {Koo}, {Primack}, {Chen}, {Yesuf}, {Lin}, {Barro}, {Fang}, {Pand ya},
  {Huertas-Company}, \& {Mao}}]{Luo+20}
{Luo}, Y., {Faber}, S.~M., {Rodr{\'\i}guez-Puebla}, A., {et~al.} 2020, \mnras,
  493, 1686

\bibitem[{{Mapelli} {et~al.}(2008){Mapelli}, {Moore}, \&
  {Bland-Hawthorn}}]{Mapelli+08}
{Mapelli}, M., {Moore}, B., \& {Bland-Hawthorn}, J. 2008, \mnras, 388, 697

\bibitem[{{Matthee} \& {Schaye}(2019)}]{Matthee+19}
{Matthee}, J., \& {Schaye}, J. 2019, \mnras, 484, 915

\bibitem[{{Matthews} {et~al.}(1998){Matthews}, {van Driel}, \&
  {Gallagher}}]{Matthews+98}
{Matthews}, L.~D., {van Driel}, W., \& {Gallagher}, J.~S., I. 1998, \aj, 116,
  1169

\bibitem[{{McIntosh} {et~al.}(2014){McIntosh}, {Wagner}, {Cooper}, {Bell},
  {Kere{\v{s}}}, {Bosch}, {Gallazzi}, {Haines}, {Mann}, {Pasquali}, \&
  {Christian}}]{McIntosh+14}
{McIntosh}, D.~H., {Wagner}, C., {Cooper}, A., {et~al.} 2014, \mnras, 442, 533

\bibitem[{{Mendel} {et~al.}(2013){Mendel}, {Simard}, {Ellison}, \&
  {Patton}}]{Mendel+13}
{Mendel}, J.~T., {Simard}, L., {Ellison}, S.~L., \& {Patton}, D.~R. 2013,
  \mnras, 429, 2212

\bibitem[{{Mihos} \& {Hernquist}(1996)}]{Mihos+96}
{Mihos}, J.~C., \& {Hernquist}, L. 1996, \apj, 464, 641

\bibitem[{{Moreno} {et~al.}(2021){Moreno}, {Torrey}, {Ellison}, {Patton},
  {Bottrell}, {Bluck}, {Hani}, {Hayward}, {Bullock}, {Hopkins}, \&
  {Hernquist}}]{Moreno+21}
{Moreno}, J., {Torrey}, P., {Ellison}, S.~L., {et~al.} 2021, \mnras, 503, 3113

\bibitem[{{Morselli} {et~al.}(2019){Morselli}, {Popesso}, {Cibinel}, {Oesch},
  {Montes}, {Atek}, {Illingworth}, \& {Holden}}]{Morselli+19}
{Morselli}, L., {Popesso}, P., {Cibinel}, A., {et~al.} 2019, \aap, 626, A61

\bibitem[{{Morselli} {et~al.}(2020){Morselli}, {Rodighiero}, {Enia},
  {Corbelli}, {Casasola}, {Rodr{\'\i}guez-Mu{\~n}oz}, {Renzini}, {Tacchella},
  {Baronchelli}, {Bianchi}, {Cassata}, {Franceschini}, {Mancini}, {Negrello},
  {Popesso}, \& {Romano}}]{Morselli+20}
{Morselli}, L., {Rodighiero}, G., {Enia}, A., {et~al.} 2020, \mnras, 496, 4606

\bibitem[{{Naab} \& {Ostriker}(2017)}]{Naab+17}
{Naab}, T., \& {Ostriker}, J.~P. 2017, \araa, 55, 59

\bibitem[{{Nelson} {et~al.}(2015){Nelson}, {Genel}, {Vogelsberger}, {Springel},
  {Sijacki}, {Torrey}, \& {Hernquist}}]{Nelson+15}
{Nelson}, D., {Genel}, S., {Vogelsberger}, M., {et~al.} 2015, \mnras, 448, 59

\bibitem[{{Noeske} {et~al.}(2007){Noeske}, {Weiner}, {Faber}, {Papovich},
  {Koo}, {Somerville}, {Bundy}, {Conselice}, {Newman}, {Schiminovich}, {Le
  Floc'h}, {Coil}, {Rieke}, {Lotz}, {Primack}, {Barmby}, {Cooper}, {Davis},
  {Ellis}, {Fazio}, {Guhathakurta}, {Huang}, {Kassin}, {Martin}, {Phillips},
  {Rich}, {Small}, {Willmer}, \& {Wilson}}]{Noeske+07}
{Noeske}, K.~G., {Weiner}, B.~J., {Faber}, S.~M., {et~al.} 2007, \apjl, 660,
  L43

\bibitem[{{Noll} {et~al.}(2009){Noll},{Burgarella},{Giovannoli}, {Buat},{Marcillac}, {Mu{\~n}oz-Mateos}}]{Noll+09}
{Noll}, S., {Burgarella}, D., {Giovannoli}, E, {et~al.} 2009 {\aap}, 472, 455

\bibitem[{{Omand} {et~al.}(2014){Omand}, {Balogh}, \& {Poggianti}}]{Omand+14}
{Omand}, C. M.~B., {Balogh}, M.~L., \& {Poggianti}, B.~M. 2014, \mnras, 440,
  843

\bibitem[{{Pan} {et~al.}(2019){Pan}, {Lin}, {Hsieh}, {Barrera-Ballesteros},
  {S{\'a}nchez}, {Hsu}, {Keenan}, {Tissera}, {Boquien}, {Dai}, {Knapen},
  {Riffel}, {Argudo-Fern{\'a}ndez}, {Xiao}, \& {Yuan}}]{Pan+19}
{Pan}, H.-A., {Lin}, L., {Hsieh}, B.-C., {et~al.} 2019, \apj, 881, 119

\bibitem[{Pandey \& Sarkar(2016)}]{Pandey+17}
Pandey, B., \& Sarkar, S. 2016, \mnras, 467, L6

\bibitem[{{Patton} {et~al.}(2016){Patton}, {Qamar}, {Ellison}, {Bluck},
  {Simard}, {Mendel}, {Moreno}, \& {Torrey}}]{Patton+16}
{Patton}, D.~R., {Qamar}, F.~D., {Ellison}, S.~L., {et~al.} 2016, \mnras, 461,
  2589

\bibitem[{{Pawlik} {et~al.}(2016){Pawlik}, {Wild}, {Walcher}, {Johansson},
  {Villforth}, {Rowlands}, {Mendez-Abreu}, \& {Hewlett}}]{Pawlik+16}
{Pawlik}, M.~M., {Wild}, V., {Walcher}, C.~J., {et~al.} 2016, \mnras, 456, 3032

\bibitem[{{Peng} \& {Maiolino}(2014)}]{Peng+14}
{Peng}, Y.-j., \& {Maiolino}, R. 2014, \mnras, 443, 3643

\bibitem[{{Peng} {et~al.}(2010){Peng}, {Lilly}, {Kova{\v{c}}}, {Bolzonella},
  {Pozzetti}, {Renzini}, {Zamorani}, {Ilbert}, {Knobel}, {Iovino}, {Maier},
  {Cucciati}, {Tasca}, {Carollo}, {Silverman}, {Kampczyk}, {de Ravel},
  {Sanders}, {Scoville}, {Contini}, {Mainieri}, {Scodeggio}, {Kneib}, {Le
  F{\`e}vre}, {Bardelli}, {Bongiorno}, {Caputi}, {Coppa}, {de la Torre},
  {Franzetti}, {Garilli}, {Lamareille}, {Le Borgne}, {Le Brun}, {Mignoli},
  {Perez Montero}, {Pello}, {Ricciardelli}, {Tanaka}, {Tresse}, {Vergani},
  {Welikala}, {Zucca}, {Oesch}, {Abbas}, {Barnes}, {Bordoloi}, {Bottini},
  {Cappi}, {Cassata}, {Cimatti}, {Fumana}, {Hasinger}, {Koekemoer},
  {Leauthaud}, {Maccagni}, {Marinoni}, {McCracken}, {Memeo}, {Meneux}, {Nair},
  {Porciani}, {Presotto}, \& {Scaramella}}]{Peng+10}
{Peng}, Y.-j., {Lilly}, S.~J., {Kova{\v{c}}}, K., {et~al.} 2010, \apj, 721, 193

\bibitem[{{Postman} \& {Geller}(1984)}]{Postman+84}
{Postman}, M., \& {Geller}, M.~J. 1984, \apj, 281, 95

\bibitem[{{Puglisi} {et~al.}(2019){Puglisi}, {Daddi}, {Liu}, {Bournaud},
  {Silverman}, {Circosta}, {Calabr{\`o}}, {Aravena}, {Cibinel}, {Dannerbauer},
  {Delvecchio}, {Elbaz}, {Gao}, {Gobat}, {Jin}, {Le Floc'h}, {Magdis},
  {Mancini}, {Riechers}, {Rodighiero}, {Sargent}, {Valentino}, \&
  {Zanisi}}]{Puglisi+19}
{Puglisi}, A., {Daddi}, E., {Liu}, D., {et~al.} 2019, \apjl, 877, L23

\bibitem[{{Reichard} {et~al.}(2008){Reichard}, {Heckman}, {Rudnick},
  {Brinchmann}, \& {Kauffmann}}]{Reichard+08}
{Reichard}, T.~A., {Heckman}, T.~M., {Rudnick}, G., {Brinchmann}, J., \&
  {Kauffmann}, G. 2008, \apj, 677, 186

\bibitem[{{Reichard} {et~al.}(2009){Reichard}, {Heckman}, {Rudnick},
  {Brinchmann}, {Kauffmann}, \& {Wild}}]{Reichard+09}
{Reichard}, T.~A., {Heckman}, T.~M., {Rudnick}, G., {et~al.} 2009, \apj, 691,
  1005

\bibitem[{{Richter} \& {Sancisi}(1994)}]{Richter+94}
{Richter}, O.~G., \& {Sancisi}, R. 1994, \aap, 290, L9

\bibitem[{{Rix} \& {Zaritsky}(1995)}]{Rix+95}
{Rix}, H.-W., \& {Zaritsky}, D. 1995, \apj, 447, 82

\bibitem[{{Rodriguez-Gomez} {et~al.}(2015){Rodriguez-Gomez}, {Genel},
  {Vogelsberger}, {Sijacki}, {Pillepich}, {Sales}, {Torrey}, {Snyder},
  {Nelson}, {Springel}, {Ma}, \& {Hernquist}}]{Rodriguez-Gomez+15}
{Rodriguez-Gomez}, V., {Genel}, S., {Vogelsberger}, M., {et~al.} 2015, \mnras,
  449, 49

\bibitem[{{Rodriguez-Gomez} {et~al.}(2019){Rodriguez-Gomez}, {Snyder}, {Lotz},
  {Nelson}, {Pillepich}, {Springel}, {Genel}, {Weinberger}, {Tacchella},
  {Pakmor}, {Torrey}, {Marinacci}, {Vogelsberger}, {Hernquist}, \&
  {Thilker}}]{Rodriguez-Gomez+19}
{Rodriguez-Gomez}, V., {Snyder}, G.~F., {Lotz}, J.~M., {et~al.} 2019, \mnras,
  483, 4140

\bibitem[{{Rodr{\'\i}guez-Puebla} {et~al.}(2016){Rodr{\'\i}guez-Puebla},
  {Primack}, {Behroozi}, \& {Faber}}]{Rodriguez-Puebla+16}
{Rodr{\'\i}guez-Puebla}, A., {Primack}, J.~R., {Behroozi}, P., \& {Faber},
  S.~M. 2016, \mnras, 455, 2592

\bibitem[{{Rowe} {et~al.}(2015){Rowe}, {Jarvis}, {Mandelbaum}, {Bernstein},
  {Bosch}, {Simet}, {Meyers}, {Kacprzak}, {Nakajima}, {Zuntz}, {Miyatake},
  {Dietrich}, {Armstrong}, {Melchior}, \& {Gill}}]{Rowe+15}
{Rowe}, B.~T.~P., {Jarvis}, M., {Mandelbaum}, R., {et~al.} 2015, Astronomy and
  Computing, 10, 121

\bibitem[{{Rudnick} {et~al.}(2000){Rudnick}, {Rix}, \&
  {Kennicutt}}]{Rudnick+00}
{Rudnick}, G., {Rix}, H.-W., \& {Kennicutt}, Robert~C., J. 2000, \apj, 538, 569

\bibitem[{{Sachdeva} {et~al.}(2020){Sachdeva}, {Ho}, {Li}, \&
  {Shankar}}]{Sachdeva+20}
{Sachdeva}, S., {Ho}, L.~C., {Li}, Y.~A., \& {Shankar}, F. 2020, \apj, 899, 89

\bibitem[{{Saha} {et~al.}(2007){Saha}, {Combes}, \& {Jog}}]{Saha+07}
{Saha}, K., {Combes}, F., \& {Jog}, C.~J. 2007, \mnras, 382, 419

\bibitem[{{Saintonge} {et~al.}(2017){Saintonge}, {Catinella}, {Tacconi},
  {Kauffmann}, {Genzel}, {Cortese}, {Dav{\'e}}, {Fletcher},
  {Graci{\'a}-Carpio}, {Kramer}, {Heckman}, {Janowiecki}, {Lutz}, {Rosario},
  {Schiminovich}, {Schuster}, {Wang}, {Wuyts}, {Borthakur}, {Lamperti}, \&
  {Roberts-Borsani}}]{Saintonge+17}
{Saintonge}, A., {Catinella}, B., {Tacconi}, L.~J., {et~al.} 2017, \apjs, 233,
  22

\bibitem[{{Salim} {et~al.}(2018){Salim}, {Boquien}, \& {Lee}}]{Salim+18}
{Salim}, S., {Boquien}, M., \& {Lee}, J.~C. 2018, \apj, 859, 11

\bibitem[{{Salim} {et~al.}(2016){Salim}, {Lee}, {Janowiecki}, {da Cunha},
  {Dickinson}, {Boquien}, {Burgarella}, {Salzer}, \& {Charlot}}]{Salim+16}
{Salim}, S., {Lee}, J.~C., {Janowiecki}, S., {et~al.} 2016, \apjs, 227, 2

\bibitem[{{S{\'a}nchez}(2020)}]{Sanchez+20}
{S{\'a}nchez}, S.~F. 2020, \araa, 58, 99

\bibitem[{{S{\'a}nchez Almeida} \& {Dalla Vecchia}(2018)}]{Sanchez+18}
{S{\'a}nchez Almeida}, J., \& {Dalla Vecchia}, C. 2018, \apj, 859, 109

\bibitem[{{S{\'a}nchez Almeida} {et~al.}(2014){S{\'a}nchez Almeida},
  {Elmegreen}, {Mu{\~n}oz-Tu{\~n}{\'o}n}, \& {Elmegreen}}]{Sanchez+14}
{S{\'a}nchez Almeida}, J., {Elmegreen}, B.~G., {Mu{\~n}oz-Tu{\~n}{\'o}n}, C.,
  \& {Elmegreen}, D.~M. 2014, \aapr, 22, 71

\bibitem[{{Sancisi} {et~al.}(2008){Sancisi}, {Fraternali}, {Oosterloo}, \& {van
  der Hulst}}]{Sancisi+08}
{Sancisi}, R., {Fraternali}, F., {Oosterloo}, T., \& {van der Hulst}, T. 2008,
  \aapr, 15, 189

\bibitem[{{Sanders} \& {Mirabel}(1996)}]{Sanders+96}
{Sanders}, D.~B., \& {Mirabel}, I.~F. 1996, \araa, 34, 749

\bibitem[{{Sarkar} \& {Pandey}(2020)}]{Sarkar+20}
{Sarkar}, S., \& {Pandey}, B. 2020, \mnras, 497, 4077

\bibitem[{{Scott} {et~al.}(2017){Scott}, {Brough}, {Croom}, {Davies}, {van de
  Sande}, {Allen}, {Bland-Hawthorn}, {Bryant}, {Cortese}, {D'Eugenio},
  {Federrath}, {Ferreras}, {Goodwin}, {Groves}, {Konstantopoulos}, {Lawrence},
  {Medling}, {Moffett}, {Owers}, {Richards}, {Robotham}, {Tonini}, \&
  {Yi}}]{Scott+17}
{Scott}, N., {Brough}, S., {Croom}, S.~M., {et~al.} 2017, \mnras, 472, 2833

\bibitem[{{Shangguan} {et~al.}(2019){Shangguan}, {Ho}, {Li}, {Zhuang}, {Xie},
  \& {Li}}]{Shangguan+19}
{Shangguan}, J., {Ho}, L.~C., {Li}, R., {et~al.} 2019, \apj, 870, 104

\bibitem[{{Shen} {et~al.}(2003){Shen}, {Mo}, {White}, {Blanton}, {Kauffmann},
  {Voges}, {Brinkmann}, \& {Csabai}}]{Shen+03}
{Shen}, S., {Mo}, H.~J., {White}, S. D.~M., {et~al.} 2003, \mnras, 343, 978

\bibitem[{{Silverman} {et~al.}(2015){Silverman}, {Daddi}, {Rodighiero},
  {Rujopakarn}, {Sargent}, {Renzini}, {Liu}, {Feruglio}, {Kashino}, {Sanders},
  {Kartaltepe}, {Nagao}, {Arimoto}, {Berta}, {B{\'e}thermin}, {Koekemoer},
  {Lutz}, {Magdis}, {Mancini}, {Onodera}, \& {Zamorani}}]{Silverman+15}
{Silverman}, J.~D., {Daddi}, E., {Rodighiero}, G., {et~al.} 2015, \apjl, 812,
  L23

\bibitem[{{Simard} {et~al.}(2011){Simard}, {Mendel}, {Patton}, {Ellison}, \&
  {McConnachie}}]{Simard+11}
{Simard}, L., {Mendel}, J.~T., {Patton}, D.~R., {Ellison}, S.~L., \&
  {McConnachie}, A.~W. 2011, \apjs, 196, 11

\bibitem[{{Simard} {et~al.}(2002){Simard}, {Willmer}, {Vogt}, {Sarajedini},
  {Phillips}, {Weiner}, {Koo}, {Im}, {Illingworth}, \& {Faber}}]{Simard+02}
{Simard}, L., {Willmer}, C. N.~A., {Vogt}, N.~P., {et~al.} 2002, \apjs, 142, 1

\bibitem[{{Slonim} {et~al.}(2001){Slonim}, {Somerville}, {Tishby}, \&
  {Lahav}}]{Slonim+01}
{Slonim}, N., {Somerville}, R., {Tishby}, N., \& {Lahav}, O. 2001, \mnras, 323,
  270

\bibitem[{{Snyder} {et~al.}(2015){Snyder}, {Torrey}, {Lotz}, {Genel},
  {McBride}, {Vogelsberger}, {Pillepich}, {Nelson}, {Sales}, {Sijacki},
  {Hernquist}, \& {Springel}}]{Snyder+15}
{Snyder}, G.~F., {Torrey}, P., {Lotz}, J.~M., {et~al.} 2015, \mnras, 454, 1886

\bibitem[{{Somerville} \& {Dav{\'e}}(2015)}]{Somerville+15}
{Somerville}, R.~S., \& {Dav{\'e}}, R. 2015, \araa, 53, 51

\bibitem[{{Speagle} {et~al.}(2014){Speagle}, {Steinhardt}, {Capak}, \&
  {Silverman}}]{Speagle+14}
{Speagle}, J.~S., {Steinhardt}, C.~L., {Capak}, P.~L., \& {Silverman}, J.~D.
  2014, \apjs, 214, 15

\bibitem[{{Stark} {et~al.}(2013){Stark}, {Kannappan}, {Wei}, {Baker}, {Leroy},
  {Eckert}, \& {Vogel}}]{Stark+13}
{Stark}, D.~V., {Kannappan}, S.~J., {Wei}, L.~H., {et~al.} 2013, \apj, 769, 82

\bibitem[{{Steffen} {et~al.}(2021){Steffen}, {Fu}, {Comerford}, {Dai}, {Feng},
  {Gross}, \& {Xue}}]{Steffen+21}
{Steffen}, J.~L., {Fu}, H., {Comerford}, J.~M., {et~al.} 2021, \apj, 909, 120

\bibitem[{{Strateva} {et~al.}(2001){Strateva}, {Ivezi{\'c}}, {Knapp},
  {Narayanan}, {Strauss}, {Gunn}, {Lupton}, {Schlegel}, {Bahcall}, {Brinkmann},
  {Brunner}, {Budav{\'a}ri}, {Csabai}, {Castander}, {Doi}, {Fukugita},
  {Gy{\H{o}}ry}, {Hamabe}, {Hennessy}, {Ichikawa}, {Kunszt}, {Lamb}, {McKay},
  {Okamura}, {Racusin}, {Sekiguchi}, {Schneider}, {Shimasaku}, \&
  {York}}]{Strateva+01}
{Strateva}, I., {Ivezi{\'c}}, {\v{Z}}., {Knapp}, G.~R., {et~al.} 2001, \aj,
  122, 1861

\bibitem[{{Tacchella} {et~al.}(2016){Tacchella}, {Dekel}, {Carollo},
  {Ceverino}, {DeGraf}, {Lapiner}, {Mand elker}, \& {Primack
  Joel}}]{Tacchella+16}
{Tacchella}, S., {Dekel}, A., {Carollo}, C.~M., {et~al.} 2016, \mnras, 457,
  2790

\bibitem[{{Tacconi} {et~al.}(2020){Tacconi}, {Genzel}, \&
  {Sternberg}}]{Tacconi+20}
{Tacconi}, L.~J., {Genzel}, R., \& {Sternberg}, A. 2020, \araa, 58, 157

\bibitem[{{Tacconi} {et~al.}(2018){Tacconi}, {Genzel}, {Saintonge}, {Combes},
  {Garc{\'\i}a-Burillo}, {Neri}, {Bolatto}, {Contini}, {F{\"o}rster Schreiber},
  {Lilly}, {Lutz}, {Wuyts}, {Accurso}, {Boissier}, {Boone}, {Bouch{\'e}},
  {Bournaud}, {Burkert}, {Carollo}, {Cooper}, {Cox}, {Feruglio}, {Freundlich},
  {Herrera-Camus}, {Juneau}, {Lippa}, {Naab}, {Renzini}, {Salome}, {Sternberg},
  {Tadaki}, {{\"U}bler}, {Walter}, {Weiner}, \& {Weiss}}]{Tacconi+18}
{Tacconi}, L.~J., {Genzel}, R., {Saintonge}, A., {et~al.} 2018, \apj, 853, 179

\bibitem[{{Thomas} \& {Davies}(2006)}]{Thomas+06}
{Thomas}, D., \& {Davies}, R.~L. 2006, \mnras, 366, 510

\bibitem[{{Thomas} {et~al.}(2010){Thomas}, {Maraston}, {Schawinski}, {Sarzi},
  \& {Silk}}]{Thomas+10}
{Thomas}, D., {Maraston}, C., {Schawinski}, K., {Sarzi}, M., \& {Silk}, J.
  2010, \mnras, 404, 1775

\bibitem[{{Thorp} {et~al.}(2019){Thorp}, {Ellison}, {Simard}, {S{\'a}nchez}, \&
  {Antonio}}]{Thorp+19}
{Thorp}, M.~D., {Ellison}, S.~L., {Simard}, L., {S{\'a}nchez}, S.~F., \&
  {Antonio}, B. 2019, \mnras, 482, L55

\bibitem[{{Tinker}(2020)}]{Tinker+20}
{Tinker}, J.~L. 2020, arXiv e-prints, arXiv:2010.02946

\bibitem[{{Torrey} {et~al.}(2018){Torrey}, {Vogelsberger}, {Hernquist},
  {McKinnon}, {Marinacci}, {Simcoe}, {Springel}, {Pillepich}, {Naiman},
  {Pakmor}, {Weinberger}, {Nelson}, \& {Genel}}]{Torrey+18}
{Torrey}, P., {Vogelsberger}, M., {Hernquist}, L., {et~al.} 2018, \mnras, 477,
  L16

\bibitem[{{Torrey} {et~al.}(2019){Torrey}, {Vogelsberger}, {Marinacci},
  {Pakmor}, {Springel}, {Nelson}, {Naiman}, {Pillepich}, {Genel}, {Weinberger},
  \& {Hernquist}}]{Torrey+19}
{Torrey}, P., {Vogelsberger}, M., {Marinacci}, F., {et~al.} 2019, \mnras, 484,
  5587

\bibitem[{{Tremonti} {et~al.}(2004){Tremonti}, {Heckman}, {Kauffmann},
  {Brinchmann}, {Charlot}, {White}, {Seibert}, {Peng}, {Schlegel}, {Uomoto},
  {Fukugita}, \& {Brinkmann}}]{Tremonti+04}
{Tremonti}, C.~A., {Heckman}, T.~M., {Kauffmann}, G., {et~al.} 2004, \apj, 613,
  898

\bibitem[{{Trujillo} {et~al.}(2001){Trujillo}, {Graham}, \&
  {Caon}}]{Trujillo+01}
{Trujillo}, I., {Graham}, A.~W., \& {Caon}, N. 2001, \mnras, 326, 869

\bibitem[{{van de Voort} {et~al.}(2017){van de Voort}, {Bah{\'e}}, {Bower},
  {Correa}, {Crain}, {Schaye}, \& {Theuns}}]{vandeVoort+17}
{van de Voort}, F., {Bah{\'e}}, Y.~M., {Bower}, R.~G., {et~al.} 2017, \mnras,
  466, 3460

\bibitem[{{van de Voort} {et~al.}(2011){van de Voort}, {Schaye}, {Booth},
  {Haas}, \& {Dalla Vecchia}}]{vandeVoort+11}
{van de Voort}, F., {Schaye}, J., {Booth}, C.~M., {Haas}, M.~R., \& {Dalla
  Vecchia}, C. 2011, \mnras, 414, 2458

\bibitem[{{van den Bosch}(2016)}]{Vandenbosch+16}
{van den Bosch}, R. C.~E. 2016, \apj, 831, 134

\bibitem[{{van der Wel} {et~al.}(2009){van der Wel}, {Bell}, {van den Bosch},
  {Gallazzi}, \& {Rix}}]{vanderWel+09}
{van der Wel}, A., {Bell}, E.~F., {van den Bosch}, F.~C., {Gallazzi}, A., \&
  {Rix}, H.-W. 2009, \apj, 698, 1232

\bibitem[{{van Dokkum} {et~al.}(2015){van Dokkum}, {Nelson}, {Franx}, {Oesch},
  {Momcheva}, {Brammer}, {F{\"o}rster Schreiber}, {Skelton}, {Whitaker}, {van
  der Wel}, {Bezanson}, {Fumagalli}, {Illingworth}, {Kriek}, {Leja}, \&
  {Wuyts}}]{vanDokkum+15}
{van Dokkum}, P.~G., {Nelson}, E.~J., {Franx}, M., {et~al.} 2015, \apj, 813, 23

\bibitem[{{van Eymeren} {et~al.}(2011){van Eymeren}, {J{\"u}tte}, {Jog},
  {Stein}, \& {Dettmar}}]{vanEymeren+11}
{van Eymeren}, J., {J{\"u}tte}, E., {Jog}, C.~J., {Stein}, Y., \& {Dettmar},
  R.~J. 2011, \aap, 530, A30

\bibitem[{{vanDokkum} {et~al.}(2014){vanDokkum}, {Bezanson}, {van der Wel},
  {Nelson}, {Momcheva}, {Skelton}, {Whitaker}, {Brammer}, {Conroy},
  {F{\"o}rster Schreiber}, {Fumagalli}, {Kriek}, {Labb{\'e}}, {Leja},
  {Marchesini}, {Muzzin}, {Oesch}, \& {Wuyts}}]{vanDokkum+14}
{vanDokkum}, P.~G., {Bezanson}, R., {van der Wel}, A., {et~al.} 2014, \apj,
  791, 45

\bibitem[{{Veilleux} {et~al.}(2002){Veilleux}, {Kim}, \&
  {Sanders}}]{Veilleux+02}
{Veilleux}, S., {Kim}, D.~C., \& {Sanders}, D.~B. 2002, \apjs, 143, 315

\bibitem[{{Violino} {et~al.}(2018){Violino}, {Ellison}, {Sargent}, {Coppin},
  {Scudder}, {Mendel}, \& {Saintonge}}]{Violino+18}
{Violino}, G., {Ellison}, S.~L., {Sargent}, M., {et~al.} 2018, \mnras, 476,
  2591

\bibitem[{{Vulcani} {et~al.}(2018){Vulcani}, {Poggianti}, {Moretti}, {Mapelli},
  {Fasano}, {Fritz}, {Jaff{\'e}}, {Bettoni}, {Gullieuszik}, \&
  {Bellhouse}}]{Vulcani+18}
{Vulcani}, B., {Poggianti}, B.~M., {Moretti}, A., {et~al.} 2018, \apj, 852, 94

\bibitem[{{Wake} {et~al.}(2012){Wake}, {van Dokkum}, \& {Franx}}]{Wake+12}
{Wake}, D.~A., {van Dokkum}, P.~G., \& {Franx}, M. 2012, \apjl, 751, L44

\bibitem[{{Wang} \& {Lilly}(2020{\natexlab{a}})}]{WangLilly20a}
{Wang}, E., \& {Lilly}, S.~J. 2020{\natexlab{a}}, \apj, 895, 25

\bibitem[{{Wang} \& {Lilly}(2020{\natexlab{b}})}]{WangLilly20b}
{Wang}, E., \& {Lilly}, S.~J. 2020{\natexlab{b}}, \apj, 892, 87

\bibitem[{{Wang} {et~al.}(2020){Wang}, {Catinella}, {Saintonge}, {Pan},
  {Serra}, \& {Shao}}]{Wang+20}
{Wang}, J., {Catinella}, B., {Saintonge}, A., {et~al.} 2020, \apj, 890, 63

\bibitem[{{Wang} {et~al.}(2011){Wang}, {Kauffmann}, {Overzier}, {Catinella},
  {Schiminovich}, {Heckman}, {Moran}, {Haynes}, {Giovanelli}, \&
  {Kong}}]{Wang+11}
{Wang}, J., {Kauffmann}, G., {Overzier}, R., {et~al.} 2011, \mnras, 412, 1081

\bibitem[{{Wang} {et~al.}(2012){Wang}, {Kauffmann}, {Overzier}, {Tacconi},
  {Kong}, {Saintonge}, {Catinella}, {Schiminovich}, {Moran}, \&
  {Johnson}}]{Wang+12}
{Wang}, J., {Kauffmann}, G., {Overzier}, R., {et~al.} 2012, \mnras, 423, 3486

\bibitem[{Waskom(2021)}]{Waskom21}
Waskom, M.~L. 2021, Journal of Open Source Software, 6, 3021

\bibitem[{{Wechsler} \& {Tinker}(2018)}]{Wechsler+18}
{Wechsler}, R.~H., \& {Tinker}, J.~L. 2018, \araa, 56, 435

\bibitem[{Wei \& Simko(2021)}]{corrplot21}
Wei, T., \& Simko, V. 2021, R package 'corrplot': Visualization of a
  Correlation Matrix, (Version 0.90)

\bibitem[{{Wetzel} {et~al.}(2012){Wetzel}, {Tinker}, \& {Conroy}}]{Wetzel+12}
{Wetzel}, A.~R., {Tinker}, J.~L., \& {Conroy}, C. 2012, \mnras, 424, 232

\bibitem[{{Whitaker} {et~al.}(2017){Whitaker}, {Bezanson}, {van Dokkum},
  {Franx}, {van der Wel}, {Brammer}, {F{\"o}rster-Schreiber}, {Giavalisco},
  {Labb{\'e}}, {Momcheva}, {Nelson}, \& {Skelton}}]{Whitaker+17}
{Whitaker}, K.~E., {Bezanson}, R., {van Dokkum}, P.~G., {et~al.} 2017, \apj,
  838, 19

\bibitem[{{Whitney} {et~al.}(2021){Whitney}, {Ferreira}, {Conselice}, \&
  {Duncan}}]{Whitney+21}
{Whitney}, A., {Ferreira}, L., {Conselice}, C.~J., \& {Duncan}, K. 2021, arXiv
  e-prints, arXiv:2105.01675

\bibitem[{{Wilcots} \& {Prescott}(2004)}]{Wilcots+04}
{Wilcots}, E.~M., \& {Prescott}, M. K.~M. 2004, \aj, 127, 1900

\bibitem[{{Woo} {et~al.}(2015){Woo}, {Dekel}, {Faber}, \& {Koo}}]{Woo+15}
{Woo}, J., {Dekel}, A., {Faber}, S.~M., \& {Koo}, D.~C. 2015, \mnras, 448, 237

\bibitem[{{Wright} {et~al.}(2021){Wright}, {Lagos}, {Power}, \&
  {Correa}}]{Wright+21}
{Wright}, R.~J., {Lagos}, C. d.~P., {Power}, C., \& {Correa}, C.~A. 2021,
  \mnras, 504, 5702

\bibitem[{{Wuyts} {et~al.}(2011){Wuyts}, {F{\"o}rster Schreiber}, {van der
  Wel}, {Magnelli}, {Guo}, {Genzel}, {Lutz}, {Aussel}, {Barro}, {Berta},
  {Cava}, {Graci{\'a}-Carpio}, {Hathi}, {Huang}, {Kocevski}, {Koekemoer},
  {Lee}, {Le Floc'h}, {McGrath}, {Nordon}, {Popesso}, {Pozzi}, {Riguccini},
  {Rodighiero}, {Saintonge}, \& {Tacconi}}]{Wuyts+11}
{Wuyts}, S., {F{\"o}rster Schreiber}, N.~M., {van der Wel}, A., {et~al.} 2011,
  \apj, 742, 96

\bibitem[{{Yang} {et~al.}(2021){Yang}, {Xu}, {He}, {Gu}, {Katsianis}, {Meng},
  {Shi}, {Zou}, {Zhang}, {Liu}, {Wang}, {Dong}, {Lu}, {Li}, {Chen}, {Wang},
  {Mo}, {Fu}, {Guo}, {Leauthaud}, {Luo}, {Zhang}, \& {Zu}}]{Yang+21}
{Yang}, X., {Xu}, H., {He}, M., {et~al.} 2021, \apj, 909, 143

\bibitem[{{Yesuf} {et~al.}(2020){Yesuf}, {Faber}, {Koo}, {Woo}, {Primack}, \&
  {Luo}}]{Yesuf+20a}
{Yesuf}, H.~M., {Faber}, S.~M., {Koo}, D.~C., {et~al.} 2020, \apj, 889, 14

\bibitem[{{Yesuf} {et~al.}(2014){Yesuf}, {Faber}, {Trump}, {Koo}, {Fang},
  {Liu}, {Wild}, \& {Hayward}}]{Yesuf+14}
{Yesuf}, H.~M., {Faber}, S.~M., {Trump}, J.~R., {et~al.} 2014, \apj, 792, 84

\bibitem[{{Yesuf} \& {Ho}(2020)}]{Yesuf+20b}
{Yesuf}, H.~M., \& {Ho}, L.~C. 2020, \apj, 901, 42

\bibitem[{{Yozin} \& {Bekki}(2014)}]{Yozin+14}
{Yozin}, C., \& {Bekki}, K. 2014, \mnras, 439, 1948

\bibitem[{{Yu} \& {Ho}(2019)}]{YuHo19}
{Yu}, S.-Y., \& {Ho}, L.~C. 2019, \apj, 871, 194

\bibitem[{{Yu}, {Ho} \& {Wang}(2021)}]{Yu+21}
{Yu}, S.-Y., {Ho}, L.~C, \& {Wang}, J.  2021, \apj, 917, 88

\bibitem[{{Zaritsky} \& {Rix}(1997)}]{Zaritsky+97}
{Zaritsky}, D., \& {Rix}, H.-W. 1997, \apj, 477, 118

\bibitem[{{Zaritsky} {et~al.}(2013){Zaritsky}, {Salo}, {Laurikainen},
  {Elmegreen}, {Athanassoula}, {Bosma}, {Comer{\'o}n}, {Erroz-Ferrer},
  {Elmegreen}, {Gadotti}, {Gil de Paz}, {Hinz}, {Ho}, {Holwerda}, {Kim},
  {Knapen}, {Laine}, {Laine}, {Madore}, {Meidt}, {Menendez-Delmestre},
  {Mizusawa}, {Mu{\~n}oz-Mateos}, {Regan}, {Seibert}, \& {Sheth}}]{Zaritsky+13}
{Zaritsky}, D., {Salo}, H., {Laurikainen}, E., {et~al.} 2013, \apj, 772, 135

\bibitem[{{Zolotov} {et~al.}(2015){Zolotov}, {Dekel}, {Mandelker}, {Tweed},
  {Inoue}, {DeGraf}, {Ceverino}, {Primack}, {Barro}, \& {Faber}}]{Zolotov+15}
{Zolotov}, A., {Dekel}, A., {Mandelker}, N., {et~al.} 2015, \mnras, 450, 2327

\end{thebibliography}

\appendix

\section{Correlations, Scores and Ranks of Structural Variables for Different Subsamples}

\begin{figure*}
\includegraphics[width=0.47\linewidth]{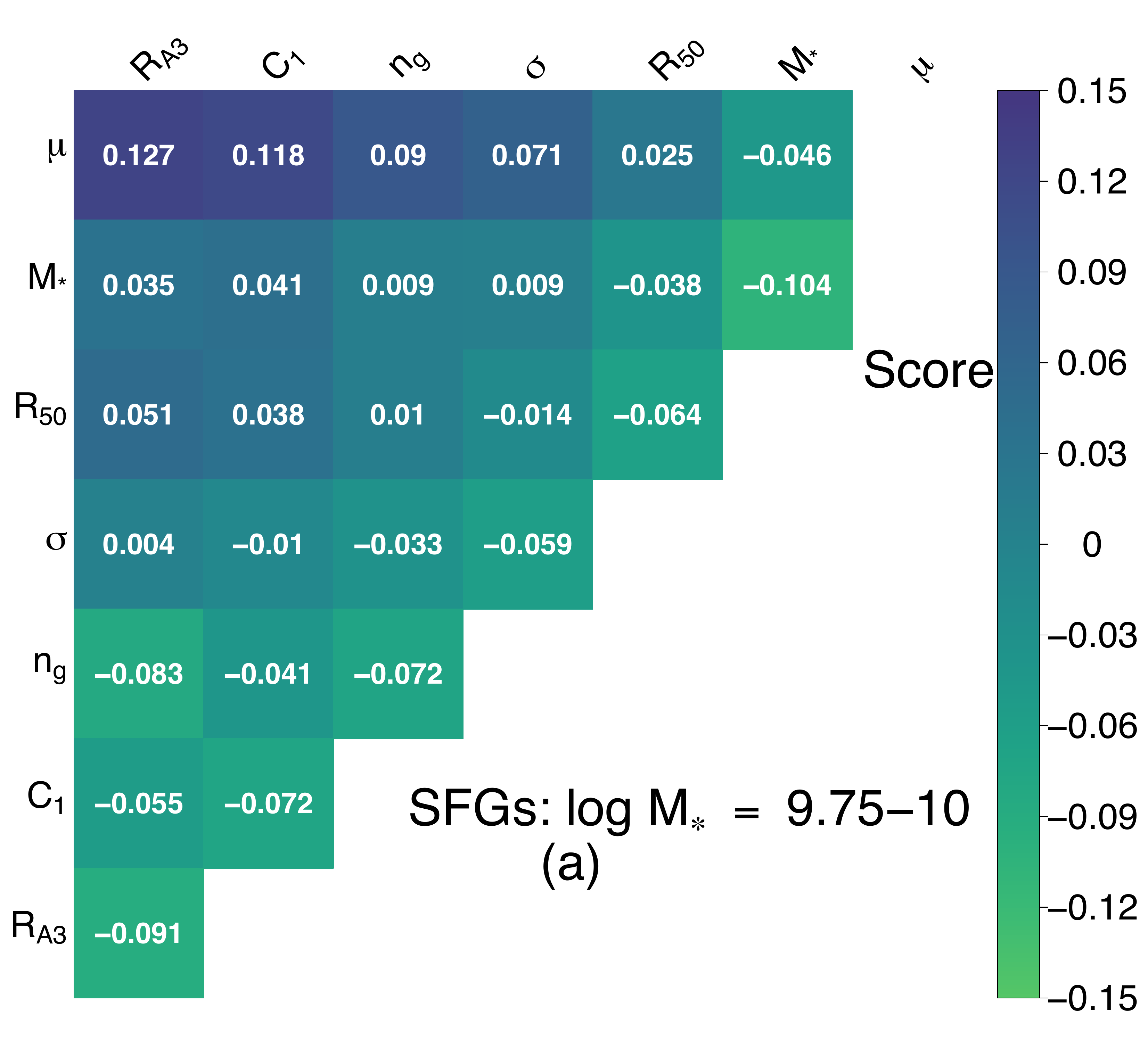}
\includegraphics[width=0.47\linewidth]{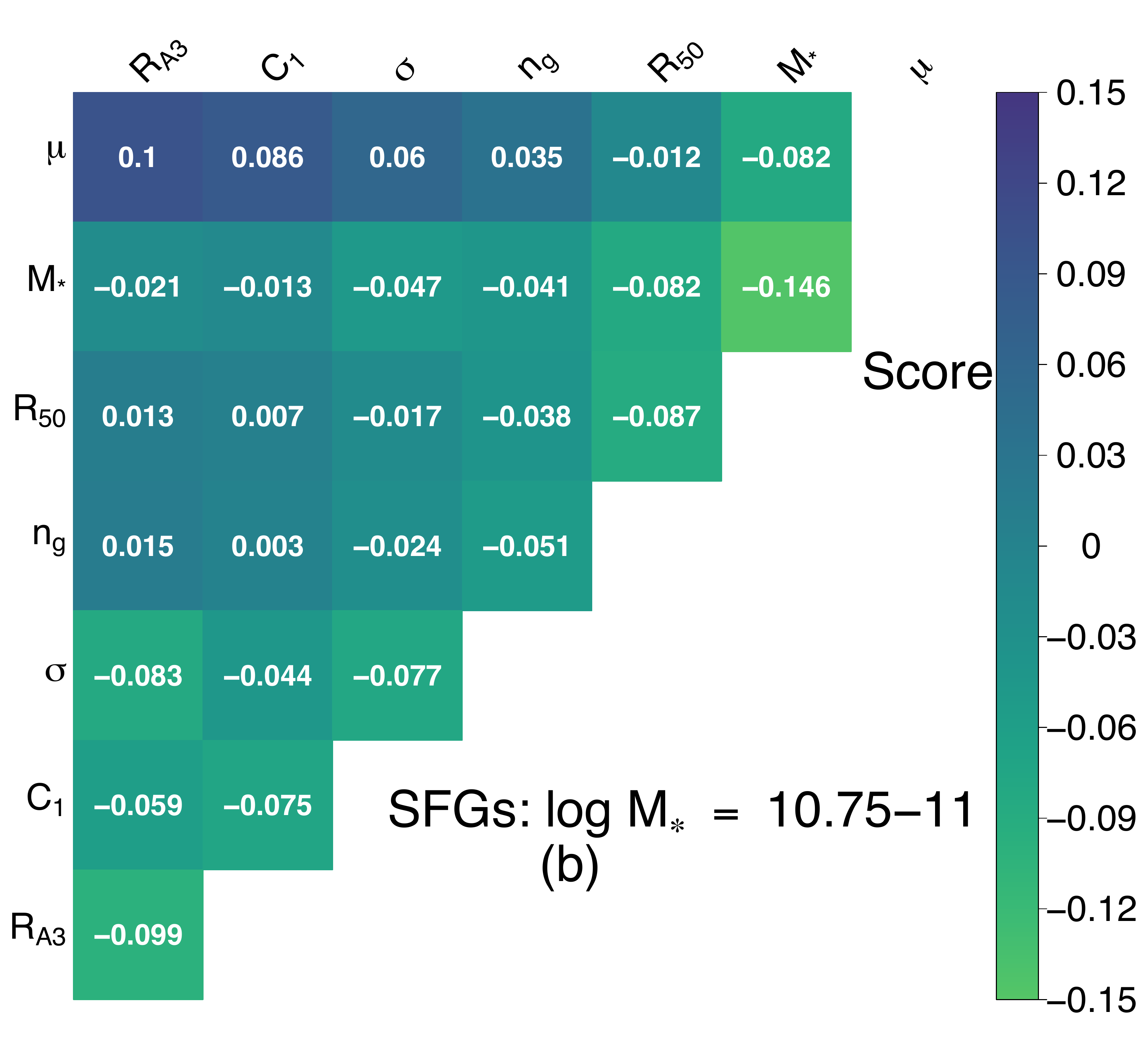}
\includegraphics[width=0.47\linewidth]{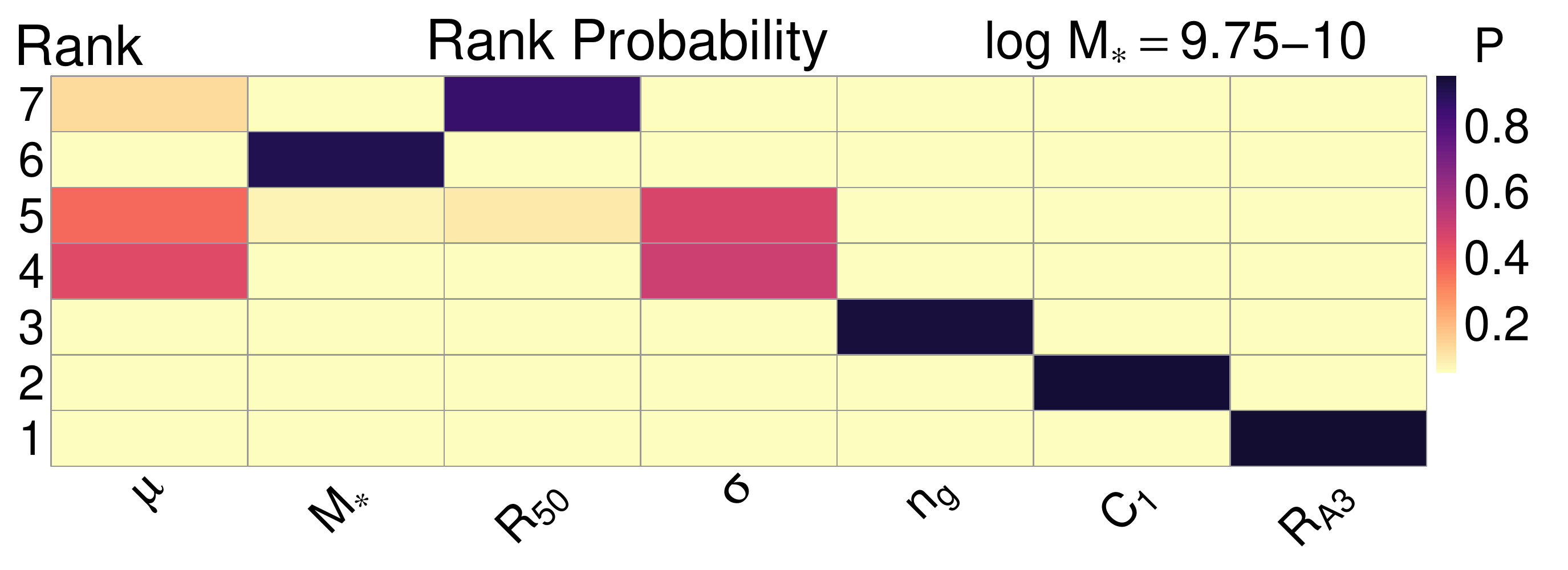}
\hfill
\includegraphics[width=0.47\linewidth]{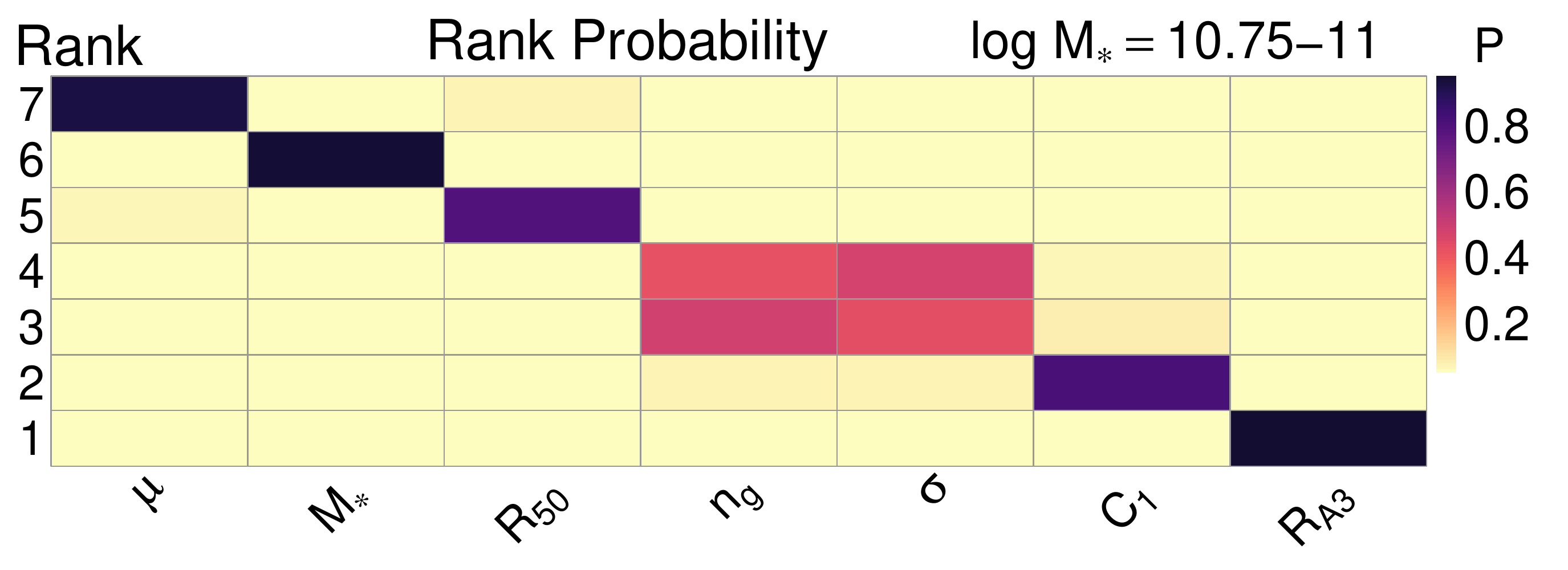}
\caption{Similar to Figure~\ref{fig:score_SFMz012} but for the backward variable selection. Here the important variables are found at the bottom. $R_{A3}$ is still the most important variable. The ranks for the other variables slightly change from those of the forward selection. \label{fig:score_backward}}
\end{figure*}

\begin{figure*}
\includegraphics[width=0.47\linewidth]{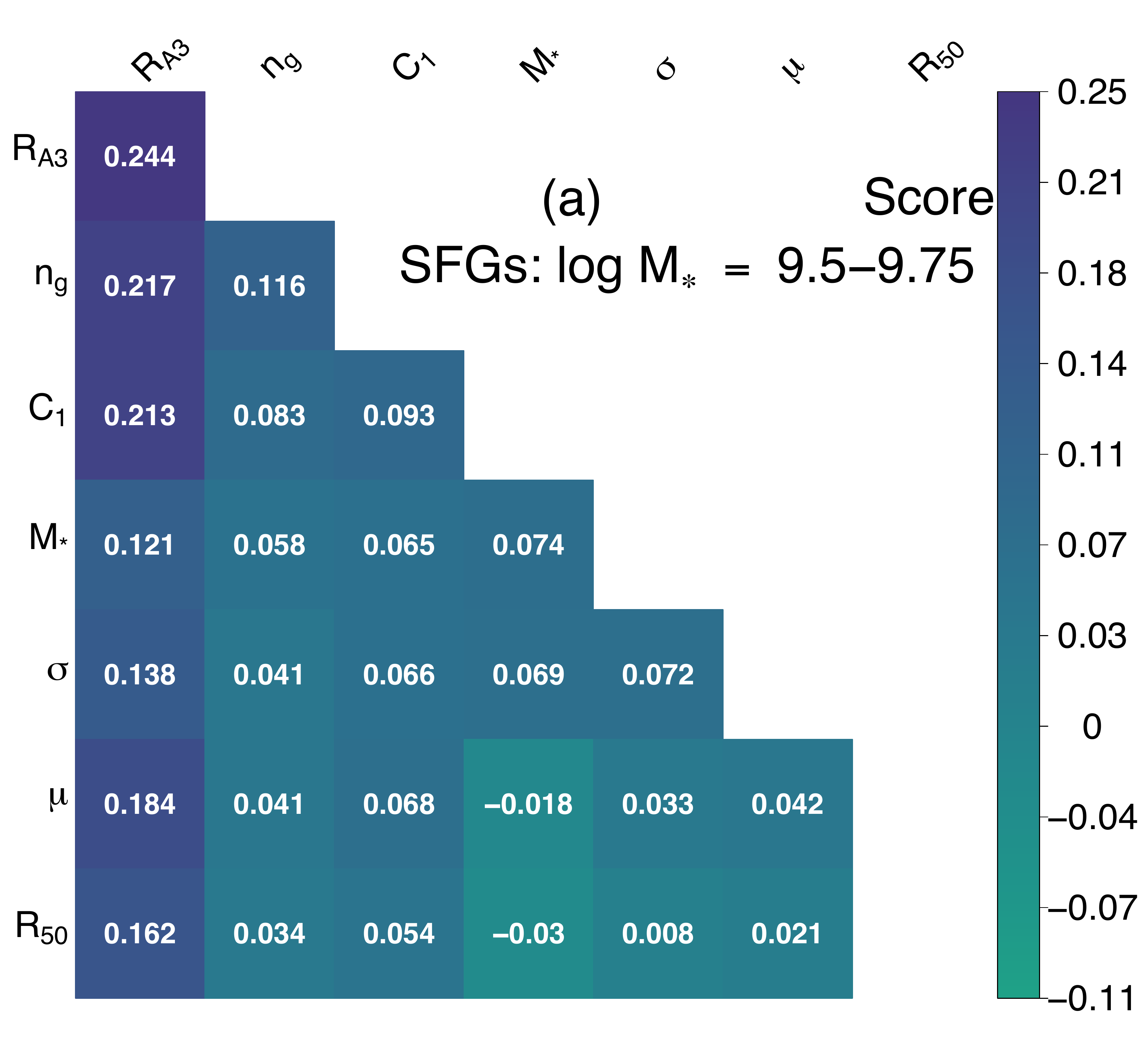}
\includegraphics[width=0.47\linewidth]{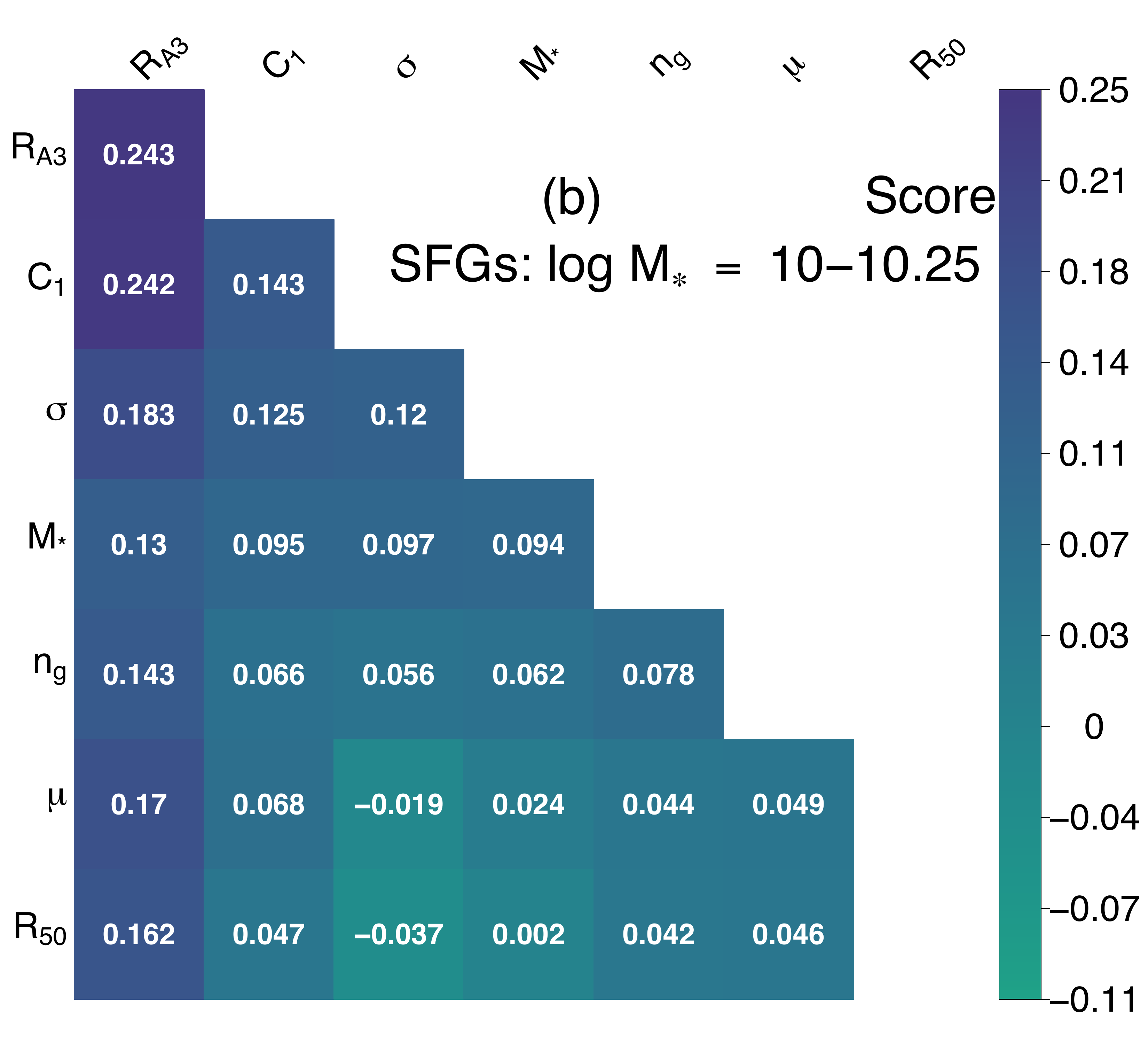}
\includegraphics[width=0.47\linewidth]{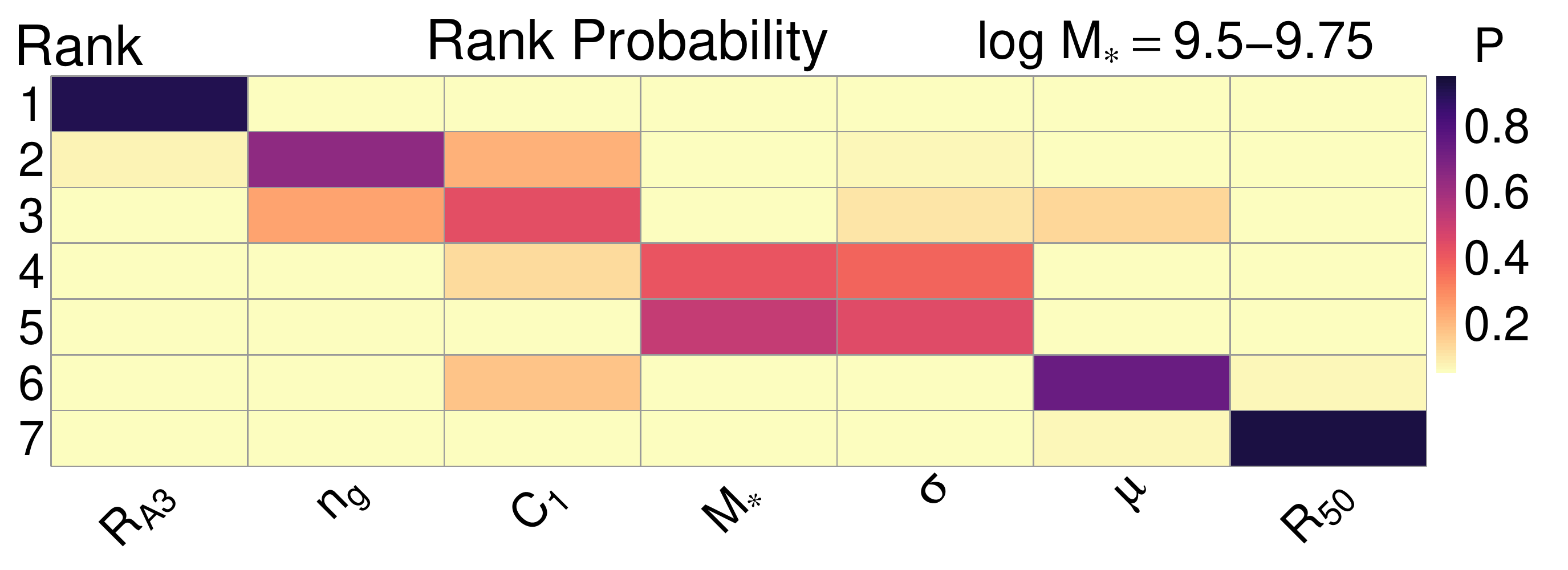}
\includegraphics[width=0.47\linewidth]{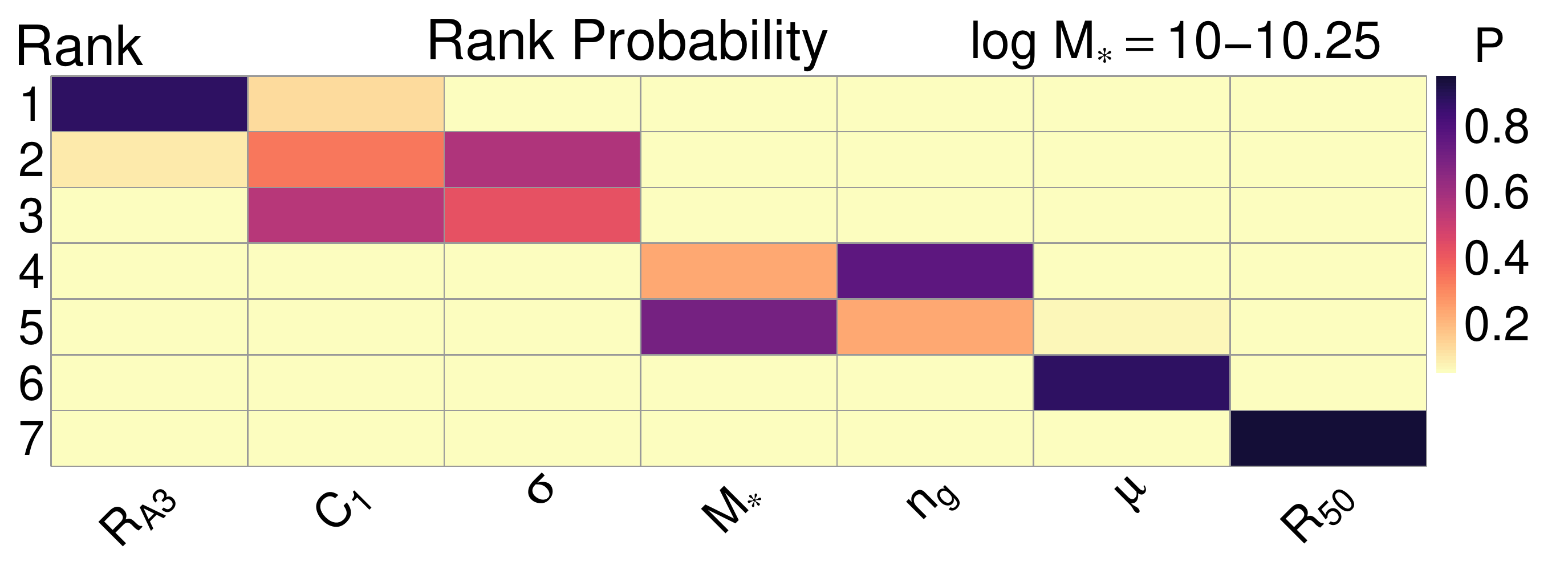}
\includegraphics[width=0.47\linewidth]{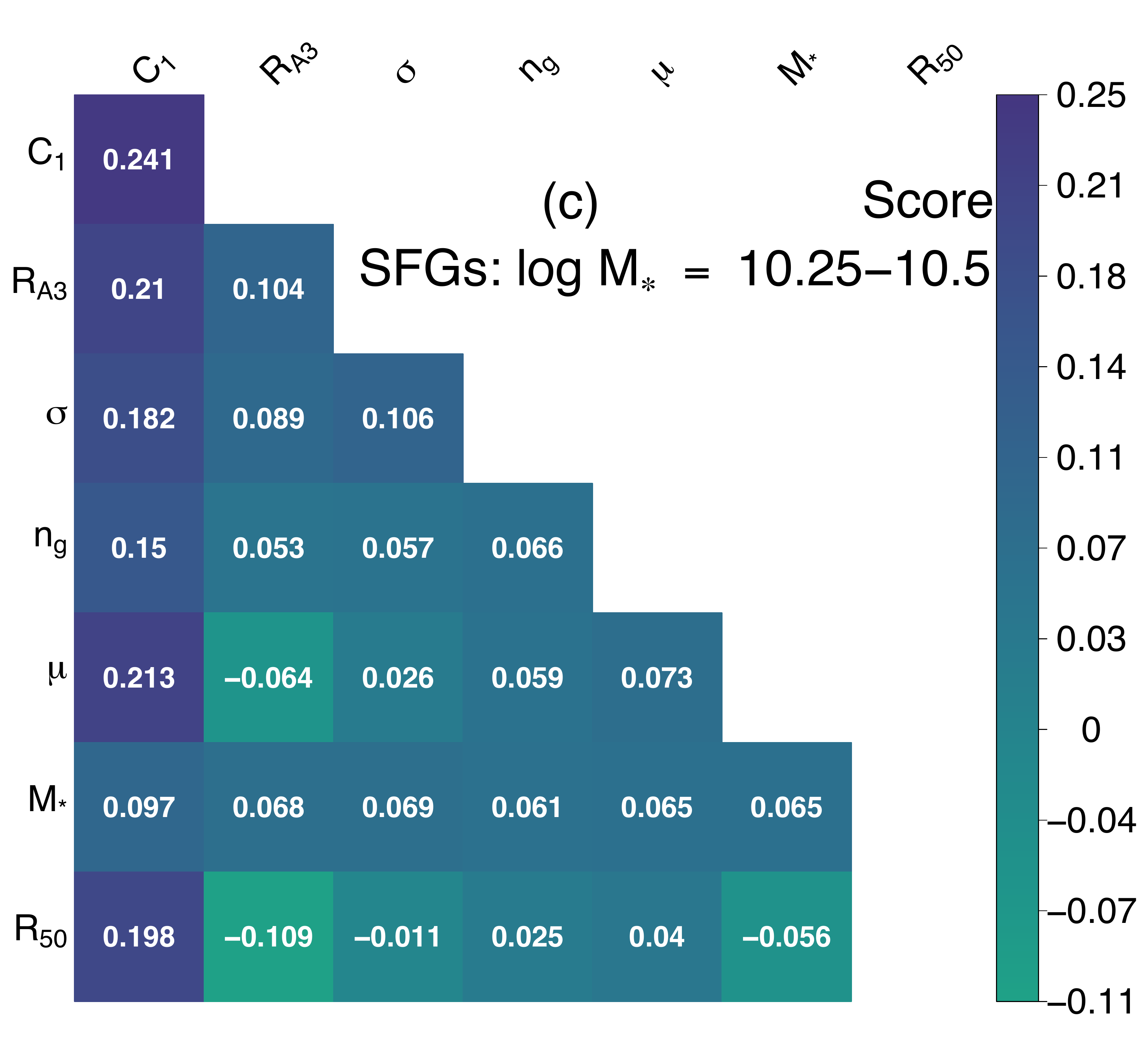}
\includegraphics[width=0.47\linewidth]{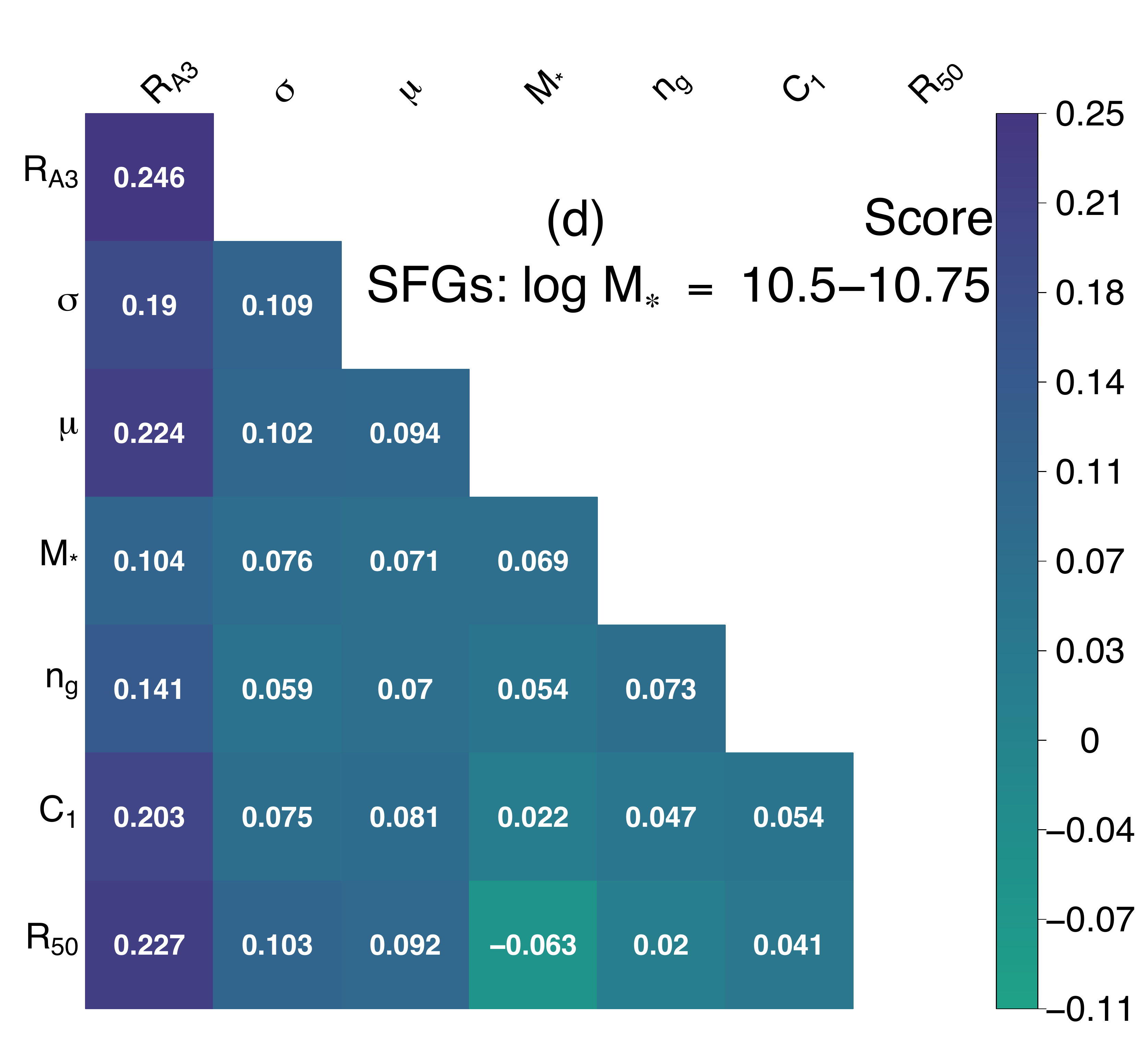}
\includegraphics[width=0.47\linewidth]{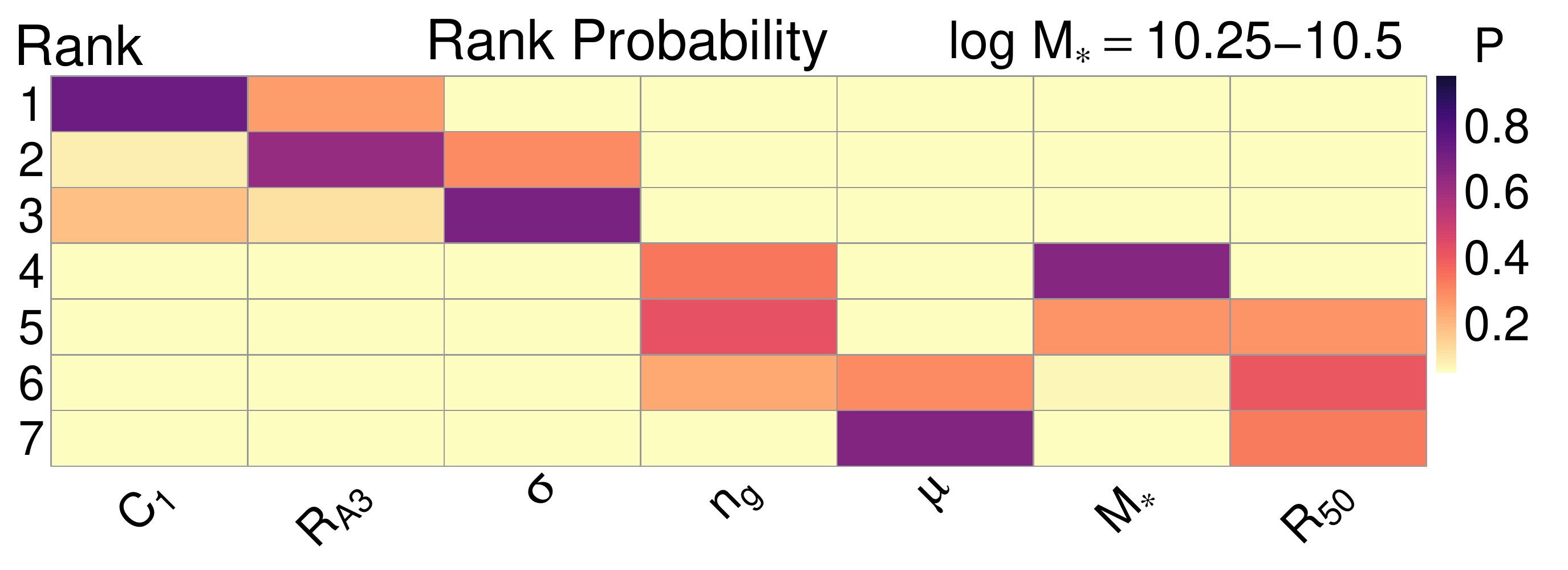}
\hfill
\includegraphics[width=0.45\linewidth]{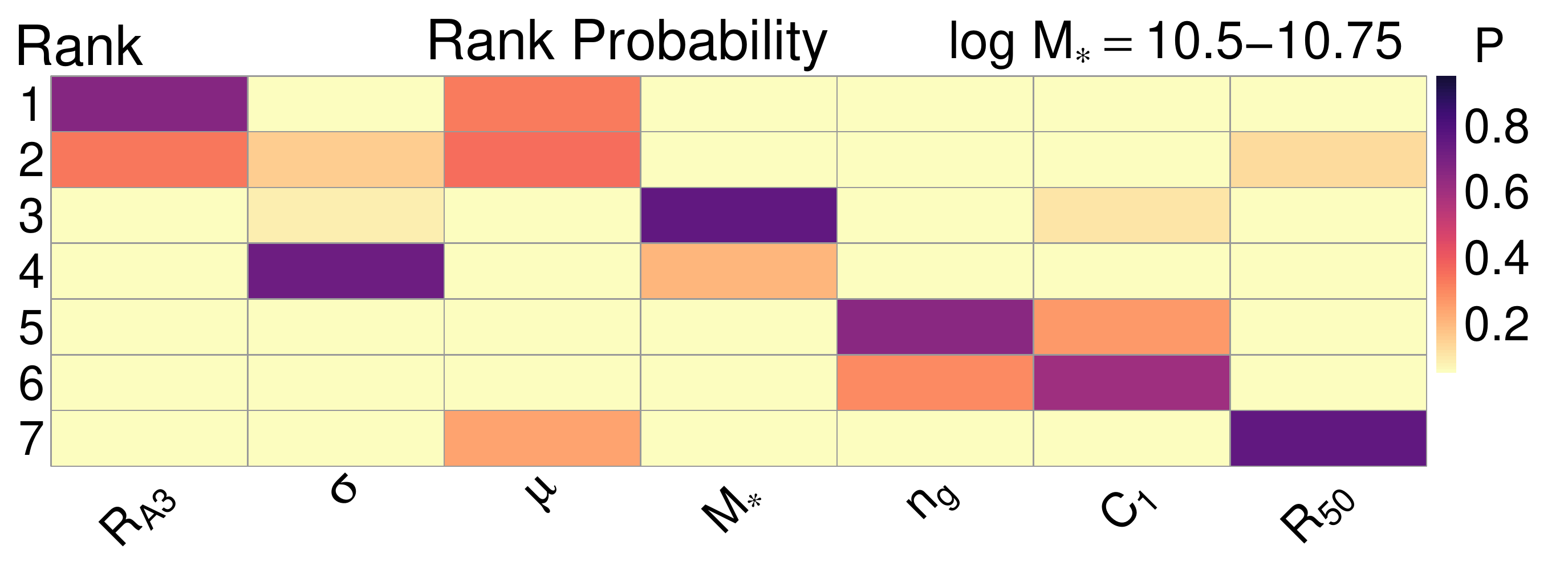}
\caption{Similar to Figure~\ref{fig:score_SFMz012} but for different $M_\star$ ranges. $R_{A3}$ is the highest ranked variable except in panel (c). \label{fig:score_SFMz012App}}
\end{figure*}

In the main text, Figure~\ref{fig:score_SFMz012} presented ranks and scores of variables for predicting \delSSFR for only two narrow mass ranges and using the forward variable selection. Figure~\ref{fig:score_backward} shows a similar figure for the backward selection for the two mass ranges. The backward selection prunes the full set of variables by minimizing equation~\ref{eq:score} and  gives similar results as the forward selection. Furthermore, Figure~\ref{fig:score_SFMz012App} similarly presents the results for additional four stellar mass ranges for the forward selection. $R_{A3}$ is the highest ranked variable in the three of the four narrow mass subsamples presented Figure~\ref{fig:score_SFMz012App}. In general, the bulge-related parameters $C_1$, $\sigma$, or $n_g$ are among the top three variables.

\begin{figure*}
\includegraphics[width=0.98\linewidth]{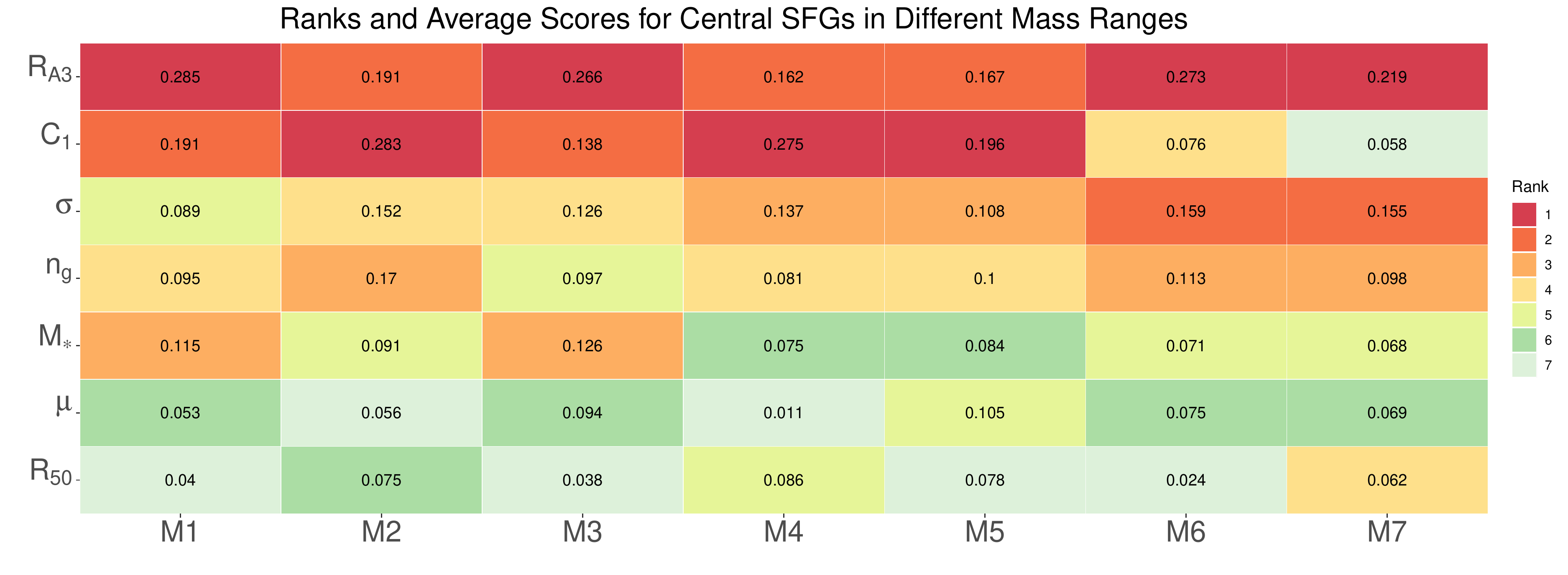}
\caption{The ranks of the variables for predicting \delSSFR. Similar to Figure~\ref{fig:rankM}, but for central galaxies only. The stellar mass ranges from M1:\,$\log\,(M_\star/M_\odot) = 9.5-9.75$ to M7:\,$\log\,(M_\star/M_\odot) = 11-11.25$ with increases mass of 0.25\,dex. \label{fig:rankMcent}}
\end{figure*}

\begin{figure*}
\includegraphics[width=0.8\linewidth]{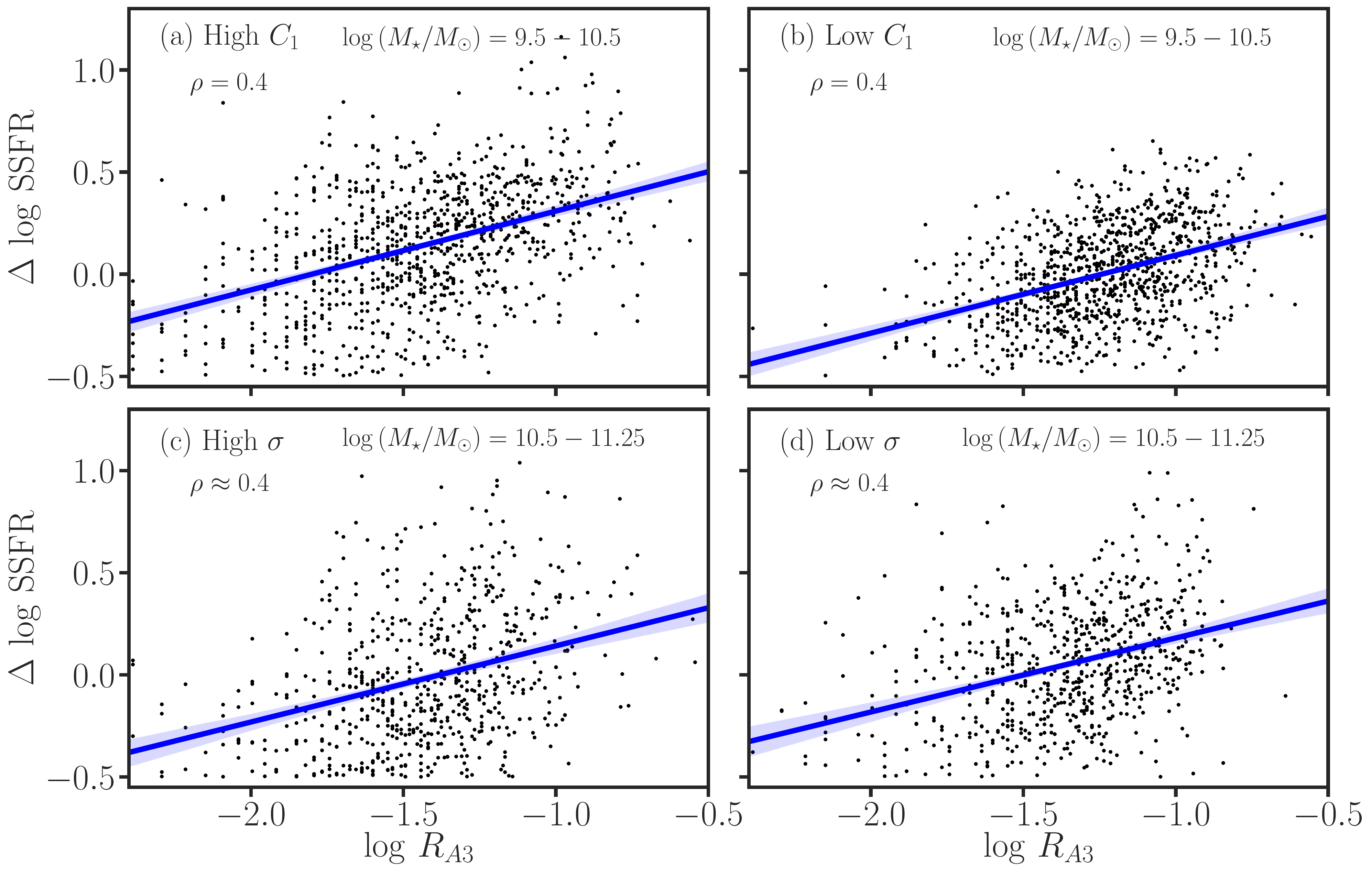}
\caption{The relationship between $R_{A3}$ and \delSSFR for SFGs binned by $M_\star$ and $C_1$ or $\sigma$. All subsamples show a correlation between  $R_{A3}$ and \delSSFR $(\rho \approx 0.4)$, but SFGs with high $C_1 > 0.3$ or high $\sigma > 110$\,km\,s$^{-1}$ have higher \delSSFR than those with low $C_1$ or low $\sigma$ in a given $M_\star$ range. \label{fig:RA3_delSSFR}}
\end{figure*}

\begin{figure*}
\includegraphics[width=0.8\linewidth]{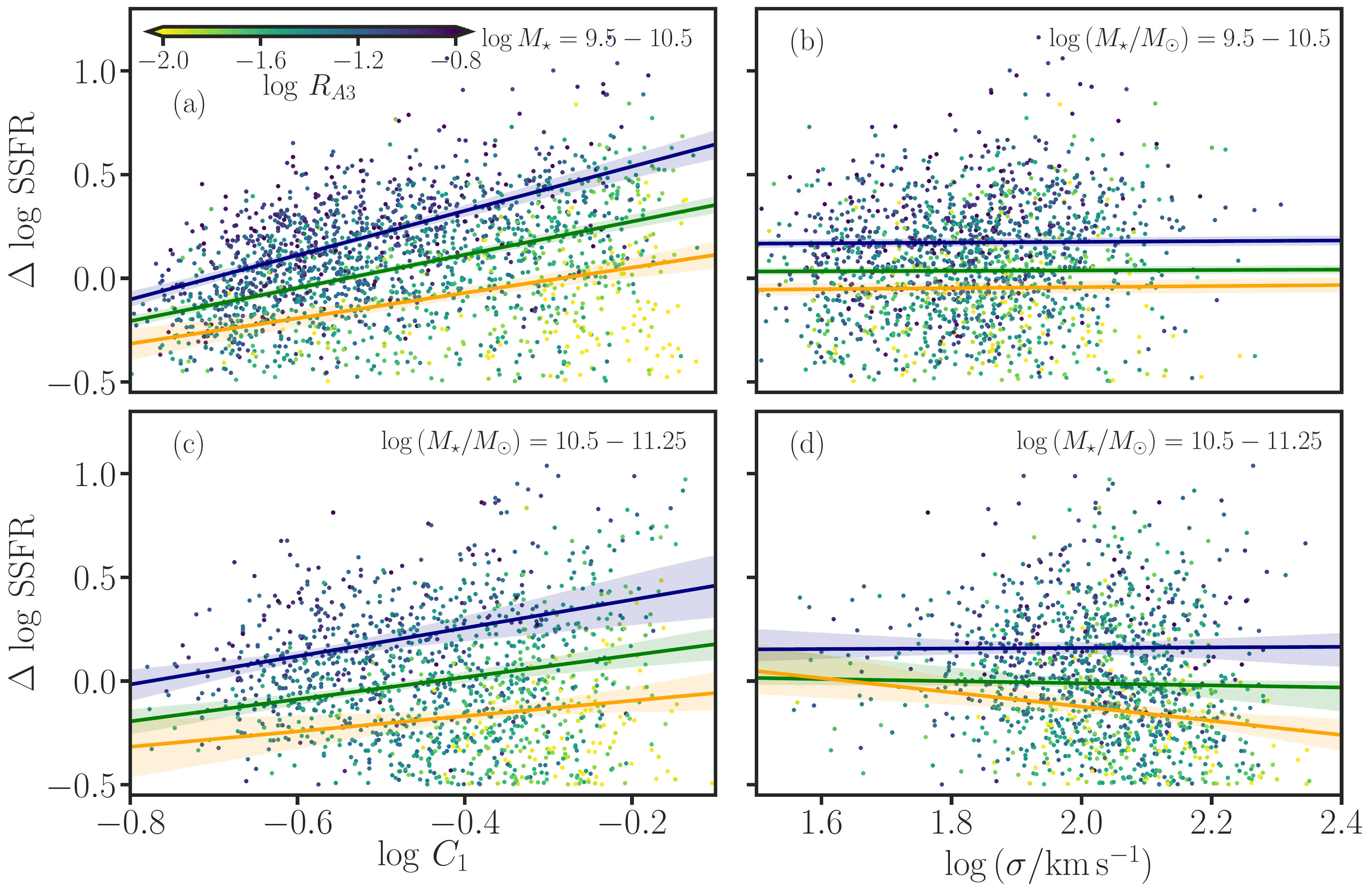}
\caption{The relationship between \delSSFR and $C_1$ or $\sigma$, for SFGs binned by $M_\star$ and color-coded by $R_{A3}$. At the a given $M_\star$ range and $R_{A3}$, there is a significant dependence on $C_1$ or $\sigma$. We plotted linear regression fits for the illustrative purpose; some of the relationships are non-linear and heteroscedastic (variable dispersion with $C_1$ or $\sigma$). \label{fig:C1sig_delSSFR}}
\end{figure*}

Similar to Figure~\ref{fig:rankM}, Figure~\ref{fig:rankMcent} shows the summary of the ranks of the variables as a function of $M_\star$, now excluding satellite SFGs. We use the satellite probability, $P_\mathrm{sat} < 0.5$, from the catalog by \citet{Tinker+20}. The results in the two figures are similar. We defer detailed analysis of the environmental effects on \delSSFR for future work. Figures~\ref{fig:RA3_delSSFR} \& ~\ref{fig:C1sig_delSSFR} further visualize the trends of \delSSFR with $R_{A3}$, $C_1$, and $\sigma$. Using a simple linear regression, we show clearly that \delSSFR depends on these three variables.

\end{document}